\documentclass[11pt,a4paper]{article} 
\pdfoutput=1

\usepackage{jheppub}                  

\usepackage{graphicx}
\usepackage[T1]{fontenc}
\usepackage[normalem]{ulem}	

\def\lsim{\raise0.3ex\hbox{$\;<$\kern-0.75em\raise-1.1ex
\hbox{$\sim\;$}}}
\def\gsim{\raise0.3ex\hbox{$\;>$\kern-0.75em\raise-1.1ex
\hbox{$\sim\;$}}}

\def\thetitle{ 
Physics of parameter correlations around the solar-scale enhancement in neutrino theory with unitarity violation \\
}
\title{\thetitle}
\hypersetup{pdftitle={\thetitle}}
\author{Ivan Martinez-Soler$^{a,b,c}$}
\author{Hisakazu Minakata$^{d}$}

\affiliation{
  $^a$Theoretical Physics Department, Fermi National Accelerator Laboratory, P.O. Box 500, Batavia IL 60510, USA \\
  $^b$Department of Physics and Astronomy, Northwestern University, Evanston, IL 60208, USA \\
  $^c$Colegio de F\'isica Fundamental e Interdisciplinaria de las Am\'ericas (COFI), 254 Norzagaray street, San Juan, Puerto Rico 00901 \\
  $^d$Center for Neutrino Physics, Department of Physics, Virginia Tech, Blacksburg, Virginia 24061, USA \\
}

\emailAdd{ivan.martinezsoler@northwestern.edu}
\emailAdd{minakata71@vt.edu}
\date{\today}

\abstract{
We discuss physics of the three neutrino flavor transformation with non-unitary mixing matrix, with particular attention to the correlation between the $\nu$SM- and the $\alpha$ parameters which represent effect of unitarity violating (UV) new physics. Toward the goal, a new perturbative framework is created to illuminate the effect of non-unitarity in region of the solar-scale enhanced oscillations. We refute the skepticism about the physical reality of the $\nu$SM CP $\delta$ - $\alpha$ parameter phase correlation by analysis with the SOL convention of $U_{\text{\tiny MNS}}$ in which $e^{ \pm i \delta}$ is attached to $s_{12}$. Then, a comparative study between the solar- and atmospheric-scale oscillation regions allowed by the framework reveals a dynamical $\delta-$(blobs of the $\alpha$ parameters) correlation in the solar oscillation region, in sharp contrast to the ``chiral'' type phase correlation $[e^{- i \delta } \bar{\alpha}_{\mu e}, e^{ - i \delta} \bar{\alpha}_{\tau e}, \bar{\alpha}_{\tau \mu}]$ in the PDG convention seen in the atmospheric oscillation region. An explicit perturbative calculation to first order in the $\nu_{\mu} \rightarrow \nu_{e}$ channel allows us to decompose the UV related part of the probability into the unitary evolution part and the genuine non-unitary part. We observe that the effect of non-unitarity tends to cancel between these two parts, as well as between the different $\alpha_{\beta \gamma}$ parameters.

}

\begin{document} 

\begin{flushright}
NUHEP-TH/19-09, 
FERMILAB-PUB-19-397-T
\end{flushright}

\maketitle

\section{Introduction}
\label{sec:introduction}

The discovery of neutrino oscillation and hence neutrino mass \cite{Kajita:2016cak,McDonald:2016ixn} under the framework of three-generation lepton flavor mixing \cite{Maki:1962mu} created a new field of research in particle physics. It led to construction of the next-generation accelerator and underground experiments with the massive detectors, Hyper-Kamiokande~\cite{Abe:2018uyc} and DUNE \cite{Abi:2020evt}. They are going to establish CP violation due to the lepton Kobayashi-Maskawa phase \cite{Kobayashi:1973fv}, possible lepton counterpart of the quark CP violation \cite{Christenson:1964fg}. They will also determine the neutrino mass ordering at high confidence level by utilizing the earth matter effect \cite{Wolfenstein:1977ue,Mikheev:1986gs}. Of course, the flagship projects will be challenged by the ongoing \cite{Abe:2017aap,Abe:2019vii,Acero:2019ksn} and the other upcoming experiments, for example, ESS$\nu$SB \cite{Baussan:2013zcy}, JUNO \cite{An:2015jdp}, T2KK\footnote{
A possible acronym for the setting, ``Tokai-to-Kamioka observatory-Korea neutrino observatory'', updated from the one used in ref.~\cite{Kajita:2006bt}.} \cite{Abe:2016ero}, INO \cite{Kumar:2017sdq}, IceCube-Gen2/PINGU \cite{TheIceCube-Gen2:2016cap}, and KM3NeT/ORCA \cite{Adrian-Martinez:2016zzs}, which compete for the same goals. 

Toward establishing the three-flavor mixing scheme, in particular in the absence of confirmed anomaly beyond the neutrino-mass embedded Standard Model ($\nu$SM),\footnote{
For possible candidates of the anomalies which suggest physics beyond the $\nu$SM see e.g., ref.~\cite{Diaz:2019fwt}.
}
one of the most important topics in the future would be the high-precision paradigm test. In this context, leptonic unitarity test, either by closing the unitarity triangle \cite{Farzan:2002ct}, or by an alternative method of constraining the models of unitarity violation (UV)\footnote{
It is appropriate to mention that in the physics literature UV usually means ``ultraviolet''. But, in this paper UV is used as an abbreviation for ``unitarity violation'' or ``unitarity violating''.
} 
at high-energy \cite{Antusch:2006vwa,Escrihuela:2015wra}, or at low-energy scales \cite{Fong:2016yyh,Fong:2017gke,Blennow:2016jkn} are extensively discussed. It includes the subsequent developments, for example in~\cite{FernandezMartinez:2007ms,Goswami:2008mi,Antusch:2009pm,Antusch:2009gn,Antusch:2014woa,Ge:2016xya,Fernandez-Martinez:2016lgt,Dutta:2016vcc,Escrihuela:2016ube}. A summary of the current constraints on UV is given e.g., in refs.~\cite{Blennow:2016jkn,Parke:2015goa}.

It was observed that in the $3 \times 3$ active neutrino subspace the evolution of the system can be formulated in the same footing in low-scale as well as high-scale UV scenarios \cite{Fong:2017gke,Blennow:2016jkn}. Nonetheless, dynamics of the three neutrino system with non-unitary mixing in matter has not been investigated in a sufficient depth. Apart from numerically implemented calculation done in some of the aforementioned references, only a very limited effort was devoted for analytical understanding of the system so for. It has a sharp contrast to the fact that great amount of efforts were devoted to understand the three-flavor neutrino oscillation.\footnote{
Here, we give a cautious remark that when the term ``neutrino oscillation'' is used in this paper, or often in many other literatures, it may imply not only the original meaning, but also something beyond such as ``neutrino flavor transformation'', or ``neutrino flavor conversion'', depending upon the contexts.
}
A general result known to us so far is the exact $S$ matrix with non-unitarity in matter with constant density \cite{Fong:2017gke} calculated by using the KTY-type construction \cite{Kimura:2002wd}. It allows us to obtain the exact expression of the oscillation probability with non-unitarity. 

In a previous paper \cite{Martinez-Soler:2018lcy}, we have started a systematic investigation of analytic structure of the three neutrino evolution in matter with non-unitarity. 
We have used so called the $\alpha$ parametrization \cite{Escrihuela:2015wra} to implement non-unitarity in the three neutrino system. Using a perturbative framework dubbed as the ``helio-UV perturbation theory'' (a UV extended version of~\cite{Minakata:2015gra}) with the two kind of expansion parameters, the helio-to-terrestial ratio $\epsilon \approx \Delta m^2_{21} / \Delta m^2_{31}$ and the $\alpha$ parameters, we computed the oscillation probability valid to first-order in the expansion parameters. The region of validity of the perturbative framework spans the one around the atmospheric-scale enhanced oscillations which covers the relevant region for the ongoing and the next generation long-baseline (LBL) accelerator neutrino oscillation experiments. Possibility of application to the data from the near future facilities and the currently almost un-understood properties of the system may justify the examination even though it is to first order in expansion. 

To our view, the most significant observation in ref.~\cite{Martinez-Soler:2018lcy} is that the $\nu$SM CP phase $\delta$ and the complex $\alpha$ parameters have an the intriguing phase correlation of the form $[e^{- i \delta } \bar{\alpha}_{\mu e}, e^{ - i \delta} \bar{\alpha}_{\tau e}, \bar{\alpha}_{\tau \mu}]$ in the Particle Data Group (PDG) convention of $U_{\text{\tiny MNS}}$ \cite{Tanabashi:2018oca}. What is unique in the phase correlation is that it universally holds in all the oscillation channels as well as unitary and non-unitary parts of the oscillation probability. One should note that the definition of the $\alpha$ parameters, and consequently the precise form of the correlation between the CP phases, depends on the phase convention of the lepton flavor mixing MNS matrix.\footnote{
In the ATM phase convention of $U_{\text{\tiny MNS}}$ in which $e^{ \pm i \delta}$ is attached to $s_{23}$, the phase correlation takes the form $[e^{- i \delta } \alpha_{\mu e}, \alpha_{\tau e}, e^{i \delta} \alpha_{\tau \mu}]$.
}

A puzzling feature of the $\delta$ - $\alpha$ parameter phase correlation in ref.~\cite{Martinez-Soler:2018lcy} is that it disappears in the SOL convention of $U_{\text{\tiny MNS}}$ in which $e^{ \pm i \delta}$ is attached to $s_{12}$. It triggered a skepticism of the nature of phase correlation, which may allow the following two alternative interpretations: 
\begin{enumerate}

\item
Existence of the SOL phase convention of $U_{\text{\tiny MNS}}$ in which $\delta$ and $\alpha$ phase correlation is absent implies that the CP phase correlation is not physical, but an artifact of inadequate choice of $U_{\text{\tiny MNS}}$ phase convention. 

\item
Physics must be $U_{\text{\tiny MNS}}$ convention independent. In all the other convention of $U_{\text{\tiny MNS}}$ except for the SOL, there exists $\delta-\alpha$ parameter phase correlation. Therefore, the existence of phase correlation is generic and it must be physical. 

\end{enumerate}
\noindent
If the interpretation 1 and the reasoning behind it are correct, $\delta$ and $\alpha$ phase correlation must be absent under the SOL convention of $U_{\text{\tiny MNS}}$ everywhere in the allowed kinematical regions. Conversely, if we see a non-vanishing phase correlation in the oscillation probability calculated with the SOL convention somewhere, it implies that the interpretation 1 cannot be true. We will show throughout this paper that the interpretation 2 holds by investigation of the system in region of the solar-scale enhanced oscillation. 

\section{The goal of this paper by itself, and in combining a companion work~\cite{Martinez-Soler:2018lcy}} 
\label{sec:purpose}

In this paper, we discuss physics of neutrino flavor transformation in region of solar-scale enhanced oscillation.\footnote{
The feature of merely replacing the atmospheric oscillation to the solar one may trigger the question to us: ``Are you attempting another experiment replacing copper with iron?''. At this stage, we would like to say that a mere change in the field of exercise brings new insights to us because the system is so rich in dynamics with the extra nine UV parameters introduced into the $\nu$SM system. In sections~~\ref{sec:correlation} and~\ref{sec:numerical-examination}, the readers will see our clear-cut full answer to this question. }
We will try to achieve the two goals:

\begin{itemize}

\item 
To examine the system of the three-flavor neutrinos in the SOL convention ($e^{ \pm i \delta}$ attached to $s_{12}$) of $U_{\text{\tiny MNS}}$ in region of the enhanced solar oscillation, which will testify for physical reality of the correlation between $\nu$SM phase - UV $\alpha$ parameter phases.

\item 
To understand the $\nu$SM - UV parameter correlation in more generic context and in wider kinematical region by combining the results of this and the previous works ~\cite{Martinez-Soler:2018lcy}. 

\end{itemize}

A few words for the examination of the ``solar region'' are ready. We feel that an immense need exists for the real understanding of parameter correlation in theories with non-unitarity, in particular outside the region investigated in ref.~\cite{Martinez-Soler:2018lcy}. The natural ``field of research'' for this purpose is the region of solar-scale enhanced oscillation, the unique place for enhancement other than the atmospheric one in our world of the three generation leptons. The feature can be seen clearly in the ``terrestrial-friendly'' region of $E$ vs. $L$ plot e.g., in refs.~\cite{Martinez-Soler:2019nhb,Minakata:2019gyw} whose latter also serves for a brief summary of recent activities on atmospheric neutrinos at low energies. We note that it has been the target of investigation for a long time, see e.g., \cite{Peres:2003wd,Peres:2009xe,Akhmedov:2008qt,Razzaque:2014vba}, and possibly others that we may miss, mainly in the context of atmospheric neutrino observation at low energies. It should also be mentioned that this topic is now receiving the renewed interest \cite{Minakata:2019gyw,Martinez-Soler:2019nhb} given the new possibilities of gigantic detectors such as JUNO \cite{Settanta:2019ecp}, DUNE \cite{Kelly:2019itm}, and Hyper-K \cite{Abe:2015zbg}. 
Thus, the second goal of this paper is to achieve a deeper understanding of parameter correlation by combining knowledges in regions of the atmospheric-scale and the solar-scale enhanced oscillations.

Very recently, we have formulated a perturbative framework in the $\nu$SM, dubbed as the ``solar resonance perturbation theory'' \cite{Martinez-Soler:2019nhb}, whose validity is around the very region of our interest. We extend this perturbative framework to include the effect of UV, by treating the $\alpha$ parameters as the additional expansion parameters. Using the framework, we investigate dynamics of the three neutrino evolution with non-unitary mixing matrix under the constant matter density approximation, with particular attention to the parameter correlation. We will show that the system displays a rich, new phenomenon of clustering of the $\nu$SM and the UV variables. 

Nonetheless, we find it not sufficient to rely on analytic treatment based on perturbation theory to extract the characteristic feature of the system due to a new and intricate feature of the parameter correlation. For this reason we rely also on exact numerical analyses as well as the perturbative formula we derive in this paper to elucidate physics of the parameter correlation in region of enhanced solar-scale oscillation. It will be particularly illuminating when our analysis is done in a style of comparative study between the solar- and atmospheric-scale oscillation regions, as will be done in section~\ref{sec:numerical-examination}. We hope that such understanding will eventually help analyzing data for leptonic unitarity test.  

In section~\ref{sec:parameter-correlation}, we introduce the concept of parameter correlations by describing a pedagogical example of the three-neutrino system with the non-standard interactions (NSI). In section~\ref{sec:formulation-solar-res-Ptheory}, we give a step-by-step formulation of the perturbative framework which to be utilized in analyzing features of the three-neutrino evolution with non-unitary mixing matrix. The prescription for computing $S$ matrix elements is given with the help of the tilde basis $\tilde{S}$ matrix elements summarized in appendix~\ref{sec:tilde-S-summary}. In section~\ref{sec:general-formula-P}, a general formula for the oscillation probability is derived, and applied to computation of the appearance probability in the $\nu_{\mu} \rightarrow \nu_{e}$ channel. This section together with appendix~\ref{sec:Pmue-rest} contains the explicit expression of the oscillation probability in the $\nu_{\mu} \rightarrow \nu_{e}$ channel to first order in expansion parameters. In section~\ref{sec:correlation}, we discuss the characteristic features of the correlation between the $\nu$SM CP phase and UV $\alpha$ parameters in the region of validity of our perturbative framework. In section~\ref{sec:numerical-examination}, physics of neutrino flavor transformation with UV is discussed paying a particular attention to parameter correlation, contrasting between the regions of the solar- and atmospheric-scale enhanced oscillations. In section~\ref{sec:conclusion}, we give the concluding remarks.

\section{Parameter correlation in neutrino oscillation with beyond-$\nu$SM extended settings} 
\label{sec:parameter-correlation}

It may be useful to start the description of this paper by briefly recollecting some known features of parameter correlation in neutrino oscillation, in particular, in an extended setting that includes physics beyond the $\nu$SM. In this context, a general framework that is most frequently discussed is the one which includes the neutrinos' non-standard interactions (NSI) \cite{Wolfenstein:1977ue}
\begin{eqnarray} 
H_{ \text{NSI} } = 
\frac{a}{2E} 
\left[
\begin{array}{ccc}
\varepsilon_{e e} & \varepsilon_{e \mu}  & \varepsilon_{e \tau}  \\
\varepsilon_{e \mu}^* & \varepsilon_{\mu \mu}  & \varepsilon_{\mu \tau}  \\
\varepsilon_{e \tau}^* & \varepsilon_{\mu \tau}^*  & \varepsilon_{\tau \tau} 
\end{array}
\right], 
\label{NSI-Hamiltonian}
\end{eqnarray}
in the flavor basis Hamiltonian, where the $\varepsilon$ parameters describe flavor dependent strengths of NSI and $a$ denotes the matter potential, see \eqref{matt-potential}. We discuss only so called the ``propagation NSI''. For a review of physics of NSI in wider contexts, see e.g., refs.~\cite{Ohlsson:2012kf,Miranda:2015dra,Farzan:2017xzy}. We note that inclusion of the NSI Hamiltonian \eqref{NSI-Hamiltonian} brings the extra nine parameters into the $\nu$SM Hamiltonian with six degrees of freedom, the two $\Delta m^2$, the three mixing angles, and the unique CP phase, under the influence of the matter potential background. 

\subsection{Emergence of collective variables involving $\nu$SM and NSI parameters} 
\label{sec:collective-variables}

With more than doubled, a large number of the parameters, it is conceivable that dynamics of neutrino oscillation naturally involves rich correlations among these variables.\footnote{
They include the correlations between the NSI variables themselves. The examples include the $\varepsilon_{e e}$ - $\varepsilon_{e \tau}$ - $\varepsilon_{\tau \tau}$ correlation discussed in refs.~\cite{Friedland:2004ah,Friedland:2005vy}.  }
Here, we discuss only a particular type of correlation uncovered in ref.~\cite{Kikuchi:2008vq} because, we believe, it illuminates the point. In this reference, the authors formulated a perturbative framework of the system with NSI by using the three (the latter two assumed to be) small expansion parameters, $\epsilon \equiv \Delta m^2_{21} / \Delta m^2_{31}$, $s_{13} \equiv \sin \theta_{13}$, and the $\varepsilon$ parameters. 
They derived the formulas of the oscillation probability to second order (third order in $\nu_{\mu} \rightarrow \nu_{e}$ channel) in the expansion parameters, which is nothing but an extension of the Cervera {\it et.~al.} formulas \cite{Cervera:2000kp}\footnote{
The Cervera {\it et.~al.} formula is the most commonly used probability formula in the standard three flavor mixing in matter for many purposes, e.g., in the discussion of parameter degeneracy \cite{BurguetCastell:2001ez,Barger:2001yr,Minakata:2002qi}. 
}
to include NSI. In this calculation the PDG convention of $U_{\text{\tiny MNS}}$ \cite{Tanabashi:2018oca} is used.

An interesting and unexpected feature of the NSI-extended formulas is the emergence of the two sets of ``collective variables''
\begin{eqnarray} 
\Theta_{13} &\equiv& 
s_{13} \frac{ \Delta m^2_{31} }{a} 
+ e^{ i \delta} 
\left( s_{23} \varepsilon_{e \mu} + c_{23}  \varepsilon_{e \tau} \right), 
\nonumber \\ 
\Theta_{12} &\equiv& 
\left( c_{12} s_{12} \frac{ \Delta m^2_{21} }{a} 
+ c_{23} \varepsilon_{e \mu} - s_{23} \varepsilon_{e \tau} 
\right) e^{i \delta}, 
\label{Theta-Xi-def}
\end{eqnarray}
where an overall $e^{- i \delta}$ is factored out from the matrix element $S_{e \mu}$ to make the $s_{13}$ term $\delta$ free through which $e^{ i \delta}$ dependences in $\Theta_{12}$ in eq.~\eqref{Theta-Xi-def} results. 
That is, if we replace $s_{13} \frac{ \Delta m^2_{31} }{a}$ and $c_{12} s_{12} \frac{ \Delta m^2_{21} }{a}$ in the original formulas by $\Theta_{13}$ and $\Theta_{12}$, respectively, the extended second-order formulas with full inclusion of NSI effects automatically appear~\cite{Kikuchi:2008vq}. In fact, the procedure works for the third order formula for $P(\nu_{\mu} \rightarrow \nu_{e})$ as well. We note that the second order computation of ref.~\cite{Kikuchi:2008vq} includes the $\nu_{\mu} - \nu_{\tau}$ sector, and the additional corrective variables are identified. But, for simplicity, we do not discuss them here and refer the interested readers ref.~\cite{Kikuchi:2008vq}. 

Appearance of the cluster variables composed of the $\nu$SM and NSI parameters in \eqref{Theta-Xi-def} implies that there exists strong correlations between the $\nu$SM variables $s_{13}$ - $\delta$ and the NSI $\varepsilon_{e \mu}$ - $\varepsilon_{e \tau}$ parameters in such a way that they form the collective variable $\Theta_{13}$ to convert the Cervera {\it et.~al.} formula to the NSI-extended version. The similar statement can be made for the other cluster variable $\Theta_{12}$ as well. The NSI-extended second order formula derived in this way serves for understanding the $s_{13}$ - $\varepsilon_{e \mu}$ confusion uncovered in ref.~\cite{Huber:2002bi} in a more complete manner in such a way that the effects of $\varepsilon_{e \tau}$ and CP phase $\delta$ are also included. It also predicts occurrence of the similar correlation among the variables to produce the collective variable $\Theta_{12}$, whose feature could be confirmed by experiments at low energies, $\frac{ \Delta m^2_{21} }{a} \sim \mathcal{O} (1)$, the possibility revisited recently \cite{Minakata:2019gyw,Martinez-Soler:2019nhb}.

Therefore, there is nothing strange in the parameter correlations among the $\nu$SM and new physics parameters. It appears that the phenomenon arises generically, at least under the environment that the matter effect is comparable to the vacuum effect. 

\subsection{Dynamical nature of the parameter correlation}
\label{sec:dynamical-nature}

We must point out, however, that the features of the parameter correlation depend on the values of the parameters involved, and also on the kinematical region of neutrino energy and baseline with background matter density. Therefore, depending upon the region of validity of the perturbative framework which is used to derive the correlation, the form of parameter correlation changes. We call all these features collectively as the {\em ``dynamical nature''} of the parameter correlation.\footnote{
One must be aware that our terminology of ``dynamical'' correlation may be different from those used in condensed matter physics or many body theory. In our case the correlated parameters are not the dynamical variables in quantum theory and there is no direct interactions between them. 
}

We want to see explicitly whether a change in features of the correlation occurs when the values of the parameters involved are varied, or its effect is incorporated into the framework of perturbation theory. For this purpose let us go back to the collective variable correlation in \eqref{Theta-Xi-def}. We know now the value of $\theta_{13}$ is larger than what was assumed at the time the Cervera {\it et.~al.} formula was derived \cite{Tanabashi:2018oca}. The latest value from Daya Bay is $s_{13} = 0.148$ \cite{Adey:2018zwh}, which is of the order of $\sqrt{\epsilon} = 0.176$. Then, we need higher order corrections of $s_{13}$ up to the fourth order terms to match to the second order accuracy in $\epsilon$ \cite{Minakata:2009sr,Asano:2011nj}. When it is carried out with inclusion of NSI \cite{Asano:2011nj}, it is seen that part of the additional terms generated do not fit to the form of collective variables given in \eqref{Theta-Xi-def}. Therefore, when we make $\theta_{13}$ larger, the parameter correlation which produced the collective variables \eqref{Theta-Xi-def} is started to dissolve. 

Thus, the analysis of this particular example reveals the dynamical nature of the parameter correlation in the neutrino propagation with NSI. 
We expect that overseeing the results of computations of the oscillation probabilities in this and the previous papers \cite{Martinez-Soler:2018lcy} would reveal the similar dynamical behavior of the parameter correlation in the three-flavor neutrino evolution in matter with non-unitary mixing. 

\subsection{Phase correlation through NSI-UV parameter correspondence?}
\label{sec:correspondence}

Can we extract information of the $\delta$ - $\alpha$ parameter phase correlation from the collective variables \eqref{Theta-Xi-def}? The answer is {\it Yes} if we assume a ``uniform chemical composition model'' of the matter. As far as the propagation NSI is concerned there is a one to one mapping between NSI $\varepsilon$ parameters and the UV $\alpha$ parameters, as noticed by Blennow {\it et.~al.} \cite{Blennow:2016jkn} under the assumption $N_{n} = N_{e}$, an equal neutron and proton number densities in charge-neutral medium. Of course, an extension to the more generic case of $N_{e} = r N_{n}$ \cite{Martinez-Soler:2018lcy} can be easily done without altering the conclusion. For the purpose of the present discussion, one also has to ``approve'' the procedure by which the $e^{i \delta}$ dependence of the collective variables \eqref{Theta-Xi-def} is fixed. That is, removing an overall phase from the matrix element $S_{e \mu}$ to make the $s_{13}$ term $\delta$ free, as done in ref.~\cite{Kikuchi:2008vq}. 

Assuming that the two conditions above are met, it leads to the collective variables in \eqref{Theta-Xi-def} written by the UV $\alpha$ parameters, 
\begin{eqnarray} 
\Theta_{13} &=& 
s_{13} \frac{ \Delta m^2_{31} }{a} 
+ \frac{1}{2}  
\left\{ s_{23} \left( \bar{\alpha}_{\mu e} e^{ - i \delta} \right)^* 
+ c_{23} \left( \bar{\alpha}_{\tau e} e^{ - i \delta} \right)^* \right\}, 
\nonumber \\ 
\Theta_{12} &=& 
c_{12} s_{12} e^{i \delta} \frac{ \Delta m^2_{21} }{a} 
+ \frac{1}{2} 
\left\{ c_{23} \left( \bar{\alpha}_{\mu e} e^{ - i \delta} \right)^* 
- s_{23} \left( \bar{\alpha}_{\tau e} e^{ - i \delta} \right)^* \right\}, 
\label{Theta-Xi-UV}
\end{eqnarray}
where we have to use the $\alpha$ parameters defined in the PDG convention of $U_{\text{\tiny MNS}}$. The emerged correlation between $\delta$ and the $\alpha$ parameters is consistent with the canonical phase combination obtained in ref.~\cite{Martinez-Soler:2018lcy} in the PDG convention. For the relationships between the $\alpha$ parameters with the various $U_{\text{\tiny MNS}}$ conventions, see section~\ref{sec:mass-basis}.
It is not unreasonable because the regions of validity of the perturbative frameworks in refs.~\cite{Kikuchi:2008vq} and~\cite{Martinez-Soler:2018lcy} overlaps.

\section{Formulating perturbation theory around the solar-scale enhancement with non-unitarity} 
\label{sec:formulation-solar-res-Ptheory}

Physics discussion in this paper necessitates a new analytical framework to illuminate the effect of non-unitary mixing matrix in region of the solar-scale enhanced oscillations, the UV extended version of the ``solar-resonance perturbation theory'' \cite{Martinez-Soler:2019nhb}. 

\subsection{Neutrino evolution in the vacuum mass eigenstate basis }
\label{sec:mass-basis}

As is customary in our formulation of the three active neutrino evolution in matter with unitarity violation (UV) \cite{Martinez-Soler:2018lcy}, we start from the evolution equation in the vacuum mass eigenstate basis, whose justification is given in refs.~\cite{Fong:2017gke,Blennow:2016jkn}.\footnote{
In a nutshell, the equation \eqref{check-evolution} with \eqref{check-H-def} describes evolution of the active three neutrinos in the $3 \times 3$ sub-space in the $(3+N_{s})$ model (as a model for low-scale UV) \cite{Fong:2016yyh,Fong:2017gke}, or just the three neutrino system in high-scale UV, see e.g.,~\cite{Blennow:2016jkn}.
}
With use of the ``check basis'' for the vacuum mass eigenstate basis, it takes the form of Schr\"odinger equation 
\begin{eqnarray}
i \frac{d}{dx} \check{\nu} = \check{H} \check{\nu} 
\label{check-evolution}
\end{eqnarray}
with Hamiltonian 
\begin{eqnarray} 
\check{H} \equiv 
\frac{1}{2E} 
\left\{  
\left[
\begin{array}{ccc}
0 & 0 & 0 \\
0 & \Delta m^2_{21} & 0 \\
0 & 0 & \Delta m^2_{31} \\
\end{array}
\right] + 
N^{\dagger} \left[
\begin{array}{ccc}
a - b & 0 & 0 \\
0 & -b & 0 \\
0 & 0 & -b \\
\end{array}
\right] N 
\right\} 
\label{check-H-def}
\end{eqnarray}
where $E$ is neutrino energy and $\Delta m^2_{ji} \equiv m^2_{j} - m^2_{i}$. 
A usual phase redefinition of neutrino wave function is done to leave only the mass squared differences. 
$N$ denotes the non-unitary flavor mixing matrix which relates the flavor neutrino states to the vacuum mass eigenstates as 
\begin{eqnarray} 
\nu_{\beta} = N_{\beta i} \check{\nu}_{i}.
\label{N-def}
\end{eqnarray}
where
$\beta$ (and the other Greek indices) runs over $e, \mu, \tau$, while the mass eigenstate index $i$ (and the other Latin indices) runs over $1,2,$ and $3$. It must be noticed that the neutrino evolution described by eq.~\eqref{check-evolution} is unitary, as is obvious from the hermitian Hamiltonian \eqref{check-H-def}. How the apparent inconsistency between the unitary evolution and the non-unitarity of the flavor basis $S$ matrix will be resolved in section~\ref{sec:basis-relations}, one of the points of emphasis in ref.~\cite{Martinez-Soler:2018lcy}. Notice that due to limited number of appropriate symbols the notations for the various basis may not be always the same in our series of papers. 

The functions $a(x)$ and $b(x)$ in (\ref{tilde-H-def}) denote the Wolfenstein matter potential \cite{Wolfenstein:1977ue} due to CC and NC reactions, respectively. 
\begin{eqnarray} 
a &=&  
2 \sqrt{2} G_F N_e E \approx 1.52 \times 10^{-4} \left( \frac{Y_e \rho}{\rm g\,cm^{-3}} \right) \left( \frac{E}{\rm GeV} \right) {\rm eV}^2, 
\nonumber \\
b &=& \sqrt{2} G_F N_n E = \frac{1}{2} \left( \frac{N_n}{N_e} \right) a. 
\label{matt-potential}
\end{eqnarray}
Here, $G_F$ is the Fermi constant, $N_e$ and $N_n$ are the electron and neutron number densities in matter. $\rho$ and $Y_e$ denote, respectively, the matter density and number of electron per nucleon in matter. 
We define the following notations for simplicity to be used in the discussions hereafter in this paper:  
\begin{eqnarray}
\Delta_{ji} \equiv \frac{\Delta m^2_{ji}}{2E},
\hspace{8mm}
\Delta_{a} \equiv \frac{ a }{2E}, 
\hspace{8mm}
\Delta_{b} \equiv \frac{ b }{2E}. 
\label{Delta-def}
\end{eqnarray}

For simplicity and clarity we will work with the uniform matter density approximation in this paper. But, it is not difficult to extend our treatment to varying matter density case if adiabaticity holds.

Throughout this paper, due to the reasoning mentioned in section~\ref{sec:introduction}, we use the SOL convention of the $U_{\text{\tiny MNS}}$ matrix, the standard $3 \times 3$ unitary flavor mixing matrix 
\begin{eqnarray} 
&& U_{\text{\tiny SOL}} 
= \left[
\begin{array}{ccc}
1 & 0 &  0  \\
0 & c_{23} & s_{23} \\
0 & - s_{23} & c_{23} \\
\end{array}
\right] 
\left[
\begin{array}{ccc}
c_{13}  & 0 & s_{13} \\
0 & 1 & 0 \\
- s_{13} & 0 & c_{13} \\
\end{array}
\right] 
\left[
\begin{array}{ccc}
c_{12} & s_{12} e^{ i \delta}  &  0  \\
- s_{12} e^{- i \delta} & c_{12} & 0 \\
0 & 0 & 1 \\
\end{array}
\right] \equiv 
U_{23} U_{13} U_{12}, 
\label{MNS-SOL}
\end{eqnarray}
where we have used the obvious notations $s_{ij} \equiv \sin \theta_{ij}$ etc. and $\delta$ denotes the lepton KM phase \cite{Kobayashi:1973fv}, or the $\nu$SM CP violating phase. 
Our terminology ``SOL''  is because the phase factor $e^{ \pm i \delta}$ is attached to the ``solar angle'' $s_{12}$. It is physically equivalent to the commonly used PDG convention \cite{Tanabashi:2018oca} in which the phase factor is attached to $s_{13}$. 

We use the $\alpha$ parametrization of non-unitary mixing matrix \cite{Escrihuela:2015wra} defined in the $U_{\text{\tiny SOL}}$ convention 
\begin{eqnarray} 
N &=& 
\left( \bf{1} - \tilde{\alpha} \right) U_{\text{\tiny SOL}} = 
\left\{ 
\bf{1} - 
\left[ 
\begin{array}{ccc}
\tilde{\alpha}_{ee} & 0 & 0 \\
\tilde{\alpha}_{\mu e} & \tilde{\alpha}_{\mu \mu}  & 0 \\
\tilde{\alpha}_{\tau e}  & \tilde{\alpha}_{\tau \mu} & \tilde{\alpha}_{\tau \tau} \\
\end{array}
\right] 
\right\}
U_{\text{\tiny SOL}}
\label{alpha-matrix-def}
\end{eqnarray}
As seen in \eqref{alpha-matrix-def}, and discussed in detail in ref.~\cite{Martinez-Soler:2018lcy}, the definition of the $\alpha$ matrix depends on the phase convention of the flavor mixing matrix $U_{\text{\tiny MNS}}$. In consistent with the notation used in ref.~\cite{Martinez-Soler:2018lcy}, we denote the $\alpha$ matrix elements in the SOL convention as $\tilde{\alpha}_{\beta \gamma}$.

The other convention of the MNS matrix which is heavily used in ref.~\cite{Martinez-Soler:2018lcy} is the ``ATM'' convention in which $e^{ \pm i \delta}$ is attached to the ``atmospheric angle'' $s_{23}$: 
\begin{eqnarray}
&& U_{\text{\tiny ATM}} = 
\left[
\begin{array}{ccc}
1 & 0 &  0  \\
0 & c_{23} & s_{23} e^{ i \delta} \\
0 & - s_{23} e^{- i \delta} & c_{23} \\
\end{array}
\right] 
\left[
\begin{array}{ccc}
c_{13}  & 0 &  s_{13} \\
0 & 1 & 0 \\
- s_{13} & 0 & c_{13}  \\
\end{array}
\right] 
\left[
\begin{array}{ccc}
c_{12} & s_{12}  &  0  \\
- s_{12} & c_{12} & 0 \\
0 & 0 & 1 \\
\end{array}
\right]. 
\label{MNS-ATM}
\end{eqnarray}
The $\alpha$ parameters defined in the ATM and PDG conventions of $U_{\text{\tiny MNS}}$ are denoted as $\alpha_{\beta \gamma}$ and 
$\bar{\alpha}_{\beta \gamma}$, respectively, in ref.~\cite{Martinez-Soler:2018lcy}, the notation we follow in this paper. 
Then, we recapitulate here the relationships between the $\alpha$ parameters defined with the PDG ($\bar{\alpha}$), ATM ($\alpha$) and the SOL ($\tilde{\alpha}$) conventions of $U_{\text{\tiny MNS}}$:
\begin{eqnarray} 
&& \tilde{\alpha}_{\mu e} 
= \bar{\alpha}_{\mu e} e^{ - i \delta } = \alpha_{\mu e} e^{ - i \delta }, 
\nonumber \\
&& \tilde{\alpha}_{\tau e} 
= \bar{\alpha}_{\tau e} e^{ - i \delta } = \alpha_{\tau e}, 
\nonumber \\
&& \tilde{\alpha}_{\tau \mu} 
= \bar{\alpha}_{\tau \mu} = \alpha_{\tau \mu} e^{ i \delta }.
\label{alpha-bar-alpha-tilde-alpha}
\end{eqnarray}
where we note that the diagonal $\alpha$ parameters are equal among the three conventions.

\subsection{Region of validity, expansion parameters, and the target sensitivity} 
\label{sec:region-validity} 

In this section, we aim at constructing the perturbative framework which is valid at around the solar oscillation maximum, $\Delta m^2_{21} L / 4 E \sim \mathcal{O} (1)$. Given the formula 
\begin{eqnarray}
\frac{ \Delta m^2_{21} L}{4 E} 
&=&
0.953
\left(\frac{\Delta m^2_{21}}{7.5 \times 10^{-5}\mbox{eV}^2}\right)
\left(\frac{L}{1000 \mbox{km}}\right)
\left(\frac{E}{100 \mbox{MeV}}\right)^{-1}, 
\label{kinematic1}
\end{eqnarray}
it implies neutrino energy $E=( 1 - 5 ) \times 100$ MeV and baseline $L= ( 1 - 10 ) \times 1000$ km. In this region, the matter potential is comparable in size to the vacuum effect represented by $\Delta m^2_{21}$, 
\begin{eqnarray} 
\frac{ a }{ \Delta m^2_{21} } 
&=& 
%
0.609 
\left(\frac{ \Delta m^2_{21} }{ 7.5 \times 10^{-5}~\mbox{eV}^2}\right)^{-1}
\left(\frac{\rho}{3.0 \,\text{g/cm}^3}\right) \left(\frac{E}{200~\mbox{MeV}}\right) 
\sim \mathcal{O} (1).
\label{a/Dm2solar}
\end{eqnarray}
Hence, our perturbative framework must fully take into account the MSW effect caused by the earth matter effect. A more detailed discussion of the region of validity without UV effect is given in ref.~\cite{Martinez-Soler:2019nhb}. 

As in the solar resonance perturbation theory we will have the ``effective'' expansion parameter in the $\nu$SM sector, $A_{ \text{exp} } = c_{13} s_{13} \left( a / \Delta m^2_{31} \right) \sim 10^{-3}$, as will be discussed in section~\ref{sec:effective-exp-parameter}. The reason for having such a very small expansion parameter is due to the special structure of our perturbative Hamiltonian. 

To formulate our perturbative framework with UV, we use $\tilde{\alpha}_{\beta \gamma}$ defined in eq.~\eqref{alpha-matrix-def} as the extra expansion parameters. That is, we assume that deviation from unitarity is small. Therefore, $\tilde{\alpha}_{\beta \gamma} \ll 1$ holds for all $\beta$ and $\gamma$. Though we follow basically the same procedure as in ref.~\cite{Martinez-Soler:2019nhb}, we give a step-by-step presentation of the formulation because of the additional complexities associated with inclusion of UV, and to make this paper self-contained. 

What would be a reachable or a possible target sensitivity to $\tilde{\alpha}_{\beta \gamma}$ in the context of unitarity test? For the sake of rough estimation, we assume momentarily a perfect knowledges of the $\nu$SM mixing parameters. Then, let us ask: Which level of sensitivity to UV $\tilde{\alpha}_{\beta \gamma}$ parameters could one expect given the accuracy of measurement of $\Delta P_{\beta \alpha} \equiv P(\nu_{\beta} \rightarrow \nu_{\alpha}) - P(\nu_{\beta} \rightarrow \nu_{\alpha})_{ \nu\text{SM} }$ (see eq.~\eqref{Pmue-UV-part}) is of order, for example, $10^{-2}$ or $10^{-4}$? Notice that $\Delta P_{\beta \alpha}$ is the non-unitary contribution to the oscillation probability. The former number is more or less the situation at the current time or in the near future, while the latter is taken arbitrarily as an expectation in a foreseeable feature. Since $\Delta P_{\beta \alpha} \sim \tilde{\alpha}_{\beta \gamma}$, we would expect the constraints on $\tilde{\alpha}_{\beta \gamma}$ parameters of the order of $10^{-2}$, or $10^{-4}$, respectively. 

We note that once the accuracy of measurement reached a ``perturbative regime'' the first-order UV correction is sufficient, as far as qualitative discussions are concerned. The second order computation yields terms of the order of $\tilde{\alpha}_{\beta \gamma}^2 \sim10^{-4}$, or $\sim10^{-8}$, respectively, far beyond the accuracy of $\Delta P_{\beta \alpha}$ measurement in each era. This is the reason why we restrict ourselves to the first-order formulas in this and the companion papers \cite{Martinez-Soler:2018lcy}.

In low-scale UV scenarios, the probability leaking term as well as the flux ``mis-normalization'' term in the appearance channels are of order $\sim \vert W \vert^4$ where $W$ denotes collectively the active-sterile mixing matrix elements \cite{Fong:2017gke}. Due to unitarity in the whole $3 + N_{ \text{sterile} }$ space, $\tilde{\alpha}_{\beta \gamma}$ must be of order $\simeq W^2$. Then, the leaking and the mis-normalization terms are of order $\tilde{\alpha}_{\beta \gamma}^2 \sim10^{-4}$ or $10^{-8}$ in the above two regimes, respectively, which are far too small compared to the accuracy of $\Delta P_{\beta \alpha}$ measurement in each era. It constitutes one of the serious problems in their determination.

\subsection{Transformation to the tilde basis}
\label{sec:tilde-basis} 

We transform to a different basis to formulate our perturbation theory for solar-scale enhancement. It is the tilde basis 
\begin{eqnarray} 
\tilde{\nu}_{i} = \left( U_{12} \right)_{ij} \check{\nu}_{j} 
\label{tilde-basis-def}
\end{eqnarray}
with Hamiltonian 
\begin{eqnarray} 
\tilde{H} = U_{12} \check{H} U_{12}^{\dagger}, 
\hspace{10mm}
\text{or}
\hspace{10mm}
\check{H} = U_{12}^{\dagger} \tilde{H} U_{12}. 
\label{tilde-H-def}
\end{eqnarray}
Notice that the term ``tilde basis'' has no connection to our notation of $\tilde{\alpha}$ parameters in the SOL convention. The Hamiltonian in the tilde basis is given by 
\begin{eqnarray}
\tilde{H}
= \tilde{H}_{ \nu\text{SM} } 
+ \tilde{H}_\text{ UV }^{(1)} + \tilde{H}_\text{ UV }^{(2)} 
\label{tilde-H-explicit}
\end{eqnarray}
where each term of the right-hand side of \eqref{tilde-H-explicit} is given by 
\begin{eqnarray} 
&& \tilde{H}_{ \nu\text{SM} } = 
\left[
\begin{array}{ccc}
s^2_{12} \Delta_{21} & 
c_{12} s_{12} e^{ i \delta} \Delta_{21} & 
0 \\
c_{12} s_{12} e^{- i \delta} \Delta_{21} & 
c^2_{12} \Delta_{21} & 0 \\
0 & 0 & \Delta_{31} \\
\end{array}
\right]
+ 
\left[
\begin{array}{ccc}
c^2_{13} \Delta_{a} & 0 & c_{13} s_{13} \Delta_{a} \\
0 & 0 & 0 \\
c_{13} s_{13} \Delta_{a} & 0 & s^2_{13} \Delta_{a} \\
\end{array}
\right], 
\label{tilde-H-SM}
\end{eqnarray}
\begin{eqnarray}
\tilde{H}_\text{ UV }^{(1)} &=& 
\Delta_{b} 
U_{13}^{\dagger} U_{23}^{\dagger} 
\left[ 
\begin{array}{ccc}
2 \tilde{\alpha}_{ee} \left( 1 - \frac{ \Delta_{a} }{ \Delta_{b} } \right) & \tilde{\alpha}_{\mu e}^* & \tilde{\alpha}_{\tau e}^* \\
\tilde{\alpha}_{\mu e} & 2 \tilde{\alpha}_{\mu \mu}  & \tilde{\alpha}_{\tau \mu}^* \\
\tilde{\alpha}_{\tau e}  & \tilde{\alpha}_{\tau \mu} & 2 \tilde{\alpha}_{\tau \tau} \\
\end{array}
\right] 
U_{23} U_{13},
\label{tilde-H-UV-1st}
\end{eqnarray}
\begin{eqnarray}
\tilde{H}_\text{ UV }^{(2)} &=& 
- \Delta_{b} 
U_{13}^{\dagger} U_{23}^{\dagger} 
\left[
\begin{array}{ccc}
\tilde{\alpha}_{ee}^2 \left( 1 - \frac{ \Delta_{a} }{ \Delta_{b} } \right) + |\tilde{\alpha}_{\mu e}|^2 + |\tilde{\alpha}_{\tau e}|^2 & 
\tilde{\alpha}_{\mu e}^* \tilde{\alpha}_{\mu \mu} + \tilde{\alpha}_{\tau e}^* \tilde{\alpha}_{\tau \mu} & 
\tilde{\alpha}_{\tau e}^* \tilde{\alpha}_{\tau \tau} \\
\tilde{\alpha}_{\mu e} \tilde{\alpha}_{\mu \mu} + \tilde{\alpha}_{\tau e} \tilde{\alpha}_{\tau \mu}^* & 
\tilde{\alpha}_{\mu \mu}^2 + |\tilde{\alpha}_{\tau \mu}|^2 & 
\tilde{\alpha}_{\tau \mu}^* \tilde{\alpha}_{\tau \tau} \\
\tilde{\alpha}_{\tau e} \tilde{\alpha}_{\tau \tau} & 
\tilde{\alpha}_{\tau \mu} \tilde{\alpha}_{\tau \tau} & 
\tilde{\alpha}_{\tau \tau}^2 \\
\end{array}
\right] 
U_{23} U_{13}. 
\nonumber \\ 
\label{tilde-H-UV-2nd}
\end{eqnarray}

In this paper, we restrict ourselves to the perturbative correction to first order in the expansion parameters. There is a number of reasons for this limitation. It certainly simplifies our discussion of the $\delta - \alpha$ parameter phase correlation, though we will make a brief comment on effect of $\tilde{H}_\text{ UV }^{(2)}$ to the correlation in section~\ref{sec:delta-alpha}. Unfortunately, the expression of the first order UV correction to the oscillation probability is sufficiently complex at this order, as we will see in sections~\ref{sec:P-mue-int-UV}, ~\ref{sec:P-mue-ext-UV} and appendix~\ref{sec:Pmue-rest}. We do not consider our restriction to first order in $\tilde{\alpha}_{\beta \gamma}$ a serious limitation because the framework anticipates a precision era of neutrino experiment for unitarity test in which the condition $\tilde{\alpha}_{\beta \gamma} \ll 1$ should be justified.

\subsection{Definitions of $F$ and $K$ matrices}
\label{sec:F-K-matrix}

To make expressions of the $S$ matrix and the oscillation probability as compact as possible, it is important to introduce the new matrix notations $F$ and $K$: 
\begin{eqnarray} 
&& 
F \equiv 
\left[
\begin{array}{ccc}
F_{11} & F_{12} & F_{13} \\
F_{21} & F_{22} & F_{23} \\
F_{31} & F_{32} & F_{33} \\
\end{array}
\right] 
=
U_{23}^{\dagger} 
\left[ 
\begin{array}{ccc}
2 \tilde{\alpha}_{ee} \left( 1 - \frac{ \Delta_{a} }{ \Delta_{b} } \right) & \tilde{\alpha}_{\mu e}^* & \tilde{\alpha}_{\tau e}^* \\
\tilde{\alpha}_{\mu e} & 2 \tilde{\alpha}_{\mu \mu}  & \tilde{\alpha}_{\tau \mu}^* \\
\tilde{\alpha}_{\tau e}  & \tilde{\alpha}_{\tau \mu} & 2 \tilde{\alpha}_{\tau \tau} \\
\end{array}
\right] 
U_{23}, 
\label{Fij-def}
\end{eqnarray}
\begin{eqnarray} 
&& 
K = U_{13}^{\dagger} F U_{13} \equiv 
\left[
\begin{array}{ccc}
K_{11} & K_{12} & K_{13} \\
K_{21} & K_{22} & K_{23} \\
K_{31} & K_{32} & K_{33} \\
\end{array}
\right] 
\nonumber \\
&&
\hspace{-8mm} 
=
\left[
\begin{array}{ccc}
c^2_{13} F_{11} + s^2_{13} F_{33} - c_{13} s_{13} \left( F_{13} + F_{31} \right) & 
c_{13} F_{12} - s_{13} F_{32} & 
c^2_{13} F_{13} - s^2_{13} F_{31} + c_{13} s_{13} \left( F_{11} - F_{33} \right) \\
c_{13} F_{21} - s_{13} F_{23} & 
F_{22} & 
s_{13} F_{21} + c_{13} F_{23} \\
c^2_{13} F_{31} - s^2_{13} F_{13} + c_{13} s_{13} \left( F_{11} - F_{33} \right) & 
s_{13} F_{12} + c_{13} F_{32} & 
s^2_{13} F_{11} + c^2_{13} F_{33} + c_{13} s_{13} \left( F_{13} + F_{31} \right) \\
\end{array}
\right]. 
\nonumber \\
\label{Kij-def}
\end{eqnarray}
The explicit expressions of the elements $F_{ij}$ and $K_{ij}$ defined in eqs.~\eqref{Fij-def} and \eqref{Kij-def}, respectively, are given in appendix~\ref{sec:F-K-Phi-elements}. By using these notations the first order Hamiltonian in the tilde basis \eqref{tilde-H-UV-1st} can be written as 
\begin{eqnarray} 
\tilde{H}_\text{ UV }^{(1)} = \Delta_{b} K.
\label{tilde-H-UV-1st-2}
\end{eqnarray}

\subsection{Formulating perturbation theory with the hat basis}
\label{sec:formulating-P} 

We use the ``renormalized basis'' such that the zeroth-order and the perturbed Hamiltonian takes the form $\tilde{H} = \tilde{H}_{0} + \tilde{H}_{1}$. $\tilde{H}_{0}$ is given by (we discuss $\tilde{H}_{1}$ later) 
\begin{eqnarray} 
&& \tilde{H}_{0} = 
\left[
\begin{array}{ccc}
s^2_{12} \Delta_{21} + c^2_{13} \Delta_{a} & c_{12} s_{12} e^{ i \delta} \Delta_{21} & 0 \\
c_{12} s_{12} e^{ - i \delta} \Delta_{21} & c^2_{12} \Delta_{21} & 0 \\
0 & 0 & \Delta_{31} + s^2_{13} \Delta_{a}  \\
\end{array}
\right]. 
\label{tilde-hamiltonian-0th}
\end{eqnarray}

To formulate the solar-resonance perturbation theory with UV, we transform to the ``hat basis'', which diagonalizes $\tilde{H}_{0}$:
\begin{eqnarray} 
\hat{\nu}_{i}= ( U_{\varphi}^{\dagger} )_{ij} \tilde{\nu}_{j} 
\label{hat-basis}
\end{eqnarray}
with Hamiltonian 
\begin{eqnarray} 
\hat{H} = U_{\varphi}^{\dagger} \tilde{H} U_{\varphi} 
\label{H-hat-basis}
\end{eqnarray}
where $U_{\varphi}$ is parametrized as 
\begin{eqnarray} 
U_{\varphi} = 
\left[
\begin{array}{ccc}
\cos \varphi & \sin \varphi e^{ i \delta} & 0 \\
- \sin \varphi e^{ - i \delta} & \cos \varphi & 0 \\
0 & 0 & 1 \\
\end{array}
\right] 
\label{U-varphi-def}
\end{eqnarray}
$U_{\varphi}$ is determined such that $\hat{H}_{(0)}$ is diagonal, which leads to 
\begin{eqnarray} 
\cos 2 \varphi &=& 
\frac{ \cos 2\theta_{12} - c^2_{13} r_{a} }
{ \sqrt{ \left( \cos 2\theta_{12} - c^2_{13} r_{a} \right)^2 +  \sin^2 2\theta_{12} } }, 
\nonumber \\
\sin 2 \varphi &=& 
\frac{ \sin 2\theta_{12} }
{ \sqrt{ \left( \cos 2\theta_{12} - c^2_{13} r_{a} \right)^2 +  \sin^2 2\theta_{12} } }, 
\label{cos-sin-2varphi}
\end{eqnarray}
where 
\begin{eqnarray} 
r_{a} \equiv \frac{a}{\Delta m^{2}_{21}} = \frac{ \Delta_{a} }{ \Delta_{21} }.
\label{ra-def}
\end{eqnarray}
The three eigenvalues of the zeroth order Hamiltonian $\tilde{H}_{0}$ in (\ref{tilde-hamiltonian-0th}) is given by\footnote{
Notice that one can show that 
\begin{eqnarray} 
h_{1} &=& 
\frac{ \Delta_{21} }{ 2 } 
\left[
\left( 1 + c^2_{13} r_{a} \right) 
- \sqrt{ \left( \cos 2\theta_{12} - c^2_{13} r_{a} \right)^2 +  \sin^2 2\theta_{12} } 
\right],
\nonumber \\
h_{2} &=&
\frac{ \Delta_{21} }{ 2 } 
\left[
\left( 1 + c^2_{13} r_{a} \right) 
+ \sqrt{ \left( \cos 2\theta_{12} - c^2_{13} r_{a} \right)^2 +  \sin^2 2\theta_{12} } 
\right].
\label{eigenvalues2}
\end{eqnarray}
}
\begin{eqnarray} 
h_{1} &=& 
\sin^2 \left( \varphi - \theta_{12} \right) \Delta_{21} + \cos^2 \varphi c^2_{13} \Delta_{a},
\nonumber \\
h_{2} &=&
\cos^2 \left( \varphi - \theta_{12} \right) \Delta_{21} + \sin^2 \varphi c^2_{13} \Delta_{a},
\nonumber \\
h_{3} &=& 
\Delta_{31} + s^2_{13} \Delta_{a}. 
\label{eigenvalues}
\end{eqnarray}
Then, the Hamiltonian in the hat basis is given by $\hat{H} = \hat{H}_{0} + \hat{H}_{\nu\text{SM}1} + \hat{H}_{\text{UV}1}$ where 
\begin{eqnarray} 
&& \hat{H}_{0} =  
\left[
\begin{array}{ccc}
h_{1} & 0 & 0 \\
0 & h_{2} & 0 \\
0 & 0 & h_{3} \\
\end{array}
\right], 
\hspace{10mm}
\hat{H}_{1}^{\nu\text{SM}} = 
\left[
\begin{array}{ccc}
0 & 0 & c_{\varphi} c_{13} s_{13} \Delta_{a} \\
0 & 0 & s_{\varphi} c_{13} s_{13} e^{ - i \delta} \Delta_{a} \\
c_{\varphi} c_{13} s_{13} \Delta_{a} & s_{\varphi} c_{13} s_{13} e^{ i \delta} \Delta_{a} & 0 \\
\end{array}
\right], 
\nonumber \\
&& \hat{H}_{1}^{\text{UV}} = 
\Delta_{b} 
U_{\varphi}^{\dagger} K U_{\varphi}, 
\label{hat-H-0th-1st}
\end{eqnarray}
where the $K$ matrix is defined in \eqref{Kij-def}, and the simplified notations are hereafter used: $c_{\varphi} = \cos \varphi$ and $s_{\varphi} = \sin \varphi$. Notice that we have omitted the second order $\hat{H}_{\text{UV}}$, though one can easily compute it from \eqref{tilde-H-UV-2nd} if necessary. 

\subsection{Calculation of $\hat{S}$ and $\tilde{S}$ matrices}
\label{sec:hatS-tildeS}

To calculate $\hat {S} (x)$ we define $\Omega(x)$ as
\begin{eqnarray} 
\Omega(x) = e^{i \hat{H}_{0} x} \hat{S} (x).
\label{def-omega}
\end{eqnarray}
Then, $\Omega(x)$ obeys the evolution equation
\begin{eqnarray} 
i \frac{d}{dx} \Omega(x) = H_{1} \Omega(x) 
\label{omega-evolution}
\end{eqnarray}
where
\begin{eqnarray} 
H_{1} \equiv e^{i \hat{H}_{0} x} \hat{H}_{1} e^{-i \hat{H}_{0} x}. 
\label{def-H1}
\end{eqnarray}
Notice that $\hat{H}_{1} = \hat{H}_{1}^{\nu\text{SM}} + \hat{H}_{1}^{\text{UV}}$ as in \eqref{hat-H-0th-1st}.
Then, $\Omega(x)$ can be computed perturbatively as
\begin{eqnarray} 
\Omega(x) &=& 1 + 
(-i) \int^{x}_{0} dx' H_{1} (x') + 
(-i)^2 \int^{x}_{0} dx' H_{1} (x') \int^{x'}_{0} dx'' H_{1} (x'') 
+ \cdot \cdot \cdot,
\label{Omega-expansion}
\end{eqnarray}
and the $\hat{S}$ matrix is given by
\begin{eqnarray} 
\hat{S} (x) =  
e^{-i \hat{H}_{0} x} \Omega(x). 
\label{hat-Smatrix}
\end{eqnarray}
Using $\hat{H}_{1} = \hat{H}_{1}^{\nu\text{SM}} + \hat{H}_{1}^{\text{UV}}$ in \eqref{hat-H-0th-1st}, $\hat{S}$ matrix of the $\nu$SM part is given to the zeroth and the first orders in the effective expansion parameter $s_{13} \frac{ \Delta_{a} }{ h_{3} - h_{1} }$ by 
\begin{eqnarray} 
&& \hat{S}_{\nu\text{SM} }^{(0+1)} (x) =  
e^{-i \hat{H}_{0} x} \Omega_{\nu\text{SM} } (x) 
\nonumber \\
&& \hspace{-16mm} 
= 
\left[
\begin{array}{ccc}
e^{ - i h_{1} x } & 0 & c_{\varphi} c_{13} s_{13} \frac{ \Delta_{a} }{ h_{3} - h_{1} } 
\left\{ e^{ - i h_{3} x } - e^{ - i h_{1} x }  \right\} \\
0 & e^{ - i h_{2} x } & s_{\varphi} c_{13} s_{13} e^{ - i \delta} 
\frac{ \Delta_{a} }{ h_{3} - h_{2} } 
\left\{ e^{ - i h_{3} x } - e^{ - i h_{2} x } \right\} \\
c_{\varphi} c_{13} s_{13} \frac{ \Delta_{a} }{ h_{3} - h_{1} } 
\left\{ e^{ - i h_{3} x } - e^{ - i h_{1} x } \right\} & 
s_{\varphi} c_{13} s_{13} e^{ i \delta} 
\frac{ \Delta_{a} }{ h_{3} - h_{2} } 
\left\{ e^{ - i h_{3} x } - e^{ - i h_{2} x } \right\} & 
e^{ - i h_{3} x } \\
\end{array}
\right],
\nonumber \\
\label{hat-Smatrix-0th-1st}
\end{eqnarray}
where we have used the fact that $\Delta_{b}$ is spatially constant as a consequence of the uniform matter density approximation. 

Then, $\nu$SM part of the tilde basis $\tilde{S}$ matrix is given by 
\begin{eqnarray} 
&& \tilde{S}_{\nu\text{SM} }^{(0+1)} =
U_{\varphi} \hat{S}_{\nu\text{SM} }^{(0+1)} 
U_{\varphi}^{\dagger} 
= \tilde{S}_{\nu\text{SM} }^{(0)} + \tilde{S}_{\nu\text{SM} }^{(1)}, 
\end{eqnarray}
where 
\begin{eqnarray} 
&& \tilde{S}_{\nu\text{SM} }^{(0)} = 
\left[
\begin{array}{ccc}
c^2_{\varphi} e^{ - i h_{1} x } + s^2_{\varphi} e^{ - i h_{2} x } & 
c_{\varphi} s_{\varphi} e^{ i \delta} 
\left( e^{ - i h_{2} x } - e^{ - i h_{1} x } \right) & 
0 \\
c_{\varphi} s_{\varphi} e^ {- i \delta} 
\left( e^{ - i h_{2} x } - e^{ - i h_{1} x } \right) & 
s^2_{\varphi} e^{ - i h_{1} x } + c^2_{\varphi} e^{ - i h_{2} x } & 
0 \\
0 & 0 & e^{ - i h_{3} x } \\
\end{array}
\right].
\label{tilde-S-SM-0th}
\end{eqnarray}
$\tilde{S}_{\nu\text{SM} }^{(1)}$ can be written in the form 
\begin{eqnarray} 
&& \tilde{S}_{\nu\text{SM} }^{(1)} = 
\left[
\begin{array}{ccc}
0 & 0 & X \\
0 & 0 & Y e^{ - i \delta} \\
X & Y e^{ i \delta} & 0 \\
\end{array}
\right], 
\label{tilde-S-SM-1st}
\end{eqnarray}
where 
\begin{eqnarray}
&& X 
= c_{13} s_{13} 
\left\{ \frac{ \Delta_{a} }{ h_{3} - h_{1} } 
c^2_{\varphi} \left( e^{ - i h_{3} x } - e^{ - i h_{1} x } \right) 
+ \frac{ \Delta_{a} }{ h_{3} - h_{2} } 
s^2_{\varphi} \left( e^{ - i h_{3} x } - e^{ - i h_{2} x } \right) \right\}, 
\nonumber \\
&& Y 
= c_{13} s_{13}  c_{\varphi} s_{\varphi} 
\left\{ - \frac{ \Delta_{a} }{ h_{3} - h_{1} } 
\left( e^{ - i h_{3} x } - e^{ - i h_{1} x } \right) 
+ \frac{ \Delta_{a} }{ h_{3} - h_{2} } 
\left( e^{ - i h_{3} x } - e^{ - i h_{2} x } \right) \right\}. 
\label{X-Y-def}
\end{eqnarray}
Notice that $\tilde{S}_{\nu\text{SM} }$ respects the generalized T invariance. 

Now, we must compute the UV parameter related part of $\hat{S}$ and $\tilde{S}$ matrices. By remembering $\hat{H}_{1}^{\text{UV}} = \Delta_{b} U_{\varphi}^{\dagger} K U_{\varphi}$, the UV part of $H_{1}$ in \eqref{def-H1}, 
is given by 
\begin{eqnarray} 
&& H_{1}^{\text{UV} } 
= e^{i \hat{H}_{0} x} \hat{H}_{1}^{\text{UV}} e^{-i \hat{H}_{0} x} 
= \Delta_{b} e^{i \hat{H}_{0} x} U_{\varphi}^{\dagger} K U_{\varphi} 
e^{-i \hat{H}_{0} x} 
\nonumber \\
&=& 
\Delta_{b} 
U_{\varphi}^{\dagger} 
\left( U_{\varphi} e^{i \hat{H}_{0} x} U_{\varphi}^{\dagger} \right) 
K \left( U_{\varphi} e^{-i \hat{H}_{0} x} U_{\varphi}^{\dagger} \right) 
U_{\varphi}.
\label{phi-rotated-S0}
\end{eqnarray}
Due to frequent usage of the factors in the parenthesis above we give the formula for them 
\begin{eqnarray}
&& 
S^{ (\pm) }_{\varphi} \equiv 
\left( U_{\varphi} e^{ \pm i \hat{H}_{0} x} U_{\varphi}^{\dagger} \right) 
\nonumber \\
&=&
\left[
\begin{array}{ccc}
c_{\varphi}^2 e^{ \pm i h_{1} x } + s_{\varphi}^2 e^{ \pm i h_{2} x } & 
c_{\varphi} s_{\varphi} e^{ i \delta} 
\left( e^{ \pm i h_{2} x } - e^{ \pm i h_{1} x } \right) & 
0 \\
c_{\varphi} s_{\varphi} e^{ - i \delta} 
\left( e^{ \pm i h_{2} x } - e^{ \pm i h_{1} x } \right) & 
s_{\varphi}^2 e^{ \pm i h_{1} x } + c_{\varphi}^2 e^{ \pm i h_{2} x } & 
0 \\
0 & 0 & e^{ \pm i h_{3} x } \\
\end{array}
\right]. 
\end{eqnarray}
Notice that $\tilde{S}_{\nu\text{SM} }^{(0)}$ is nothing but $S^{ (-) }_{\varphi}$. Then, $H_{1}^{\text{UV} }$ takes a simple form  

\begin{eqnarray}
&& H_{1}^{\text{UV} } 
= \Delta_{b} 
U_{\varphi}^{\dagger} 
S^{ (+) }_{\varphi} K S^{ (-) }_{\varphi} 
U_{\varphi} 
\equiv
\Delta_{b} 
U_{\varphi}^{\dagger} 
\Phi U_{\varphi} 
= 
\Delta_{b} 
U_{\varphi}^{\dagger} 
\left[
\begin{array}{ccc}
\Phi_{11} & \Phi_{12} & \Phi_{13} \\
\Phi_{21} & \Phi_{22} & \Phi_{23} \\
\Phi_{31} & \Phi_{32} & \Phi_{33} \\
\end{array}
\right] 
U_{\varphi}
\label{H1UV}
\end{eqnarray}
where we have introduced another simplifying matrix notation 
$\Phi \equiv S^{ (+) }_{\varphi} K S^{ (-) }_{\varphi}$
and its elements $\Phi_{ij}$. The explicit expressions of $\Phi_{ij}$
are given in appendix~\ref{sec:F-K-Phi-elements}.

Since $U_{\varphi}$ rotation back to the tilde basis removes $U_{\varphi}^{\dagger}$ and $U_{\varphi}$ in \eqref{H1UV}, it is simpler to go directly to the calculation of the tilde basis $\tilde{S}$ matrix
\begin{eqnarray} 
&& \tilde{S} (x)_{ \text{EV} }^{(1)} = 
U_{\varphi} \hat{S} (x)_{ \text{EV} }^{(1)} U_{\varphi}^{\dagger} 
=
U_{\varphi} e^{-i \hat{H}_{0} x} \Omega(x)^{(1)}_{ \text{UV} } U_{\varphi}^{\dagger} 
= 
\Delta_{b} U_{\varphi} 
e^{-i \hat{H}_{0} x} 
U_{\varphi}^{\dagger} 
\left[ (-i) \int^{x}_{0} dx' \Phi (x') \right] 
\nonumber \\
&=& 
\Delta_{b} S^{ (-) }_{\varphi} 
(-i) \int^{x}_{0} dx' 
\left[
\begin{array}{ccc}
\Phi_{11} (x') & \Phi_{12} (x') & \Phi_{13} (x') \\
\Phi_{21} (x') & \Phi_{22} (x') & \Phi_{23} (x') \\
\Phi_{31} (x') & \Phi_{32} (x') & \Phi_{33} (x') \\
\end{array}
\right].
\label{hat-Smatrix-1st}
\end{eqnarray}
Hereafter, the subscript ``EV'' is used for the $\tilde{S}$ and $\hat{S}$ matrices to indicate that they describe unitary evolution. 
The computed results of the elements of $\tilde{S} (x)_{ \text{EV} }^{(1)}$ are given in appendix~\ref{sec:tilde-S-summary}. Notice that again $\tilde{S} (x)_{ \text{UV} }^{(1)}$ respects the generalized T invariance. 

Thus, we have computed all the tilde basis $S$ matrix elements to first order as 
\begin{eqnarray}
\tilde{S} = \tilde{S}_{\nu\text{SM} }^{(0)} + \tilde{S}_{\nu\text{SM} }^{(1)} + \tilde{S}_{\text{EV} }^{(1)}.
\end{eqnarray}
The first and the second terms are given, respectively, in \eqref{tilde-S-SM-0th} and \eqref{tilde-S-SM-1st} with \eqref{X-Y-def}, and the third in appendix~\ref{sec:tilde-S-summary}. 

\subsection{The relations between various bases and computation of the flavor basis $S$ matrix} 
\label{sec:basis-relations}

We first summarize the relationship between the flavor basis, the check
(vacuum mass eigenstate) basis, the tilde, and the hat (zeroth order
diagonalized hamiltonian) basis.
Only the unitary transformations are involved in changing from the hat
basis to the tilde basis, and from the tilde basis to the check basis:
\begin{eqnarray} 
&& \hat{H} = U^{\dagger}_{\varphi} \tilde{H} U_{\varphi}, 
\hspace{8mm}
\text{or}
\hspace{8mm}
\tilde{H} = U_{\varphi} \hat{H} U^{\dagger}_{\varphi}, 
\nonumber\\
&&
\tilde{H} = 
U_{12} \check{H} U_{12}^{\dagger}, 
\hspace{8mm}
\text{or}
\hspace{8mm}
\check{H} = U_{12}^{\dagger} \tilde{H} U_{12} 
= U_{12}^{\dagger} U_{\varphi} \hat{H} U^{\dagger}_{\varphi} U_{12} 
\label{hat-tilde-check}
\end{eqnarray}
The non-unitary transformation is involved from the check basis to the
flavor basis:
\begin{eqnarray} 
\nu_{\beta} = N_{\beta i} \check{\nu}_{i} 
= \left\{ ( 1 - \tilde{\alpha} ) U \right\}_{\beta i} \check{\nu}_{i}. 
\label{flavor-check}
\end{eqnarray}
The relationship between the flavor basis Hamiltonian $H_{ \text{flavor} }$ and the hat basis one $\hat{H}$ is
\begin{eqnarray}
H_{ \text{flavor} } 
&=& 
\left\{ ( 1 - \tilde{\alpha} ) U \right\} \check{H} \left\{ ( 1 - \tilde{\alpha} ) U \right\}^{\dagger} 
= ( 1 - \tilde{\alpha} ) U 
U_{12}^{\dagger} U_{\varphi} \hat{H} U^{\dagger}_{\varphi} U_{12} 
U^{\dagger} ( 1 - \tilde{\alpha} )^{\dagger} 
\nonumber \\
&=& 
( 1 - \tilde{\alpha} ) U_{23} U_{13} 
U_{\varphi} \hat{H} U^{\dagger}_{\varphi} 
U_{13}^{\dagger} U_{23} ^{\dagger} 
( 1 - \tilde{\alpha} )^{\dagger}. 
\label{flavor-hat}
\end{eqnarray}
Then, the flavor basis $S$ matrix is related to $\hat{S}$ and
$\tilde{S}$ matrices as
\begin{eqnarray} 
S_{ \text{flavor} } &=& 
( 1 - \tilde{\alpha} ) U_{23} U_{13} 
U_{\varphi} \hat{S} U^{\dagger}_{\varphi} 
U_{13}^{\dagger} U_{23} ^{\dagger} 
( 1 - \tilde{\alpha} )^{\dagger} 
\nonumber \\
&=&
( 1 - \tilde{\alpha} ) U_{23} U_{13} \tilde{S}
U_{13}^{\dagger} U_{23} ^{\dagger}
( 1 - \tilde{\alpha} )^{\dagger}. 
\label{S-flavor-hat}
\end{eqnarray}

Using the formula eq.~\eqref{S-flavor-hat}, it is straightforward to compute the flavor basis $S$ matrix elements. Notice, however, that $U_{13}$ is free from CP phase $\delta$ due to our choice of the SOL convention of the $U_{\text{\tiny MNS}}$ matrix in \eqref{MNS-SOL}. 

The flavor basis $S$ matrix has a structure $S_{ \text{flavor} } = ( 1 - \tilde{\alpha} ) S^{ \text{prop} } ( 1 - \tilde{\alpha} )^{\dagger}$ where $S_{ \text{prop} } \equiv U_{23} U_{13} \tilde{S} U_{13}^{\dagger} U_{23} ^{\dagger}$ describes the unitary evolution  despite the presence of non-unitary mixing \cite{Martinez-Soler:2018lcy}. The factors $(1 - \tilde{\alpha})$ and $(1 - \tilde{\alpha})^{\dagger}$, parts of the $N$ matrix which project the flavor states to the mass eigenstates and vice versa, may be interpreted as the ones analogous to the ``production NSI'' and ``detection NSI'' \cite{GonzalezGarcia:2001mp}, but a very constrained one determined by the ``propagation NSI''.

\subsection{Effective expansion parameter with and without the UV effect} 
\label{sec:effective-exp-parameter}

As announced in section~\ref{sec:region-validity}, the expression of $\tilde{S}_{\nu\text{SM} }^{(1)}$ in \eqref{tilde-S-SM-1st} with \eqref{X-Y-def} tells us that we have another expansion parameter \cite{Martinez-Soler:2019nhb} 
\begin{eqnarray}
A_{ \text{exp} } 
&\equiv& 
c_{13} s_{13} 
\biggl | \frac{ a }{ \Delta m^2_{31} } \biggr | 
= 2.78 \times 10^{-3} 
\left(\frac{ \Delta m^2_{31} }{ 2.4 \times 10^{-3}~\mbox{eV}^2}\right)^{-1}
\left(\frac{\rho}{3.0 \,\text{g/cm}^3}\right) \left(\frac{E}{200~\mbox{MeV}}\right), 
\nonumber \\
\label{expansion-parameter}  
\end{eqnarray}
which is very small. The reason for such a ``generated by the framework'' expansion parameter is the special feature of the perturbed Hamiltonian in \eqref{hat-H-0th-1st}. 

In fact, our perturbative framework is peculiar from the beginning in the sense that the key non-perturbed part of the Hamiltonian \eqref{tilde-hamiltonian-0th}, its top-left $2 \times 2$ sub-matrix, is smaller in size than the 33 element by a factor of $\sim30$, and is comparable with $\hat{H}_{1}^{\nu\text{SM}}$ in \eqref{hat-H-0th-1st}. The secret for emergence of the very small effective expansion parameter \eqref{expansion-parameter} is that the 33 element decouples in the leading order and appear in the perturbative corrections only in the energy denominator, making them  {\em smaller for the larger ratio} of $\Delta m^2_{31} / \Delta m^2_{21}$. The latter property holds because of the special structure of perturbative Hamiltonian $\hat{H}_{1}^{\nu\text{SM}}$ with non-vanishing elements only in the third row and third column. 

With inclusion of the UV Hamiltonian \eqref{tilde-H-UV-1st-2}, however, the size of the first order correction is controlled not only by $A_{ \text{exp} }$ in \eqref{expansion-parameter} but also the magnitudes of $\tilde{\alpha}_{\beta, \gamma}$. In computing the higher order corrections the energy denominator suppression does not work for all the terms because the last property, ``non-vanishing elements in the third row and third column only'', ceases to hold in the first order Hamiltonian. It can be confirmed in looking into the formulas of the oscillation probabilities in section~\ref{sec:P-mue-int-UV}, appendix~\ref{sec:Pmue-rest}, and section~\ref{sec:P-mue-ext-UV}.

\section{Neutrino oscillation probability to first order in expansion} 
\label{sec:general-formula-P}

The oscillation probability can be calculated by using the formula 
\begin{eqnarray}
P(\nu_{\beta} \rightarrow \nu_{\alpha}) = \vert ( S_{ \text{flavor} } )_{\alpha \beta} \vert^2. 
\label{probability-general}
\end{eqnarray} 
We denote the flavor basis $S$ matrices corresponding to $\tilde{S}_{\nu\text{SM} }^{(0)}$, $\tilde{S}_{\nu\text{SM} }^{(1)}$, and $\tilde{S}_{\text{UV} }^{(1)}$ as $S_{\nu\text{SM} }^{(0)}$, $S_{\nu\text{SM} }^{(1)}$, and $S_{\text{UV} }^{(1)}$, respectively, as they are related through \eqref{S-flavor-hat}. To first order we have 
\begin{eqnarray} 
S_{ \text{flavor} } &=& 
S_{\nu\text{SM} }^{(0)} + S_{\nu\text{SM} }^{(1)} + S_{\text{EV} }^{(1)}
- \tilde{\alpha} S_{\nu\text{SM} }^{(0)}
- S_{\nu\text{SM} }^{(0)} \tilde{\alpha}^{\dagger}. 
\label{S-flavor-all}
\end{eqnarray}
Then, we are ready to calculate the expressions of the oscillation
probabilities using the formula \eqref{probability-general} to first order in
the expansion parameters. 
Following ref.~\cite{Martinez-Soler:2018lcy}, we categorize $P(\nu_{\beta} \rightarrow \nu_{\alpha})$ into the three types of terms:
\begin{eqnarray} 
P(\nu_{\beta} \rightarrow \nu_{\alpha}) &=& 
P(\nu_{\beta} \rightarrow \nu_{\alpha})_{\nu\text{SM} }^{(0+1)} 
+ P(\nu_{\beta} \rightarrow \nu_{\alpha})_{ \text{ EV } }^{(1)} 
+ P(\nu_{\beta} \rightarrow \nu_{\alpha})_{ \text{ UV } }^{(1)}, 
\label{S-flavor-all}
\end{eqnarray}
where 
\begin{eqnarray} 
P(\nu_{\beta} \rightarrow \nu_{\alpha})_{\nu\text{SM} }^{(0+1)} 
&=& 
\biggl| \left( S_{\nu\text{SM} }^{(0)} \right)_{\alpha \beta} \biggr|^2 
+ 2 \mbox{Re} \left[
\left( S_{\nu\text{SM} }^{(0)} \right)_{\alpha \beta}^* 
\left( S_{\nu\text{SM} }^{(1)} \right)_{\alpha \beta} 
\right], 
\nonumber \\
P(\nu_{\beta} \rightarrow \nu_{\alpha})_{ \text{ EV } }^{(1)}
&=& 
2 \mbox{Re} \left[
\left( S_{\nu\text{SM} }^{(0)} \right)_{\alpha \beta}^* 
\left( S^{(1)}_{ \text{ EV } } \right)_{\alpha \beta} 
\right], 
\nonumber \\
P(\nu_{\beta} \rightarrow \nu_{\alpha})_{ \text{ UV } }^{(1)}
&=& 
- 2 \mbox{Re} \left[
\left( S_{\nu\text{SM} }^{(0)} \right)_{\alpha \beta}^* 
\left( \tilde{\alpha} S_{\nu\text{SM} }^{(0)} + S_{\nu\text{SM} }^{(0)} \tilde{\alpha}^{\dagger} \right)_{\alpha \beta} 
\right]. 
\label{P-three-types}
\end{eqnarray}
The subscripts ``EV'' and ``UV'' refer the unitary evolution part and the genuine non-unitary contribution, terminology defined in ref.~\cite{Martinez-Soler:2018lcy}. 

The first term in eq.~\eqref{P-three-types}, $P(\nu_{\beta} \rightarrow \nu_{\alpha})_{\nu\text{SM} }^{(0+1)} $, is already computed in ref.~\cite{Martinez-Soler:2019nhb}. Hence, we do not repeat the calculation, but urge the readers to go to this reference. Notice that use of the SOL convention of $U_{\text{\tiny MNS}}$ does not alter the expression of the oscillation probabilities. 
The rest of the terms in eq.~\eqref{P-three-types} can be computed straightforwardly by using the expressions of the tilde basis $S$ matrix which are given explicitly in appendix~\ref{sec:tilde-S-summary}, and the $\tilde{\alpha}$ matrix defined in \eqref{alpha-matrix-def}. 

Unfortunately, the resulting expression of the oscillation probability even at first order is far from simple. Therefore, to give a feeling to the readers, we will show in the next two subsections a part of the first order probability in the $\nu_{\mu} \rightarrow \nu_{e}$ channel. In section~\ref{sec:P-mue-int-UV}, one of the five terms in $P(\nu_{\mu} \rightarrow \nu_{e})_{ \text{ EV} }^{(1)}$ is given, and the whole expression of $P(\nu_{\mu} \rightarrow \nu_{e})_{ \text{ UV} }^{(1)}$ in section~\ref{sec:P-mue-ext-UV}. 
We leave the rest of the terms of $P(\nu_{\mu} \rightarrow \nu_{e})_{ \text{ EV} }^{(1)}$ to appendix~\ref{sec:Pmue-rest}. In this appendix, we also give a practical suggestion to the readers on how to compute the oscillation probabilities in the $\nu_{\mu} - \nu_{\tau}$ sector. 

For notational simplicity, we define, following ref.~\cite{Martinez-Soler:2019nhb}, the reduced Jarlskog factor in matter as 
\begin{eqnarray} 
J_{mr} &\equiv& 
c_{23} s_{23} c^2_{13} s_{13} c_{\varphi} s_{\varphi} 
=
J_r \left[ 
\left( \cos 2\theta_{12} - c^2_{13} r_{a} \right)^2 +  \sin^2 2\theta_{12} 
\right]^{ - 1/2 }, 
\label{Jmr-def}
\end{eqnarray}
which is proportional to the reduced Jarlskog factor in vacuum, $J_r \equiv c_{23} s_{23} c^2_{13} s_{13} c_{12} s_{12}$ \cite{Jarlskog:1985ht}. We have used eq.~(\ref{cos-sin-2varphi}) in the second equality in (\ref{Jmr-def}). 

In this paper, we do not discuss numerical accuracy of the first order oscillation probability because (1) the $\nu$SM part, which is controlled by $A_{ \text{exp} } \sim 10^{-3}$, is known to be very accurate already in first order~\cite{Martinez-Soler:2019nhb}, and (2) accuracy of the UV related part is trivial, the smaller the $\alpha_{\beta \gamma}$, the better the accuracy.\footnote{
If we set the target accuracy of unitarity test at a \% level, $\alpha_{\beta \gamma} \lsim 10^{-2}$. Then, within the accuracy of the $\nu$SM part, $10^{-3} \lsim \alpha_{\beta \gamma} \lsim 10^{-2}$, the second order UV corrections could play a role. But, it is of order of $\sim \alpha_{\beta \gamma}^2 \lsim 10^{-4}$, and hence it is negligible. }

\subsection{Unitary evolution part of the first order probability $P(\nu_{\mu} \rightarrow \nu_{e})_{ \text{ EV } }^{(1)}$}
\label{sec:P-mue-int-UV}

We first introduce the decomposition of $P(\nu_{\mu} \rightarrow \nu_{e})_{ \text{ EV } }^{(1)}$. After computation of all the terms, we assemble them according to the types of the $K_{ij}$ variables involved. 
See eq.~\eqref{Kij-def} and \eqref{Kij-elements} in appendix~\ref{sec:F-K-Phi-elements} for the definitions and the explicit expressions of the $K_{ij}$, respectively. For bookkeeping purpose we decompose $P(\nu_{\mu} \rightarrow \nu_{e})_{ \text{ EV } }^{(1)}$ into the following four terms:
\begin{eqnarray} 
P(\nu_{\mu} \rightarrow \nu_{e})_{ \text{ EV} }^{(1)} &=&
P(\nu_{\mu} \rightarrow \nu_{e})_{ \text{ EV} }^{(1)} \vert_{ \text{D-OD} } 
\nonumber \\
&+&
P(\nu_{\mu} \rightarrow \nu_{e})_{ \text{ EV} }^{(1)} \vert_{ \text{OD1} } 
P(\nu_{\mu} \rightarrow \nu_{e})_{ \text{ EV} }^{(1)} \vert_{ \text{OD2} } +
P(\nu_{\mu} \rightarrow \nu_{e})_{ \text{ EV} }^{(1)} \vert_{ \text{OD3} }, 
\nonumber \\
\label{P-mue-four-terms}
\end{eqnarray}
where the subscripts ``D'' and ``OD'' refer to the diagonal and the off-diagonal $K_{ij}$ variables. The organization inside each term is largely determined such that the symmetry under the transformation $\varphi \rightarrow \varphi + \frac{\pi}{2}$ is manifest. See section~\ref{sec:symmetry} for the $\varphi$ symmetry.

Here, we only present the first term in \eqref{P-mue-four-terms}, $P(\nu_{\mu} \rightarrow \nu_{e})_{ \text{EV} }^{(1)} \vert_{ \text{D-OD} }$, leaving the others to appendix~\ref{sec:Pmue-rest}: 
\begin{eqnarray} 
&& 
P(\nu_{\mu} \rightarrow \nu_{e})_{ \text{ EV} }^{(1)} \vert_{ \text{D-OD} } 
\nonumber \\
&=& 
4 J_{mr} \sin \delta \cos 2\varphi \left( K_{22} - K_{11} \right) 
( \Delta_{b} x ) \sin^2 \frac{ ( h_{2} - h_{1} ) x }{2} 
\nonumber \\
&+& 
2 \left( K_{33} - K_{11} \right) ( \Delta_{b} x ) 
\biggl[
J_{mr} \cos \delta \sin ( h_{2} - h_{1} ) x 
- 2 s^2_{23} c_{13} s_{13} 
\left\{ c^2_{\varphi} \sin ( h_{3} - h_{1} ) x 
+ s^2_{\varphi} \sin ( h_{3} - h_{2} ) x \right\} 
\biggr] 
\nonumber \\
&+& 
4 J_{mr} 
\left( K_{33} - K_{22} \right) ( \Delta_{b} x ) 
\nonumber \\
&\times&
\biggl[ 2 \cos \delta 
\sin \frac{( h_{3} - h_{2} ) x}{2} 
\sin \frac{( h_{2} - h_{1} ) x}{2} 
\sin \frac{( h_{1} - h_{3} ) x}{2} 
+ \sin \delta \left\{
\sin^2 \frac{ ( h_{3} - h_{2} ) x }{2} 
- \sin^2 \frac{ ( h_{3} - h_{1} ) x }{2} 
\right\} 
\biggr] 
\nonumber \\
&+&
\left[ \cos 2 \varphi \left( K_{22} - K_{11} \right) 
+ \sin 2 \varphi \left( K_{12} e^{- i \delta} + K_{21} e^{ i \delta} \right) \right] 
( \Delta_{b} x ) 
\nonumber \\
&\times& 
\biggl[ 
2 c_{13} c^2_{\varphi} s^2_{\varphi} 
\left\{
c^2_{23} c_{13} 
\sin ( h_{2} - h_{1} ) x 
- 4 s^2_{23} s_{13} 
\sin \frac{( h_{3} - h_{2} ) x}{2} 
\sin \frac{( h_{2} - h_{1} ) x}{2} 
\sin \frac{( h_{1} - h_{3} ) x}{2} 
\right\}
\nonumber \\
&+& 
2 J_{mr} \cos \delta 
\left[ c^2_{\varphi} \sin ( h_{3} - h_{1} ) x 
+ s^2_{\varphi} \sin ( h_{3} - h_{2} ) x \right] 
\nonumber \\
&-& 
4 J_{mr} \sin \delta 
\left\{ c^2_{\varphi} \sin^2 \frac{ ( h_{3} - h_{1} ) x }{2}
+ s^2_{\varphi} \sin^2 \frac{ ( h_{3} - h_{2} ) x }{2} 
- 4 c^2_{\varphi} s^2_{\varphi} \sin^2 \frac{ ( h_{2} - h_{1} ) x }{2} \right\} 
\biggr]
\nonumber \\
&+& 
2 \left[ \sin 2 \varphi \left( K_{22} - K_{11} \right) 
- \cos 2 \varphi \left( K_{12} e^{- i \delta} + K_{21} e^{ i \delta} \right) \right] 
\nonumber \\
&\times&
\biggl[ 
s^2_{23} c_{13} s_{13} c_{\varphi} s_{\varphi} 
\biggl\{
( \Delta_{b} x ) 
\left[ c^2_{\varphi} \sin ( h_{3} - h_{1} ) x 
+ s^2_{\varphi} \sin ( h_{3} - h_{2} ) x \right] 
\nonumber \\
&+&
2 \frac{ \Delta_{b} }{ h_{2} - h_{1} } 
\left[
\sin^2 \frac{ ( h_{3} - h_{2} ) x }{2}
- \sin^2 \frac{ ( h_{3} - h_{1} ) x }{2} 
- \cos 2 \varphi  \sin^2 \frac{ ( h_{2} - h_{1} ) x }{2} 
\right]
\biggr\}
\nonumber \\
&+&
4 J_{mr} \cos \delta c_{\varphi} s_{\varphi} 
\frac{ \Delta_{b} }{ h_{2} - h_{1} } \sin^2 \frac{ ( h_{2} - h_{1} ) x }{2} 
\biggr]
\nonumber \\
&-& 
4 c_{23} c^2_{13} c_{\varphi} s_{\varphi} 
\left[ \cos 2 \varphi \left( K_{22} - K_{11} \right) 
+ \sin 2 \varphi 
\left( K_{12} e^{- i \delta} + K_{21} e^{ i \delta} \right) \right] 
\nonumber \\
&\times& 
\frac{ \Delta_{b} }{ h_{2} - h_{1} } 
\biggl[
- s_{23} \cos \delta 
\left\{ \sin^2 \frac{ ( h_{3} - h_{2} ) x }{2}
- \sin^2 \frac{ ( h_{3} - h_{1} ) x }{2} 
- \cos 2\varphi \sin^2 \frac{ ( h_{2} - h_{1} ) x }{2} \right\} 
\nonumber \\
&+& 
c_{23} \sin 2 \varphi 
\sin^2 \frac{ ( h_{2} - h_{1} ) x }{2}
+ 2 s_{23} \sin \delta 
\sin \frac{( h_{3} - h_{2} ) x}{2} 
\sin \frac{( h_{2} - h_{1} ) x}{2} 
\sin \frac{( h_{1} - h_{3} ) x}{2} 
\biggr]
\nonumber \\
&+& 
2 J_{mr} c_{\varphi} s_{\varphi} 
\left[ \sin 2 \varphi \left( K_{22} - K_{11} \right) 
- \cos 2 \varphi \left( K_{12} e^{- i \delta} + K_{21} e^{ i \delta} \right) \right] 
( \Delta_{b} x ) 
\nonumber \\
&\times&
\biggl[
- \left\{
\cos \delta \sin ( h_{2} - h_{1} ) x 
+ 2 \sin \delta \cos 2 \varphi \sin^2 \frac{ ( h_{2} - h_{1} ) x }{2} 
\right\} 
\nonumber \\
&+& 
2 \sin \delta \left\{ \sin^2 \frac{ ( h_{3} - h_{2} ) x }{2}
- \sin^2 \frac{ ( h_{3} - h_{1} ) x }{2} 
- \cos 2\varphi \sin^2 \frac{ ( h_{2} - h_{1} ) x }{2} \right\} 
\nonumber \\
&+& 
4 \cos \delta 
\sin \frac{( h_{3} - h_{2} ) x}{2} 
\sin \frac{( h_{2} - h_{1} ) x}{2} 
\sin \frac{( h_{1} - h_{3} ) x}{2} 
\biggr].
\label{P-mue-D-OD}
\end{eqnarray}

We first note that the $\alpha$ parameter dependence is expressed through the $K_{jj}$ elements, see its definition and the expression eq.~\eqref{Kij-def} and \eqref{Kij-elements}, respectively. It is noticeable that the diagonal $K_{jj}$ elements organize themselves into the form of difference, $K_{22} - K_{11}$ type combinations, as it should be, because it comes from the rephasing invariance.\footnote{
The fact is well known in the systems with the NSI parameters $\varepsilon_{\beta \gamma}$. For a demonstration of the $\varepsilon_{\beta \beta} - \varepsilon_{\gamma \gamma}$ structure to third order in the NSI parameters, see ref.~\cite{Kikuchi:2008vq}, in particular its arXiv v1 for the  explicit form. 
}
It leads to the similar structure expressed by the diagonal $\alpha$ parameters, see section~\ref{sec:diag-alpha}. 

\subsection{Non-unitary part of the first order probability $P(\nu_{\mu} \rightarrow \nu_{e})_{ \text{ UV } }^{(1)}$} 
\label{sec:P-mue-ext-UV}

To calculate $P(\nu_{\mu} \rightarrow \nu_{e})_{ \text{ UV } }^{(1)}$ defined in the last line in eq.~\eqref{P-three-types}, we need the expressions of zeroth-order elements of $\nu$SM matrix $S_{\nu\text{SM} }^{(0)}$, which are given in appendix~\ref{sec:SM-S-0th}. They can be easily obtained from the tilde basis $S$ matrix in \eqref{tilde-S-SM-0th}. Using the $S^{(0)}$ matrix elements $P(\nu_{\mu} \rightarrow \nu_{e})_{ \text{ UV } }^{(1)}$ can be readily calculated as 
\begin{eqnarray} 
&& 
P(\nu_{\mu} \rightarrow \nu_{e})_{ \text{ UV } }^{(1)}
= 
- 2 ( \tilde{\alpha}_{ee} + \tilde{\alpha}_{\mu \mu} ) \vert S_{e \mu}^{(0)} \vert^2 
- 2 \mbox{Re} \left[ \tilde{\alpha}_{\mu e} ( S_{e e}^{(0)} )^* S_{e \mu}^{(0)} \right] 
\nonumber \\
&=&
- 2 ( \tilde{\alpha}_{ee} + \tilde{\alpha}_{\mu \mu} ) 
\biggl[
c^2_{23} c^2_{13} \sin^2 2 \varphi 
\sin^2 \frac{ ( h_{2} - h_{1} ) x }{2}
\nonumber \\
&+& 
s^2_{23} \sin^2 2\theta_{13} 
\left\{
c^2_{\varphi} \sin^2 \frac{ ( h_{3} - h_{1} ) x }{2} 
+ s^2_{\varphi} \sin^2 \frac{ ( h_{3} - h_{2} ) x }{2} 
- c^2_{\varphi} s^2_{\varphi} \sin^2 \frac{ ( h_{2} - h_{1} ) x }{2} 
\right\}
\nonumber \\
&+& 
4 J_{mr} \cos \delta 
\left\{
\cos 2 \varphi \sin^2 \frac{ ( h_{2} - h_{1} ) x }{2} 
- \sin^2 \frac{ ( h_{3} - h_{2} ) x }{2} 
+ \sin^2 \frac{ ( h_{3} - h_{1} ) x }{2} 
\right\} 
\nonumber \\
&+& 
8 J_{mr} \sin \delta 
\sin \frac{ ( h_{3} - h_{2} ) x }{2} 
\sin \frac{ ( h_{2} - h_{1} ) x }{2} 
\sin \frac{ ( h_{1} - h_{3} ) x }{2} 
\biggr]
\nonumber \\
&+& 
\mbox{Re} ( \tilde{\alpha}_{\mu e} ) 
\biggl[ 
2 c_{23} c_{13} \sin 2 \varphi 
\cos \delta 
\biggl(
c^2_{13} \cos 2 \varphi \sin^2 \frac{ ( h_{2} - h_{1} ) x }{2}
+ s^2_{13} \left\{ 
\sin^2 \frac{ ( h_{3} - h_{2} ) x }{2} - \sin^2 \frac{ ( h_{3} - h_{1} ) x }{2} 
\right\} 
\biggr)
\nonumber \\
&-& 
c_{23} c_{13} \sin 2 \varphi  
\sin \delta 
\biggl( c^2_{13} \sin ( h_{2} - h_{1} ) x 
- s^2_{13} 
\left\{ \sin ( h_{3} - h_{2} ) x - \sin ( h_{3} - h_{1} ) x \right\} 
\biggr)
\nonumber \\
&-& 
s_{23} \sin 2 \theta_{13} 
\biggl(
c^2_{13} \sin^2 2\varphi \sin^2 \frac{ ( h_{2} - h_{1} ) x }{2} 
- 2 \cos 2\theta_{13}
\left\{ c^2_{\varphi} \sin^2 \frac{ ( h_{3} - h_{1} ) x }{2}
+ s^2_{\varphi} \sin^2 \frac{ ( h_{3} - h_{2} ) x }{2} \right\} 
\biggr)
\biggr]
\nonumber \\
&-& 
\mbox{Im} (\tilde{\alpha}_{\mu e} )
\biggl[
c_{23} c_{13} \sin 2 \varphi 
\cos \delta 
\biggl( c^2_{13} \sin ( h_{2} - h_{1} ) x 
- s^2_{13} \left\{ \sin ( h_{3} - h_{2} ) x - \sin ( h_{3} - h_{1} ) x \right\} 
\biggr)
\nonumber \\
&+& 
2 c_{23} c_{13} \sin 2 \varphi 
\sin \delta 
\biggl(
c^2_{13} \cos 2 \varphi \sin^2 \frac{ ( h_{2} - h_{1} ) x }{2}
+ s^2_{13} \left\{ 
\sin^2 \frac{ ( h_{3} - h_{2} ) x }{2} - \sin^2 \frac{ ( h_{3} - h_{1} ) x }{2} 
\right\} 
\biggr)
\nonumber \\
&+&
s_{23} \sin 2\theta_{13} 
\left\{ c^2_{\varphi} \sin ( h_{3} - h_{1} ) x + s^2_{\varphi} \sin ( h_{3} - h_{2} ) x \right\} 
\biggr].
\label{Pmue-ext-UV}
\end{eqnarray}
Here, the dependence on the $\alpha_{\beta \gamma}$ is manifest. The feature of the diagonal $\alpha$ parameter correlation is vastly different from that of the unitary evolution part $P(\nu_{\mu} \rightarrow \nu_{e})_{ \text{ EV } }^{(1)}$, as will be discussed in section~\ref{sec:diag-alpha}. 

\subsection{Symmetry of the oscillation probability} 
\label{sec:symmetry}

It is observed in ref.~\cite{Martinez-Soler:2019nhb} that for each matter-dressed mixing angle $\phi$ there is an invariance under the transformation $\phi \rightarrow \phi + \frac{\pi}{2}$. $\phi$ can be $\theta_{13}$ or $\theta_{12}$ in matter.\footnote{
The $\theta_{12}$ counterpart is previously noticed in ref.~\cite{Denton:2016wmg}.
}
In our system in this paper, the oscillation probability is invariant under the transformation 
\begin{eqnarray} 
&& 
\varphi \rightarrow \varphi + \frac{\pi}{2}, 
\label{varphi-transformation-summary}
\end{eqnarray}
which induces the following transformations simultaneously 
\begin{eqnarray} 
&& 
h_{1} \rightarrow h_{2}, 
\hspace{10mm}
h_{2} \rightarrow h_{1}, 
\nonumber \\
&&
c_{\varphi} \rightarrow - s_{\varphi}, 
\hspace{6mm}
s_{\varphi} \rightarrow + c_{\varphi}, 
\hspace{6mm}
\cos 2\varphi \rightarrow - \cos 2\varphi, 
\hspace{6mm}
\sin 2\varphi \rightarrow - \sin 2\varphi.
\label{varphi-transformation}
\end{eqnarray}
Hence, $J_{mr} \rightarrow - J_{mr}$ under the transformation.

It is interesting to observe explicitly that the symmetry is respected by 
$P(\nu_{\mu} \rightarrow \nu_{e})_{ \text{ EV} }^{(1)}$ and 
$P(\nu_{\mu} \rightarrow \nu_{e})_{ \text{ UV} }^{(1)}$, whose former is given in section~\ref{sec:P-mue-int-UV} and appendix~\ref{sec:Pmue-rest}, and the latter in section~\ref{sec:P-mue-ext-UV}. 
The nature of the symmetry is identified as the ``dynamical symmetry'', not a symmetry in the Hamiltonian~\cite{Martinez-Soler:2019nhb}. Yet, it serves for a powerful consistency check of the calculation. 

\section{Dynamical correlation between $\nu$SM and the UV $\alpha$ parameters} 
\label{sec:correlation}

In this section, we discuss correlations between $\nu$SM and the UV $\alpha$ parameters, including the clustering of the latter, which are manifested in the oscillation probabilities calculated in sections~\ref{sec:P-mue-int-UV}, ~\ref{sec:P-mue-ext-UV}, and appendix~\ref{sec:Pmue-rest}.\footnote{
For most of our purposes the expressions of the flavor basis $S$ matrix, $S_{ \text{flavor} }$, are sufficiently informative, but not on the diagonal $\alpha$ parameter correlation, whose discussion requires rephasing invariant quantities. 
}

\subsection{Diagonal $\alpha$ parameter correlation} 
\label{sec:diag-alpha}

As discussed in ref.~\cite{Martinez-Soler:2018lcy}, the diagonal $\alpha$ parameters have the particular types of correlations in the evolution part of the probability 
\begin{eqnarray} 
\left( \frac{ \Delta_{a} }{ \Delta_{b} } - 1 \right) \alpha_{ee} + \alpha_{\mu \mu}, 
\hspace{8mm}
\text{and}
\hspace{8mm}
\alpha_{\mu \mu} - \alpha_{\tau \tau}, 
\label{diag-alpha}
\end{eqnarray} 
which arises due to the rephasing invariance. It becomes manifest in the would-be flavor basis $H_{ \text{wb-flavor} } \equiv U \check{H} U^{\dagger} = U_{23} \tilde{H} U_{23}^{\dagger}$. Of course, it must hold in regions of the solar-scale enhanced oscillation. In our expressions of $P(\nu_{\mu} \rightarrow \nu_{e})_{ \text{ EV } }^{(1)}$ given in section~\ref{sec:P-mue-int-UV} and in appendix~\ref{sec:Pmue-rest}, it is hidden in the diagonal $K_{jj}$ parameters in the form of $K_{jj} - K_{ii}$:\footnote{
We refer the UV paramerters, in generic contexts, as the ``$\alpha$ parameters'', but use the notation ``$\tilde{\alpha}$'' in making the statements about the formulas and the results obtained by using the SOL convention of $U_{\text{\tiny MNS}}$.  
}
\begin{eqnarray} 
K_{22} - K_{11} 
&=&
2 c^2_{13} 
\left[ \tilde{\alpha}_{ee} \left( \frac{ \Delta_{a} }{ \Delta_{b} } - 1 \right) 
+ \tilde{\alpha}_{\mu \mu} 
\right] 
- 2 ( s_{23}^2 - c_{23}^2 s^2_{13} ) 
( \tilde{\alpha}_{\mu \mu} -  \tilde{\alpha}_{\tau \tau} ) 
\nonumber \\
&-& 
2 ( 1 + s_{13}^2 ) c_{23} s_{23} \mbox{Re} \left( \tilde{\alpha}_{\tau \mu} \right) 
+ 2 c_{13} s_{13} 
\mbox{Re} \left( s_{23} \tilde{\alpha}_{\mu e} + c_{23} \tilde{\alpha}_{\tau e} \right)
\nonumber \\
K_{33} - K_{22} 
&=&
- 2 s^2_{13} \left[ 
\tilde{\alpha}_{ee} \left( \frac{ \Delta_{a} }{ \Delta_{b} } - 1 \right) 
+ \tilde{\alpha}_{\mu \mu} 
\right] 
+ 2 ( s_{23}^2 - c_{23}^2 c^2_{13} ) 
( \tilde{\alpha}_{\mu \mu} - \tilde{\alpha}_{\tau \tau} ) 
\nonumber \\
&+&
2 ( 1 + c_{13}^2 ) c_{23} s_{23} \mbox{Re} \left( \tilde{\alpha}_{\tau \mu} \right) 
+ 2 c_{13} s_{13} \mbox{Re} 
\left( s_{23} \tilde{\alpha}_{\mu e} + c_{23} \tilde{\alpha}_{\tau e} \right). 
\label{Kjj-Kii}
\end{eqnarray}
We note that $K_{33} - K_{11}$ is not independent of the above two as it is obtained by adding them. See \eqref{Kij-def} for definition of $K_{ij}$, and appendix~\ref{sec:F-K-Phi-elements} for their explicit expressions. Though the diagonal $\alpha$ parameter correlation is written in terms of the SOL convention $\tilde{\alpha}_{jj}$ variables, it is independent of the convention of $U_{\text{\tiny MNS}}$ because the variables do not depend on the convention. 

\subsection{Correlations between $\nu$SM phase $\delta$ and the $\alpha$ parameters}
\label{sec:delta-alpha}

In view of the expressions of the first order probability, its UV related but unitary part in section~\ref{sec:P-mue-int-UV} and appendix~\ref{sec:Pmue-rest}, we identify the following correlated pairs consisting of the $\delta - \alpha$ parameters, $K_{12} e^{ - i \delta }$ and $K_{23} e^{ i \delta }$, where the blobs of the $\alpha$ parameters $K_{12}$ and $K_{23}$ can be written as  
\begin{eqnarray} 
&& K_{12} e^{- i \delta} 
= 
c_{13} \left\{ c_{23} \left( \tilde{\alpha}_{\mu e} e^{ i \delta} \right)^* 
- s_{23} \left( \tilde{\alpha}_{\tau e} e^{ i \delta} \right)^* 
\right\} 
- s_{13} e^{- i \delta} 
\left[ 2 c_{23} s_{23} ( \tilde{\alpha}_{\mu \mu} - \tilde{\alpha}_{\tau \tau} ) + c_{23}^2 \tilde{\alpha}_{\tau \mu} - s_{23}^2 \tilde{\alpha}_{\tau \mu}^* \right], 
\nonumber \\
&&
K_{23} e^{ i \delta} 
= s_{13} \left\{ c_{23} \left( \tilde{\alpha}_{\mu e} e^{ i \delta} \right) 
- s_{23} \left( \tilde{\alpha}_{\tau e} e^{ i \delta} \right) \right\} 
+ c_{13} e^{ i \delta} 
\left[ 2 c_{23} s_{23} ( \tilde{\alpha}_{\mu \mu} - \tilde{\alpha}_{\tau \tau} ) + c_{23}^2 \tilde{\alpha}_{\tau \mu}^* - s_{23}^2 \tilde{\alpha}_{\tau \mu} \right]. 
\label{K12-K23}
\end{eqnarray}
Therefore, the $\delta$ - complex $\alpha$ parameter correlation {\em does exist} in the SOL convention of $U_{\text{\tiny MNS}}$, which is in marked contrast to the feature of no $\delta - \alpha$ parameter phase correlation in region of the atmospheric scale enhanced oscillation \cite{Martinez-Soler:2018lcy}. 
Notice that $K_{21} e^{ i \delta } = \left( K_{12} e^{ - i \delta } \right)^*$ and $K_{32} e^{ - i \delta } = \left( K_{23} e^{ i \delta } \right)^*$, and therefore they do not introduce correlations independent of those in \eqref{K12-K23}. In fact, the feature of the $e^{ \pm i \delta }$-$K$ blob correlation can be traced back to the form of $\Phi_{ij}$ given in appendix~\ref{sec:F-K-Phi-elements}. 

One can also conclude from the features of $\tilde{\alpha}_{\mu e}$ vs $e^{ \pm i \delta}$ correlation seen in \eqref{Kjj-Kii} and \eqref{K12-K23} there is no definite ``chiral'' combination $\tilde{\alpha}_{\mu e} e^{ i \delta}$ and/or $\tilde{\alpha}_{\tau e} e^{ i \delta}$, nor $\tilde{\alpha}_{\tau \mu} e^{ \pm i \delta}$. Consideration of the non-unitary part of the probability \eqref{Pmue-ext-UV} does not change the conclusion. 

To summarize, the feature of $\delta$ - $\alpha$ parameter correlation at around the solar scale enhanced oscillation is different from the one in region of the atmospheric scale oscillation discussed in ref.~\cite{Martinez-Soler:2018lcy}, most notably, on the following two aspects:
\begin{itemize}
\item 
The correlation between the $\nu$SM phase $\delta$ and the $\alpha$ parameters does exist in the SOL convention of $U_{\text{\tiny MNS}}$ in region of the solar scale enhanced oscillation.

\item 
But, the correlation does not have the ``chiral'' form, $\tilde{\alpha}_{\beta \gamma} e^{ \pm i \delta}$. Rather it takes the form of correlation between $e^{ \pm i \delta}$ and the blobs composed of the $\alpha$ parameters.

\end{itemize}
Since the correlation between $\delta$ and the $K_{12}$ - $K_{23}$ cluster variables lives in $\Phi$ matrix elements, which are the building block of the perturbation series, it is obvious that the correlation prevails to higher order in perturbation theory in the unitary evolution part. 

\subsection{Nature of the $\delta-\alpha$ parameter correlation: Are they
real?}
\label{sec:feature-reality}

The result in ref.~\cite{Martinez-Soler:2018lcy} shows that the SOL convention of $U_{\text{\tiny MNS}}$ is the unique case in the atmospheric-scale enhanced oscillation in which the $\delta$ - $\alpha$ parameter correlation is absent. Then, the first itemized statement above indicates that there is no $U_{\text{\tiny MNS}}$ convention in which the phase correlation is absent both at around the atmospheric- and the solar-scale enhanced oscillations. Then, we can now conclude that the $\delta - \alpha$ parameter correlations seen in this and the previous paper \cite{Martinez-Soler:2018lcy} are all physical. That is, it cannot be wiped away by a $U_{\text{\tiny MNS}}$ convention choice. 

In fact, it is very likely that, in the solar-scale enhanced region, the phase correlation exists with all the three conventions of $U_{\text{\tiny MNS}}$. 
The oscillation probability in the other $U_{\text{\tiny MNS}}$ conventions can be obtained simply by using the translation rule, eq.~\eqref{alpha-bar-alpha-tilde-alpha}.\footnote{
To close possible loophole in this statement, we performed an explicit construction of the solar resonance perturbation theory extended with the UV effect using the ATM convention of $U_{\text{\tiny MNS}}$. 
A preliminary investigation reveals that the same $\delta -$(cluster of the $\alpha$ parameters) correlation as in \eqref{K12-K23} survives, but inside $K_{ij}$ $\tilde{\alpha}_{\beta \gamma}$ must be transformed to $\alpha_{\beta \gamma}$ ($\alpha$ matrix elements in the ATM convention) by the transformation rule \eqref{alpha-bar-alpha-tilde-alpha}. It is the expected result and apparently there is no loophole in our prescription.
}
Then, we observe in the ATM and PDG conventions, even more complicated correlations between $e^{ \pm i \delta}$ and the blobs composed of the $\alpha$ parameters inside which some of the $\alpha$ parameters are attached with $e^{ \pm i \delta}$. 

One may wonder why the features of the correlation between $\alpha$ and the $\alpha$ parameters are so different between the regions of the atmospheric- and the solar-scale enhanced oscillations. But, this is entirely normal. As we have learned in section~\ref{sec:parameter-correlation}, the nature of the parameter correlation in neutrino evolution with inclusion of outside-$\nu$SM ingredients are dynamical. Their features depend on the values of the relevant parameters as well as the kinematical regions where different degrees of freedom play the dominant role. The dynamical nature of the phase correlation will be demonstrated in a visible way in section~\ref{sec:numerical-examination}.

\subsection{Clustering of the $\alpha$ parameters}
\label{sec:clustering}

In addition to the $\delta-$(blob of the $\alpha$ parameters) correlation, we observe a feature which may be called the ``clustering of the $\alpha$ parameters'' in the first order unitary evolution part of the first order oscillation probability $P(\nu_{\mu} \rightarrow \nu_{e})$ calculated in sections~\ref{sec:P-mue-int-UV} and appendix~\ref{sec:Pmue-rest}. We can identify the following  ``clustering variables'' at the level of the $\tilde{S}_{\text{EV} }^{(1)}$ matrix elements:
\begin{eqnarray} 
&&
K_{12} e^{- i \delta} + K_{21} e^{ i \delta}, 
\hspace{8mm}
c_{\varphi}^2 K_{13} - c_{\varphi} s_{\varphi} K_{23} e^{ i \delta},
\hspace{8mm}
c_{\varphi} s_{\varphi} K_{13} + c_{\varphi}^2 K_{23} e^{ i \delta}, 
\label{cluster-variables}
\end{eqnarray}
where we have not listed $(s_{\varphi}^2 K_{13} + c_{\varphi} s_{\varphi} K_{23} e^{ i \delta})$ and $(c_{\varphi} s_{\varphi} K_{13} - s_{\varphi}^2 K_{23} e^{ i \delta})$. They are not dynamically independent from the ones in \eqref{cluster-variables} because they can be generated by the symmetry transformation \eqref{varphi-transformation-summary} from the second and the third in \eqref{cluster-variables}. Also there exists the exceptional, isolated one $K_{12} e^{- i \delta}$ in eq.~\eqref{P-mue-OD1}. 

In \eqref{cluster-variables}, we did not quote the diagonal variables which come as a form of the difference, for example $(K_{22} - K_{11})$, because these combinations are enforced by rephasing invariance. But, these diagonal $\alpha$ parameter differences often come with the particular combination with the other cluster variables, e.g., as 
$\left[ \cos 2 \varphi \left( K_{22} - K_{11} \right) 
+ \sin 2 \varphi \left( K_{12} e^{- i \delta} + K_{21} e^{ i \delta} \right) \right]$, 
or $\left[ \sin 2 \varphi \left( K_{22} - K_{11} \right) 
- \cos 2 \varphi \left( K_{12} e^{- i \delta} + K_{21} e^{ i \delta} \right) \right]$. 
Moreover, the other blobs of variables $\left( c^2_{13} K_{13} - s^2_{13} K_{31} \right)$, and $\left( c^2_{13} K_{23} e^{ i \delta} - s^2_{13} K_{32} e^{ - i \delta} \right)$, which are not visible at the level of the $\tilde{S}_{\text{EV} }^{(1)}$ matrix, shows up in the oscillation probability. See eqs.~\eqref{P-mue-D-OD},  \eqref{P-mue-OD1} - \eqref{P-mue-OD3} for all the above examples of blobs. 

It appears that appearance of such cluster variables as well as the correlation between $\delta$ and the $\alpha$ parameter blobs are worth attention though we do not quite understand the cause of this phenomenon.  

\section{Physics of neutrino flavor transformation with non-unitary mixing matrix} 
\label{sec:numerical-examination}

Up to this section, we aimed at analytical understanding of the system around the region of solar-scale enhanced oscillation, for short, the ``solar region''. Likewise, we use below the simplified terminology ``atmospheric region'' for a region of enhanced atmospheric-scale oscillation. Now, we discuss physics of neutrino flavor transformation in the solar region. But, we do it in comparison with that of atmospheric region as it proves to be more revealing. We try to illuminate some new aspects of the system of the three-flavor active neutrinos with non-unitary mixing matrix by using the numerical method together with our first order formula.

We use the PDG convention of $U_{\text{\tiny MNS}}$ in all the computations in this section, because it is used in most of the analyses of neutrino flavor transformations. We also depart from our ``official'' notations $\bar{\alpha}_{\beta \gamma}$ of the $\alpha$ parameters in the PDG convention defined in section~\ref{sec:mass-basis}, and simply denote them as $\alpha_{\beta \gamma}$ in this section beyond the next subsection~\ref{sec:exact-perturbative}.

\subsection{Use of the exact and perturbative oscillation probabilities: General convention of $U_{\text{\tiny MNS}}$}
\label{sec:exact-perturbative} 

Toward the goal, we utilize the perturbative oscillation probability derived in section~\ref{sec:general-formula-P}, as well as the exact formula for the probability based on numerical integration of the evolution equation whose latter is valid even for a varied matter density.\footnote{
If the uniform matter density approximation applies one can also use the exact analytic formula for the probability derived in ref.~\cite{Fong:2017gke}. We note that the expression is reasonably simple despite its exactitude. 
}

It was pointed out in ref.~\cite{Martinez-Soler:2018lcy} that the $\alpha$ matrix depends on convention of $U_{\text{\tiny MNS}}$. By using the property, one can derive probability formula in the PDG or ATM conventions by using the substitution rule from the $\tilde{\alpha}$ parameter in the SOL convention to the $\bar{\alpha}$ parameter in the PDG, or the $\alpha$ parameter in the ATM conventions. See eq.~\eqref{alpha-bar-alpha-tilde-alpha} in section~\ref{sec:mass-basis}. 
One can also transform to a general $U_{\text{\tiny MNS}}$ convention by using the phase redefinition $U(\beta, \gamma)$ defined in ref.~\cite{Martinez-Soler:2018lcy}. Notice that the translation rule applies not only in the perturbative formulas but also in the exact formulas. 

\subsection{Overview of the effect of UV} 
\label{sec:overview}

The first step to understand the effect of non-unitarity that brought into the $\nu$SM three neutrino system is to know where and how strongly the UV $\alpha$ parameters  affect the neutrino flavor transformation. For this purpose, we turn on each $\alpha_{\beta \gamma}$ parameter one by one and calculate the non-unitary contribution to the appearance probability $\Delta P_{\mu e}$ defined by 
\begin{eqnarray}
&& 
\Delta P_{\mu e} \equiv 
P(\nu_{\mu} \rightarrow \nu_{e}) 
- P(\nu_{\mu} \rightarrow \nu_{e})_{ \nu\text{SM} } 
=
P(\nu_{\mu} \rightarrow \nu_{e})_{ \text{EV} } 
+ P(\nu_{\mu} \rightarrow \nu_{e})_{ \text{UV} } 
\label{Pmue-UV-part} 
\end{eqnarray} 
where $P(\nu_{\mu} \rightarrow \nu_{e})$ in eq.~\eqref{Pmue-UV-part} denotes the appearance probability in the $\nu_{\mu} \rightarrow \nu_{e}$ channel with the UV effect fully implemented. Both $P(\nu_{\mu} \rightarrow \nu_{e})$ and $P(\nu_{\mu} \rightarrow \nu_{e})_{ \nu\text{SM} }$ are computed numerically. In all the calculations in this section, the matter density is taken to be $\rho = 3.2~{\rm g\,cm}^{-3}$ over the entire baseline.

\begin{figure}[h!]
\begin{center}
\includegraphics[width=0.94\textwidth]{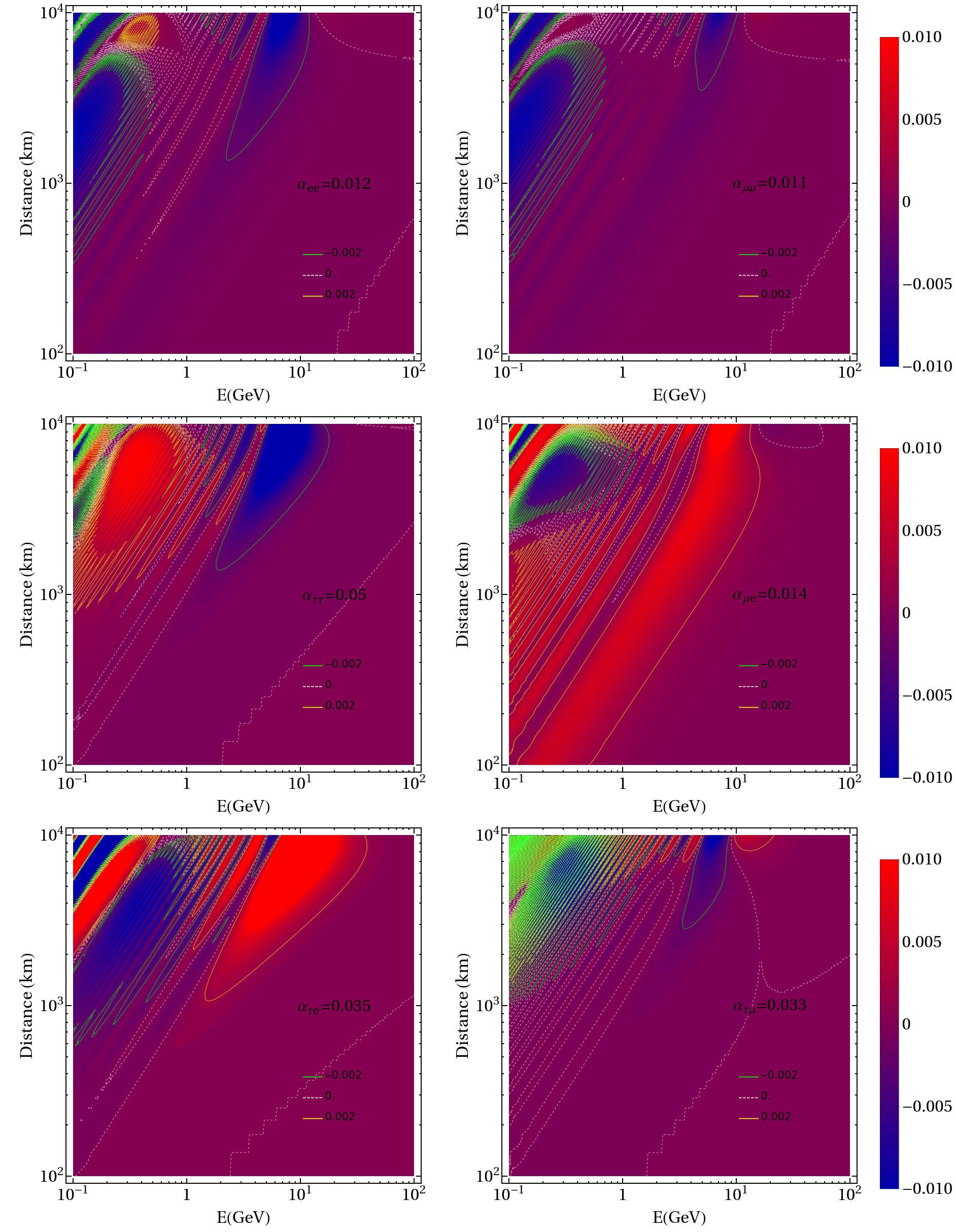}
\end{center}
\vspace{-3mm}
\caption{Plotted is $\Delta P_{\mu e} \equiv P(\nu_{\mu} \rightarrow \nu_{e}) - P(\nu_{\mu} \rightarrow \nu_{e})_{ \nu\text{SM} }$ by turning on one $\alpha_{\beta \gamma}$ at one time, in order from the top-left to the bottom-right panels, $\alpha_{e e}$, $\alpha_{\mu \mu}$, $\alpha_{\tau \tau}$, $\alpha_{\mu e}$, $\alpha_{\tau e}$, and $\alpha_{\tau \mu}$. We take the value of each $\alpha_{\beta \gamma}$ as half of the bound obtained by Blennow {\it et al.} \cite{Blennow:2016jkn} given in Table 2 in appendix A: 
$\alpha_{ee} = 0.012$, $\alpha_{\mu \mu} = 0.011$, $\alpha_{\tau \tau} = 0.05$, $\alpha_{\mu e} = 0.014$, $\alpha_{\tau e} = 0.035$, and $\alpha_{\tau \mu} = 0.033$. The matter density is taken to be $\rho = 3.2~{\rm g\,cm}^{-3}$ over the entire baseline.
} 
\label{fig:each-alpha} 
\end{figure}

In figure~\ref{fig:each-alpha} we show $\Delta P_{\mu e}$ by using color grading guided by the contour lines. In each panel we turn on one of $\alpha_{\beta \gamma}$, from the top-left to the bottom-right panels in order, $\alpha_{e e}$, $\alpha_{\mu \mu}$, $\alpha_{\tau \tau}$, $\alpha_{\mu e}$, $\alpha_{\tau e}$, and $\alpha_{\tau \mu}$.\footnote{
Notice that if the diagonal $\alpha$ parameters enter into the probability in the form $\alpha_{\beta \beta} - \alpha_{\gamma \gamma}$, only two of the three diagonal $\alpha$ parameters are independent. But, since this subtractive dependence holds only in first order in UV expansion \cite{Martinez-Soler:2018lcy}, the independent bounds exist for all three of them. It is in sharp contrast to the situation of the NSI parameters. }
In this section, we turn on only one of the $\alpha_{\beta \gamma}$ parameters in each panel, except for the top-right and bottom two panels in figure~\ref{fig:EV-vs-UV}. To have an insight into the required accuracy of the $P(\nu_{\mu} \rightarrow \nu_{e})$ measurement to improve the current bounds by a factor of 2, we take the value of each $\alpha_{\beta \gamma}$ as half of the bound obtained by Blennow {\it et al.} \cite{Blennow:2016jkn} with the positive sign.
Figure~\ref{fig:each-alpha} as a whole, displays how large is the UV effect depending upon the energy $E$ and the baseline $L$. The ``mountain ridges'' roughly follow the line of $L/E=$ constant. The atmospheric and the solar MSW enhancements are visible, respectively, at around $E \sim 10$ GeV and near the upper end of $L=10^{4}$ km and $E \simeq$ several $\times 100$ MeV and $E \simeq$ several $\times 1000$ km. 

We observe the two salient features:

\begin{itemize}
\item
$\Delta P_{\mu e}$ is at most $\simeq \pm$1\% level in all the panels in figure~\ref{fig:each-alpha}, which means a 1\% level measurement of the probability is necessary for a factor 2 improvement of the bounds.

\item
$\Delta P_{\mu e}$ changes sign depending upon which $\alpha_{\beta \gamma}$ is turned on, and on region of kinematical phase space, e.g., in the atmospheric region, or the solar region. 

\end{itemize}
\noindent
The 1\% accuracy measurement of the probability is mentioned at the end of section~\ref{sec:region-validity} in relationship with the possible target accuracy of constraining UV $\alpha$ parameters.

For the second point above, we notice in figure~\ref{fig:each-alpha} that with turning on $\alpha_{\tau \tau}$ (middle-left panel) $\Delta P_{\mu e}$ is positive in the solar region and negative in the atmospheric region. On the other hand, this tendency is reversed completely with $\alpha_{\tau e}$ (bottom-left panel), and less completely with $\alpha_{\mu e}$ (middle-right panel). In the other cases, $\Delta P_{\mu e}$ is negative in the both regions of the atmospheric-scale and solar-scale enhancement. It means that if we turn on all $\alpha_{\beta \gamma}$ at once, the effect of each element may cancel with each other at least partly. One must also take into account the fact that since we do not know a priori the sign of the $\alpha$ parameters, the pattern of the cancellation can be more complicated when all the parameters are turned on with arbitrary signs, or if phases are attached to the off-diagonal $\alpha$ parameters. It implies that (1) determination of the UV $\alpha_{\beta \gamma}$ parameters (assuming its existence) could have additional difficulties due to confusion and degeneracy caused by the cancellation between the effect of different $\alpha$ parameters, (2) the bound on UV obtained by using ``one $\alpha_{\beta \gamma}$ turned at one time'' procedure could have made the bound artificially stronger than the one obtained with the proper procedure of ``all $\alpha_{\beta \gamma}$ turned on but the rest of them marginalized''.

The features of possible cancellation between the effect of $\alpha_{\beta \gamma}$ parameters may add another difficulty to the task of identifying their effects, an already highly nontrivial one due to high precision required to measurement of the probability. Therefore, further discussion of the question of how to disentangle the effects of different alpha parameters is called for. 

\subsection{Unitary vs. non-unitary pieces of the UV related oscillation probability} 
\label{sec:Unitary-vs-UV}

The UV $\alpha$ parameter related part of the probability $\Delta P_{\mu e}$ decomposes into the two parts, the unitary evolution part $P(\nu_{\mu} \rightarrow \nu_{e})_{ \text{ EV} }$ and the genuine non-unitary part $P(\nu_{\mu} \rightarrow \nu_{e})_{ \text{ UV} }$ ~\cite{Martinez-Soler:2018lcy}, see eq.~\eqref{S-flavor-all}. Then, a natural question is which part is larger or dominating, and whether they mutually tend to add up or cancel with each other. 

\begin{figure}[h!]
\begin{center}
\vspace{3mm}
\includegraphics[width=0.48\textwidth]{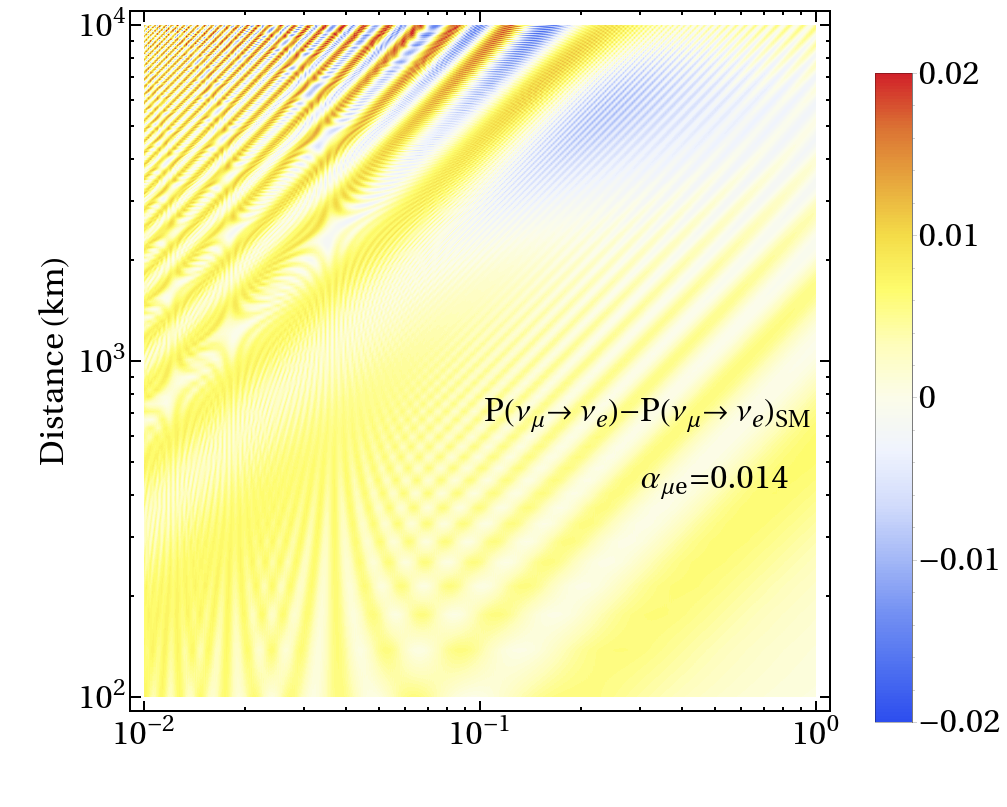}
\includegraphics[width=0.48\textwidth]{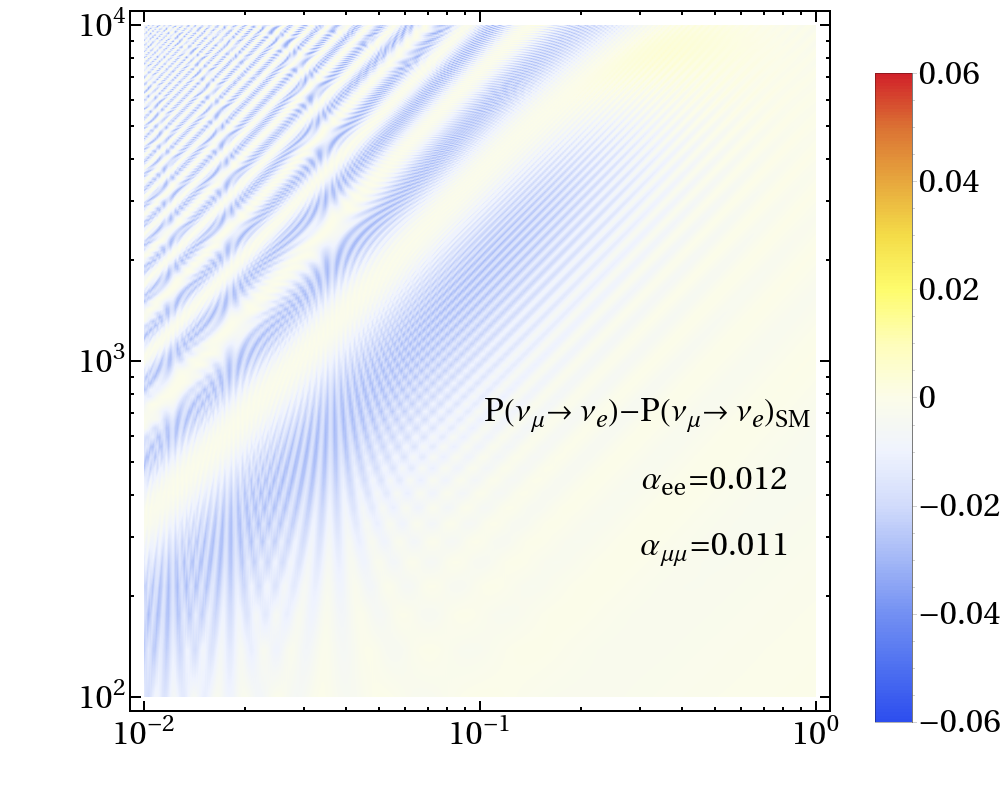}
\includegraphics[width=0.90\textwidth]{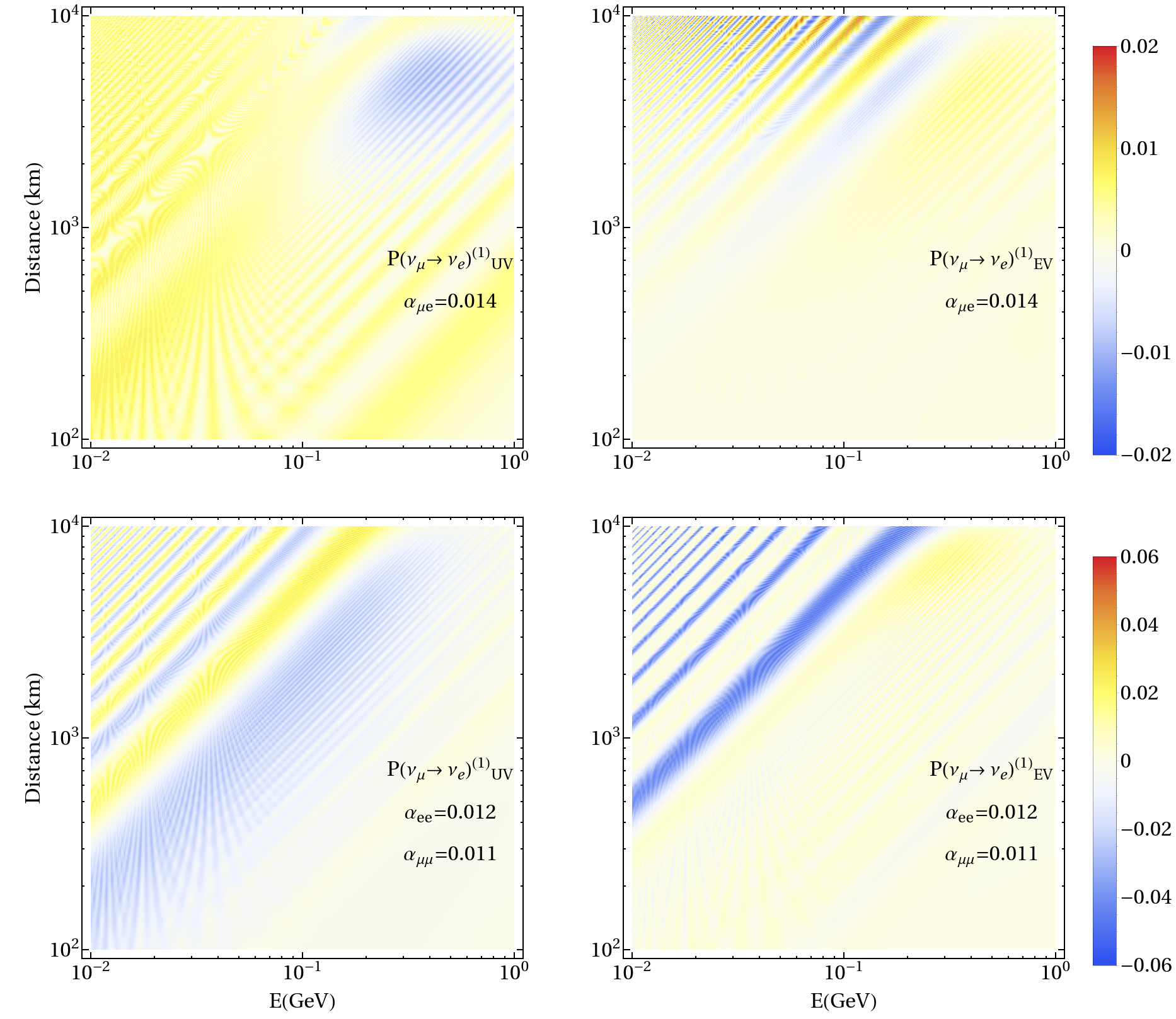}
\end{center}
\vspace{-4mm}
\caption{In the top two panels, $\Delta P_{\mu e} = P(\nu_{\mu} \rightarrow \nu_{e})_{ \text{ UV} } + P(\nu_{\mu} \rightarrow \nu_{e})_{ \text{ EV} }$ are presented by the color grading, with $\alpha_{\mu e}=0.014$ (left panel), and with $\alpha_{ee}=0.012$ and $\alpha_{\mu \mu}=0.011$ (right panel). In the middle and bottom panels $\Delta P_{\mu e}$ is decomposed to $P(\nu_{\mu} \rightarrow \nu_{e})_{ \text{ UV} }$ and $P(\nu_{\mu} \rightarrow \nu_{e})_{ \text{ EV} }$ in the left and right panels, respectively. 
} 
\label{fig:EV-vs-UV} 
\end{figure}

These questions are answered by figure~\ref{fig:EV-vs-UV}. In the top two panels the whole UV effects, $\Delta P_{\mu e} = P(\nu_{\mu} \rightarrow \nu_{e})_{ \text{ UV} } + P(\nu_{\mu} \rightarrow \nu_{e})_{ \text{ EV} }$ are presented, with $\alpha_{\mu e}=0.014$ only in the left panel, and with $\alpha_{ee}=0.012$ and $\alpha_{\mu \mu}=0.011$ in the right panel. The values of the $\alpha$ parameters are the same as used in figure~\ref{fig:each-alpha}, and hence the left panel overlaps with a part of the middle-right panel of figure~\ref{fig:each-alpha}. 

The decomposition of $\Delta P_{\mu e}$ into $P(\nu_{\mu} \rightarrow \nu_{e})_{ \text{ UV} }$ and $P(\nu_{\mu} \rightarrow \nu_{e})_{ \text{ EV} }$ is displayed in the middle ($\alpha_{\mu e}=0.014$ case) and bottom ($\alpha_{ee}=0.012$ and $\alpha_{\mu \mu}=0.011$ case) panels of figure~\ref{fig:EV-vs-UV}, respectively. We restrict ourselves into the two choices of the $\alpha$ parameters because $P(\nu_{\mu} \rightarrow \nu_{e})_{ \text{ UV} }$ in first order depends only on the two combinations, $\alpha_{\mu e}$ and $\alpha_{ee} + \alpha_{\mu \mu}$. 
$P(\nu_{\mu} \rightarrow \nu_{e})_{ \text{EV} }$ is computed by using the formula $P(\nu_{\mu} \rightarrow \nu_{e}) - P(\nu_{\mu} \rightarrow \nu_{e})_{ \nu\text{SM} } - P(\nu_{\mu} \rightarrow \nu_{e})_{ \text{UV} }^{(1)}$ with the first-order expression of $P(\nu_{\mu} \rightarrow \nu_{e})_{ \text{UV} }$, and hence $P(\nu_{\mu} \rightarrow \nu_{e})_{ \text{EV} }$ is accurate only to first order. 

An overall feature is that in wide areas in figure~\ref{fig:EV-vs-UV} $P(\nu_{\mu} \rightarrow \nu_{e})_{ \text{UV} }$ and $P(\nu_{\mu} \rightarrow \nu_{e})_{ \text{EV} }$ tend to cancel with each other. In looking into the figure more closely, however, we observe a little more intricate features. In the $\alpha_{\mu e}=0.014$ case (middle panels), above $L/E = 10^4~\mbox{km} / 230~\mbox{MeV}$ line, $P(\nu_{\mu} \rightarrow \nu_{e})_{ \text{UV} }$ contributes to lift up the probability, enhancing the yellow regions of $P(\nu_{\mu} \rightarrow \nu_{e})_{ \text{EV} }$ into the thinker ones in $\Delta P_{\mu e}$. Below the line, $P(\nu_{\mu} \rightarrow \nu_{e})_{ \text{UV} }$ is more dominating in the blue solar resonance region, but partially cancelled by $P(\nu_{\mu} \rightarrow \nu_{e})_{ \text{EV} }$. The cancellation is even more prominent in the bottom panels, the case with $\alpha_{ee} + \alpha_{\mu \mu}$ turned on. The overall feature of the color-graded contour of $\Delta P_{\mu e}$ is similar to that of $P(\nu_{\mu} \rightarrow \nu_{e})_{ \text{UV} }$, but $P(\nu_{\mu} \rightarrow \nu_{e})_{ \text{EV} }$ over-cancels the peaks of $P(\nu_{\mu} \rightarrow \nu_{e})_{ \text{UV} }$ above the $L/E = 10^4~\mbox{km} / 230~\mbox{MeV}$ line. 

The feature of cancellation is akin to, but is much more prominent compared to the one observed in the ``atmospheric region'' in ref.~\cite{Martinez-Soler:2018lcy}. Unfortunately, we cannot offer physical explanation on why the cancellation between $P(\nu_{\mu} \rightarrow \nu_{e})_{ \text{UV} }$ and $P(\nu_{\mu} \rightarrow \nu_{e})_{ \text{EV} }$ takes place, and why the feature is common to both the atmospheric and the solar regions. 
In most of the regions it acts as a partial ``hiding mechanism'' of non-unitarity since a less prominent effect is left in the observable, the appearance probability $P(\nu_{\mu} \rightarrow \nu_{e})$. To obtain the information of the genuine non-unitary part $P(\nu_{\mu} \rightarrow \nu_{e})_{ \text{UV} }$, it must be complemented by measurement of departure from unitarity, $P(\nu_{\mu} \rightarrow \nu_{e}) + P(\nu_{\mu} \rightarrow \nu_{\mu}) + P(\nu_{\mu} \rightarrow \nu_{\tau}) \neq 1$. 

\subsection{$\nu$SM - $\alpha$ parameter phase correlation: The atmospheric vs. solar regions}
\label{sec:phase-correlation-atm-solar} 

We have learned in the previous section~\ref{sec:correlation} that the features of the parameter correlation between the $\nu$SM and the UV new physics parameters in the solar region is different from the ones in the atmospheric region. A new $\delta$ - (blobs of the $\alpha$ parameters) correlation is observed. Then, it is natural to ask the question: What is the feature of $\nu$SM - UV parameter CP phase correlation in the solar region, and which characteristic difference it has from those in the atmospheric region?

To discuss correlation between $\delta$ and phases of the off-diagonal $\alpha$ parameters, we parametrize the latter as 
\begin{eqnarray}
&& 
\alpha_{\beta \gamma} = \vert \alpha_{\beta \gamma} \vert e^{ i \phi_{\beta \gamma} }, 
\label{amue-parametrize}
\end{eqnarray} 
where $\beta \gamma = \mu e, \tau e, \tau \mu$. 
To make the phase correlation visible clearly, we use $\Delta P_{\mu e} \equiv P(\nu_{\mu} \rightarrow \nu_{e}) - P(\nu_{\mu} \rightarrow \nu_{e})_{ \nu\text{SM} }$ defined in eq.~\eqref{Pmue-UV-part}, not the probability itself. 

\begin{figure}[h!]
\begin{center}
\vspace{2mm}
\includegraphics[width=0.76\textwidth]{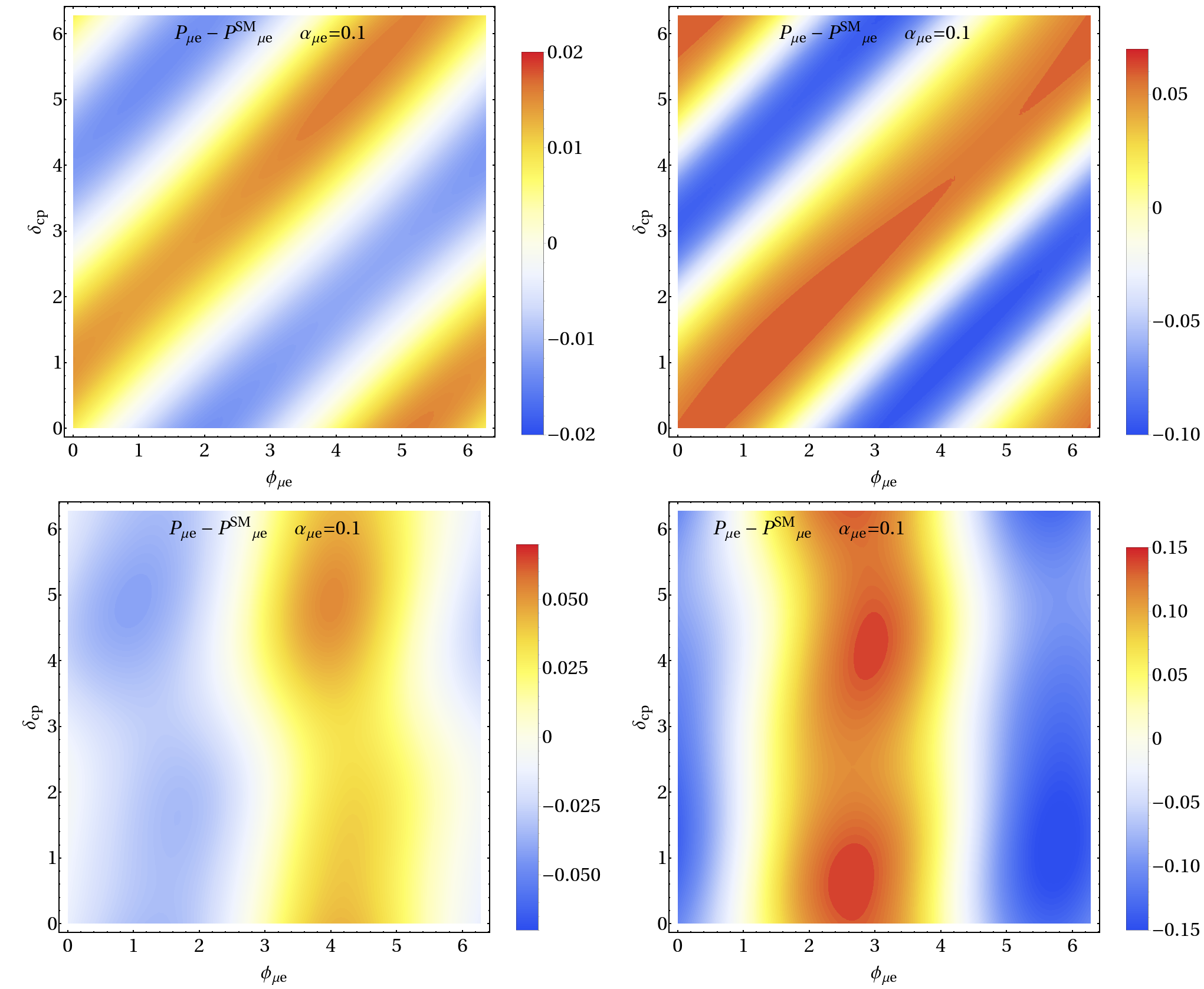}
\includegraphics[width=0.38\textwidth]{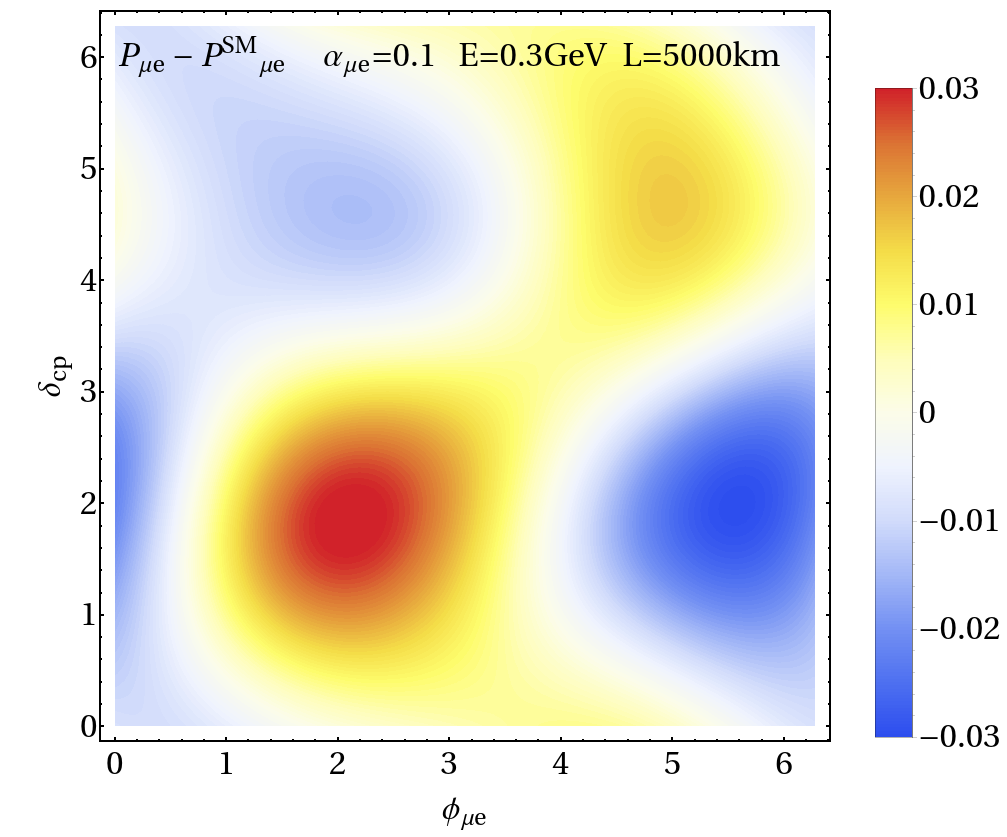}
\end{center}
\vspace{-3mm}
\caption{$\Delta P_{\mu e} \equiv P(\nu_{\mu} \rightarrow \nu_{e}) - P(\nu_{\mu} \rightarrow \nu_{e})_{ \nu\text{SM} }$ is presented in $\phi_{\mu e}$ - $\delta$ plane by color graduation, which is calculated by turning on $\alpha_{\mu e}=0.1$ only. The top two panels are in the atmospheric region with energy $E=10$ GeV, and the middle two panels are in the solar region with energy $E=200$ MeV. The baseline is taken as $L=3000$ km (left panel) and $L=12000$ km (right panel), in both the top and middle panels. The bottom panel is in the solar region with $E=300$ MeV and $L=5000$ km.
} 
\vspace{-3mm}
\label{fig:Pmue-delta-phi-mue} 
\end{figure}

\begin{figure}[h!]
\begin{center}
\vspace{1mm}
\includegraphics[width=0.70\textwidth]{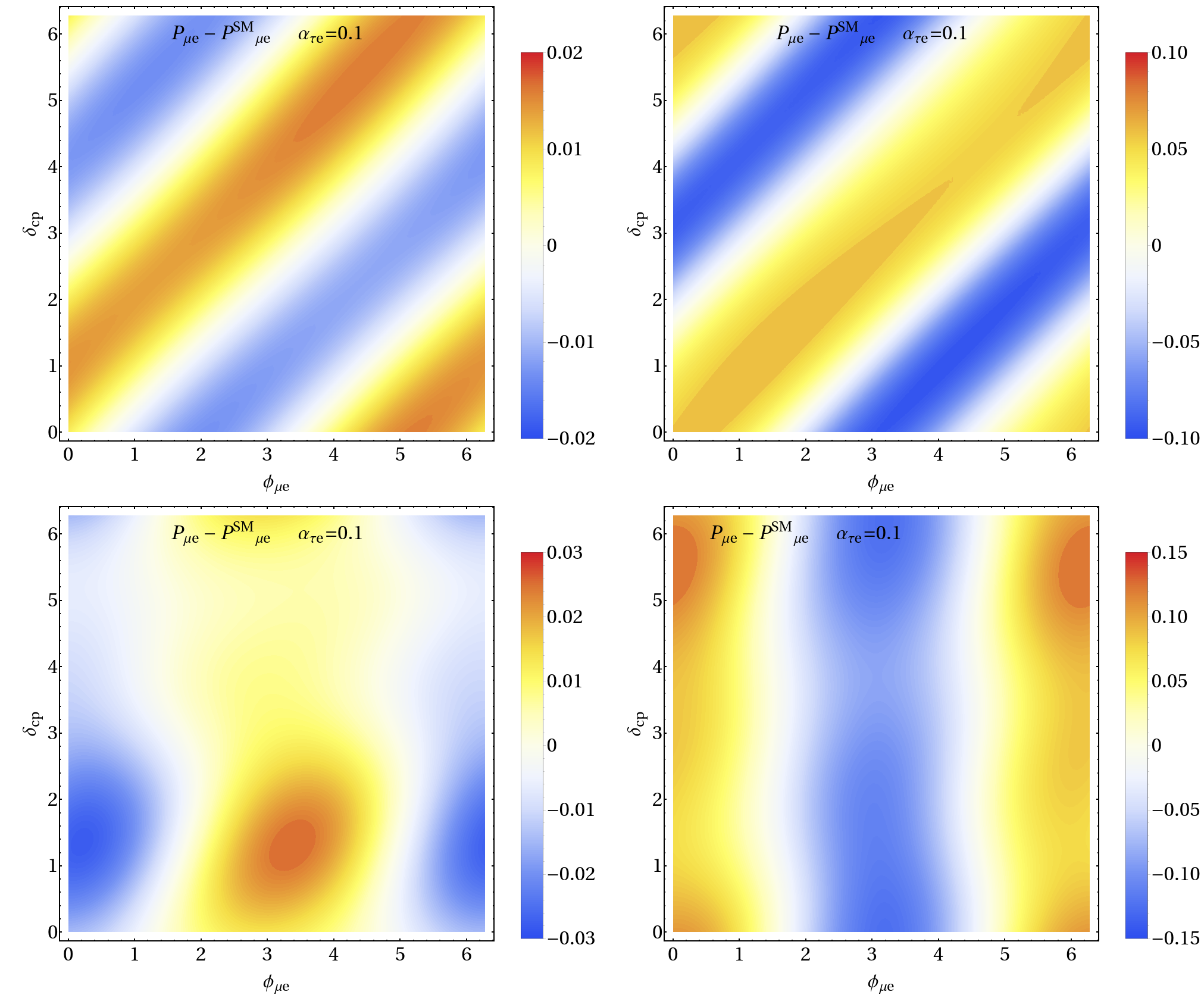}
\end{center}
\vspace{-4mm}
\caption{$\Delta P_{\mu e} \equiv P(\nu_{\mu} \rightarrow \nu_{e}) - P(\nu_{\mu} \rightarrow \nu_{e})_{ \nu\text{SM} }$ is presented in $\phi_{\mu e}$ - $\delta$ plane by color graduation, which is calculated by turning on $\alpha_{\tau e}=0.1$ only. The upper two panels are in the atmospheric region with energy $E=10$ GeV, and the lower two panels are in the solar region with energy $E=200$ MeV. The baseline is taken as $L=3000$ km (left panel) and $L=12000$ km (right panel), in both the upper and lower panels. }  
\label{fig:Pmue-delta-phi-taue} 
\end{figure}

\begin{figure}[h!]
\begin{center}
\vspace{2mm}
\includegraphics[width=0.70\textwidth]{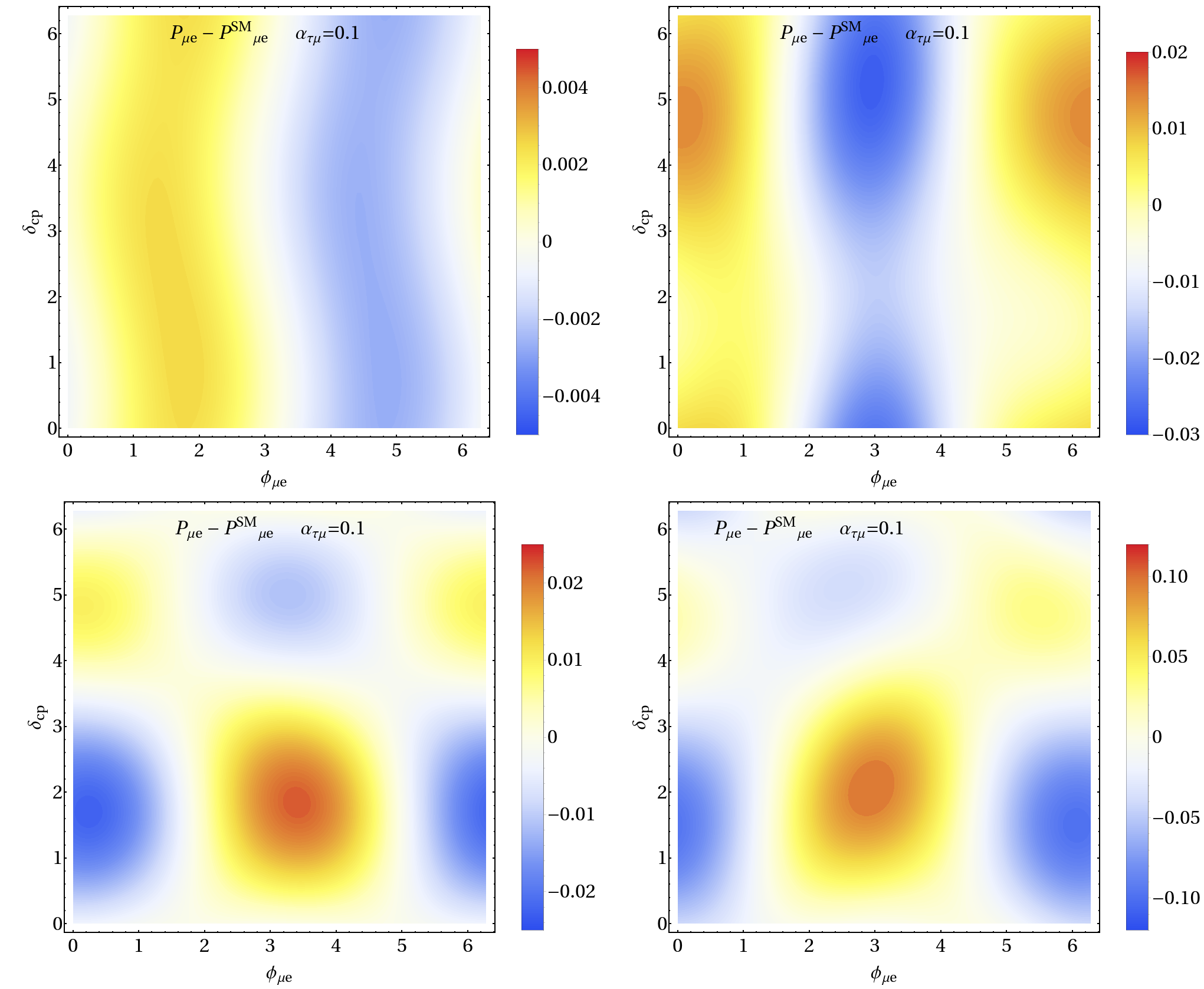}
\end{center}
\vspace{-3mm}
\caption{The same as in figure~\ref{fig:Pmue-delta-phi-taue} but with only $\alpha_{\tau \mu}=0.1$ is turned on.
} 
\vspace{-4mm}
\label{fig:Pmue-delta-phi-taumu} 
\end{figure}

In figures~\ref{fig:Pmue-delta-phi-mue}, the non-unitary contribution to the appearance probability $\Delta P_{\mu e}$ computed by turning on $\alpha_{\mu e}$ only, is presented on $\phi_{\mu e}$ - $\delta$ plane by showing the equi-contours of $\Delta P_{\mu e}$ with the color grading. In figures~\ref{fig:Pmue-delta-phi-taue} and~\ref{fig:Pmue-delta-phi-taumu}, the results of the similar exercises are presented, the case with $\alpha_{\tau e}$ turned on (figure~\ref{fig:Pmue-delta-phi-taue}), and the one with $\alpha_{\tau \mu}$ (figure~\ref{fig:Pmue-delta-phi-taumu}). 
In figures~\ref{fig:Pmue-delta-phi-mue}, ~\ref{fig:Pmue-delta-phi-taue}, and~\ref{fig:Pmue-delta-phi-taumu}, we use a large value $\alpha_{\beta \gamma} =0.1$ to enhance effects of the phase correlation, which merits higher visibility. 
The global features of the $\delta$ - $\alpha$ parameter phase correlation shown in figures~\ref{fig:Pmue-delta-phi-mue}, ~\ref{fig:Pmue-delta-phi-taue} and~\ref{fig:Pmue-delta-phi-taumu} are: 

\begin{itemize}
\item
The linear, oblique correlation seen in the case of $\alpha_{\mu e} \neq 0$ (figure~\ref{fig:Pmue-delta-phi-mue}) and $\alpha_{\tau e} \neq 0$ (figure~\ref{fig:Pmue-delta-phi-taue}) both in the atmospheric region shown in the upper panels, but no clearly visible correlation in all the other panels. 

\item
The absolute value of $\vert \Delta P_{\mu e} \vert$ is larger in the panels with baseline $L=12000$ km than those with $L=3000$ km by a factor of $\sim5$. The statement applies to all the panels including both the atmospheric and solar regions.

\end{itemize}
\noindent
Let us start from discussion of the phase correlation seen in the atmospheric region, the top two panels in figures~\ref{fig:Pmue-delta-phi-mue}, ~\ref{fig:Pmue-delta-phi-taue} and~\ref{fig:Pmue-delta-phi-taumu}. The linear, oblique correlation seen in figures~\ref{fig:Pmue-delta-phi-mue} and~\ref{fig:Pmue-delta-phi-taue}, $\phi_{\mu e}$ - $\delta$ and $\phi_{\tau e}$ - $\delta$ correlation, respectively, and no visible correlation between $\phi_{\tau \mu}$ and $\delta$\footnote{
If the probability calculated by first-order helio-UV perturbation theory~\cite{Martinez-Soler:2018lcy} is sufficiently accurate, there should be no $\delta$ dependence in the upper two panels in figure~\ref{fig:Pmue-delta-phi-taumu}, because the $\delta$ dependence would have been eliminated by the subtraction of $P(\nu_{\mu} \rightarrow \nu_{e})_{ \nu\text{SM} }$. Obviously, it is not the case. Notice that the results presented in figures~\ref{fig:Pmue-delta-phi-mue}, ~\ref{fig:Pmue-delta-phi-taue} and~\ref{fig:Pmue-delta-phi-taumu} are accurate as they do not rely on perturbative treatment.
It means that the perturbative treatment fails to provide accurate description of the probability, which is natural due to the large value 0.1 taken for $\alpha_{\tau \mu}$. In fact, the remaining $\delta$ dependence is up to a few $\times 10^{-3}$ level for $L=3000$ km, and is of order $\sim \pm0.02$ for $L=12000$ km, so that our interpretation may be valid. 
} 
shown in figure~\ref{fig:Pmue-delta-phi-taumu} (all in the upper two panels) 
is perfectly consistent with the ``canonical phase combination''~\cite{Martinez-Soler:2018lcy} 
\begin{eqnarray} 
e^{- i \delta } \alpha_{\mu e}, ~~
e^{ - i \delta} \alpha_{\tau e}, ~~
\alpha_{\tau \mu}, 
\label{C-combination-PDG}
\end{eqnarray}
which holds under the PDG convention of $U_{\text{\tiny MNS}}$. One should note the nontrivial $U_{\text{\tiny MNS}}$ convention dependence: In the ATM phase convention of $U_{\text{\tiny MNS}}$ (in which $e^{ \pm i \delta}$ is attached to $s_{23}$), the phase correlation takes the form $[e^{- i \delta } \alpha_{\mu e}, \alpha_{\tau e}, e^{i \delta} \alpha_{\tau \mu}]$~\cite{Martinez-Soler:2018lcy}. 

On the other hand, the features of the phase correlation in the solar region shown in the lower panels in figures~\ref{fig:Pmue-delta-phi-mue}, ~\ref{fig:Pmue-delta-phi-taue}, and~\ref{fig:Pmue-delta-phi-taumu} are more subtle and not easy to understand. In some panels, the equal-$\Delta P_{\mu e}$ contours are vertical, which may imply that there is no significant correlation between $\delta$ and $\alpha$ parameter phases. In the other, there exists ``circular-shaped correlation'' with positive and negative signs of $\Delta P_{\mu e}$ in the two-dimensional phase space. 
Notice that in the panels with vertical correlation and with ``circular correlation'', the $\delta$ (in-) dependence cannot be understood as a remnant of insufficient subtraction of the $\nu$SM part. It is because the values of $\Delta P_{\mu e}$ and its variation in $\phi$ or $\delta$ directions can be as large as $\sim0.1$, of the order of the $\alpha$ parameter that is turned on. The feature of the $\phi$ - $\delta$ phase correlation, in particular, coexistence of the vertical and circular shaped correlations is not understood, regrettably, by our analytic framework.\footnote{
It appears that there are some regularities which may be relevant for understanding of the phase correlation in the solar region. That is, we often observe ``red'' ($\Delta P_{\mu e} > 0$) and ``blue'' ($\Delta P_{\mu e} < 0$) vertical contours in central region, $\phi \sim \pi$. It is likely that the central ``red'' vertical correlation corresponds to the region of negative $\Delta P_{\mu e}$ in figure~\ref{fig:each-alpha}. Whereas the central ``blue'' vertical correlation corresponds to the region of positive $\Delta P_{\mu e}$ in figure~\ref{fig:each-alpha}. For the latter, we refer the lower-right panel of figure~\ref{fig:Pmue-delta-phi-taue}, and the case of $\alpha_{\tau e}=0.1$, $E=200$ MeV and $L=5000$ km.
}

With regard to the baseline dependence of the strength of the correlation, it might  be that $\vert \Delta P_{\mu e} \vert$ itself is larger at the longer baseline of $L=12000$ km among the two baselines we have chosen to display in figures~\ref{fig:Pmue-delta-phi-mue}, ~\ref{fig:Pmue-delta-phi-taue} and~\ref{fig:Pmue-delta-phi-taumu}. 

The features of the $\alpha$ parameter phase - $\nu$SM $\delta$ correlation in the atmospheric and the solar regions presented in figures~\ref{fig:Pmue-delta-phi-mue}, ~\ref{fig:Pmue-delta-phi-taue} and~\ref{fig:Pmue-delta-phi-taumu} testify that the nature of the correlation is quite dynamical, confirming our view that stated in section~\ref{sec:parameter-correlation}. Unfortunately, physical understanding of the features of the phase correlation in the solar region are not yet achieved, which calls for further studies.

\section{Concluding remarks} 
\label{sec:conclusion}

In this paper, we have attempted to achieve physics understanding of the three-flavor neutrino system with non-unitary mixing matrix. We have focused our discussion on elucidating the nature of parameter correlations in such system, in particular the one between the $\nu$SM and the UV new physics parameters. We do it in region of the solar-scale oscillations, for short the ``solar region'', in this paper. It nicely complements the one given in our previous paper~\cite{Martinez-Soler:2018lcy} which dealt with the region of atmospheric-scale oscillations, the ``atmospheric region''. 

Toward the goal, we have formulated a new perturbative framework to discuss effect of non-unitary mixing matrix in the solar region, the UV extended version of the ``solar-resonance perturbation theory'' \cite{Martinez-Soler:2019nhb}. It was necessary to resolve the question raised in ref.~\cite{Martinez-Soler:2018lcy} which casts doubt physical reality of the correlation between the $\nu$SM $\delta $ and the phases of UV $\alpha$ parameters. But, in turn, the framework serves as a powerful analytic machinery of analyzing the features of parameter correlation in the solar region.
The skepticism about the reality of the phase correlation, which is described in detail in section~\ref{sec:introduction}, is cleared up by showing that the phase correlation {\em does exist} in the solar region with the SOL ($e^{ \pm i \delta }$ attached to $s_{12}$) convention of $U_{\text{\tiny MNS}}$. See section~\ref{sec:correlation}. 

In fact, we have uncovered that the features of the $\nu$SM - UV parameter correlations are much more profound than we thought. This point can be illuminated most clearly by contrasting the atmospheric region to the solar one. In the atmospheric region, the most notable feature is the $\nu$SM $\delta$ - UV $\alpha$ parameter phase correlation of the ``chiral type'', $[e^{- i \delta } \alpha_{\mu e}, e^{ - i \delta} \alpha_{\tau e}, \alpha_{\tau \mu}]$ in the PDG convention of $U_{\text{\tiny MNS}}$~\cite{Martinez-Soler:2018lcy}. The picture no more holds in the solar region, and the correlation takes the form of $\delta$ - (blobs of the $\alpha$ parameters) correlation as we saw in section~\ref{sec:delta-alpha}. Another interesting observation in this context is that when we move the kinematical region from $E/L = 200 \mbox{MeV} / 3000 \mbox{km}$ to $E/L = 300 \mbox{MeV} / 5000 \mbox{km}$, the $\delta$ - $\phi_{\mu e}$ correlation takes vastly different forms as shown in figure~\ref{fig:Pmue-delta-phi-mue}, where $\phi_{\mu e}$ denotes the phase of $\alpha_{\mu e}$. 

We have utilized the analytic framework developed in this paper as well as the numerical method to reveal more generic feature of the effects of the UV $\alpha$ parameters. In addition to the above mentioned ones, we have observed that the effect of non-unitarity tends to cancel between the unitary evolution part (denoted as ``EV'') and the non-unitary part (denoted as ``UV'') of the probability, and between the different $\alpha_{\beta \gamma}$ parameters. See section~\ref{sec:numerical-examination}. 

One of the most intriguing features of the parameter correlation is that the form of the correlation depends also on the values of the mixing parameters. The phenomenon is briefly mentioned in section~\ref{sec:dynamical-nature} that as $\theta_{13}$ becomes larger, the correlation seen at smaller $\theta_{13}$ starts to dissolve. Since we cannot control the values of the mixing angles or $\Delta m^2$ by ourselves, the discussion might look as appealing only to an academic interest. But, we believe that it merits deepening our understanding on the mechanism and the cause of parameter correlation. We were not able to explore this point further in this paper, and a focused investigation on this issue is called for. 

All these features of the parameter correlation may be summarized by the term {\em ``dynamical nature of the parameter correlation''}.

Finally, we remark that occurrence of dynamical correlations between the parameters in systems with the many degree of freedom is very common, as discussed in section~\ref{sec:parameter-correlation}. A rich variety of correlations we encountered in our system with non-unitarity adds another example to this list. If one chooses the way of testing leptonic unitarity by setting up a class of models with UV and confrontation of them with experimental data, understanding the system with UV would be indispensable step to carry out this task. Yet we must emphasize that our understanding on the system, e.g., on the parameter correlation, is far from sufficient, generically in the system with new physics beyond the $\nu$SM.

On the experimental side, if we want to utilize the low energy region with the solar-scale enhanced oscillation, in the context of {\em precision} unitarity test, possible advantage of the Kamioka-Korea identical two-detector setup \cite{Ishitsuka:2005qi,Kajita:2006bt} may worth renewed attention. Fortunately, the construction of Hyper-K has been started, which may act as the Kamioka site detector in an extended plan of the two-detector complex \cite{Abe:2015zbg,Abe:2016ero}.

\acknowledgments

One of the authors (I.M.S.) acknowledges travel support from the Colegio de F\'isica Fundamental e Interdisciplinaria de las Am\'ericas (COFI). Fermilab is operated by the Fermi Research Alliance, LLC under contract No.~DE-AC02-07CH11359 with the United States Department of Energy. 
The other (H.M.) thanks Center for Neutrino Physics, Department of Physics, Virginia Tech for hospitality and support.

\appendix 

\section{Explicit expressions of $F_{ij}$, $K_{ij}$ and $\Phi_{ij}$}
\label{sec:F-K-Phi-elements}

The explicit expressions of the elements $F_{ij}$, $K_{ij}$ and $\Phi_{ij}$ defined, respectively, in eqs.~\eqref{Fij-def}, ~\eqref{Kij-def} and \eqref{H1UV} are given as follows: 
\begin{eqnarray} 
&& F_{11} =
2 \tilde{\alpha}_{ee} \left( 1 - \frac{ \Delta_{a} }{ \Delta_{b} } \right), 
\nonumber \\
&& F_{12} =
c_{23} \tilde{\alpha}_{\mu e}^* - s_{23} \tilde{\alpha}_{\tau e}^*,
\nonumber \\
&& F_{13} = 
s_{23} \tilde{\alpha}_{\mu e}^* + c_{23} \tilde{\alpha}_{\tau e}^*, 
\nonumber \\
&& F_{21} =
c_{23} \tilde{\alpha}_{\mu e} - s_{23} \tilde{\alpha}_{\tau e} = \left( F_{12} \right)^*
\nonumber \\
&& F_{22} =
2 \left[ 
c_{23}^2 \tilde{\alpha}_{\mu \mu} + s_{23}^2 \tilde{\alpha}_{\tau \tau} - c_{23} s_{23} \mbox{Re} \left( \tilde{\alpha}_{\tau \mu} \right) 
\right], 
\nonumber \\
&& F_{23} =
\left[ 2 c_{23} s_{23} ( \tilde{\alpha}_{\mu \mu} - \tilde{\alpha}_{\tau \tau} ) + c_{23}^2 \tilde{\alpha}_{\tau \mu}^* - s_{23}^2 \tilde{\alpha}_{\tau \mu} \right], 
\nonumber \\
&& F_{31} =
s_{23} \tilde{\alpha}_{\mu e} + c_{23} \tilde{\alpha}_{\tau e} 
= \left( F_{13} \right)^*,
\nonumber \\
&& F_{32} =
\left[ 2 c_{23} s_{23} ( \tilde{\alpha}_{\mu \mu} - \tilde{\alpha}_{\tau \tau} ) + c_{23}^2 \tilde{\alpha}_{\tau \mu} - s_{23}^2 \tilde{\alpha}_{\tau \mu}^* \right] 
= \left( F_{23} \right)^*,
\nonumber \\
&& F_{33} =
2 \left[
s_{23}^2 \tilde{\alpha}_{\mu \mu} + c_{23}^2 \tilde{\alpha}_{\tau \tau} + c_{23} s_{23} \mbox{Re} \left( \tilde{\alpha}_{\tau \mu} \right) 
\right]. 
\label{Fij-elements}
\end{eqnarray}
\begin{eqnarray} 
K_{11} &=& 
2 c^2_{13} \tilde{\alpha}_{ee} \left( 1 - \frac{ \Delta_{a} }{ \Delta_{b} } \right) 
+ 2 s^2_{13} 
\left[
s_{23}^2 \tilde{\alpha}_{\mu \mu} + c_{23}^2 \tilde{\alpha}_{\tau \tau} + c_{23} s_{23} \mbox{Re} \left( \tilde{\alpha}_{\tau \mu} \right) 
\right], 
\nonumber \\
&-& 
2 c_{13} s_{13} 
\mbox{Re} \left( s_{23} \tilde{\alpha}_{\mu e} + c_{23} \tilde{\alpha}_{\tau e} \right) 
\nonumber \\
K_{12} &=& 
c_{13} \left( c_{23} \tilde{\alpha}_{\mu e}^* - s_{23} \tilde{\alpha}_{\tau e}^* \right) 
- s_{13} 
\left[ 2 c_{23} s_{23} ( \tilde{\alpha}_{\mu \mu} - \tilde{\alpha}_{\tau \tau} ) + c_{23}^2 \tilde{\alpha}_{\tau \mu} - s_{23}^2 \tilde{\alpha}_{\tau \mu}^* \right] 
= \left( K_{21} \right)^*, 
\nonumber \\
K_{13} &=& 
2 c_{13} s_{13} 
\left[
\tilde{\alpha}_{ee} \left( 1 - \frac{ \Delta_{a} }{ \Delta_{b} } \right) 
- \left( s_{23}^2 \tilde{\alpha}_{\mu \mu} + c_{23}^2 \tilde{\alpha}_{\tau \tau} \right) 
\right]
\nonumber \\
&+& 
c^2_{13} \left( s_{23} \tilde{\alpha}_{\mu e}^* + c_{23} \tilde{\alpha}_{\tau e}^* \right) 
- s^2_{13} \left( s_{23} \tilde{\alpha}_{\mu e} + c_{23} \tilde{\alpha}_{\tau e} \right) 
- 2 c_{23} s_{23} c_{13} s_{13} 
\mbox{Re} \left( \tilde{\alpha}_{\tau \mu} \right) 
= \left( K_{31} \right)^*, 
\nonumber \\
K_{22} &=& 
2 \left[ 
c_{23}^2 \tilde{\alpha}_{\mu \mu} + s_{23}^2 \tilde{\alpha}_{\tau \tau} - c_{23} s_{23} \mbox{Re} \left( \tilde{\alpha}_{\tau \mu} \right) 
\right], 
\nonumber \\
K_{23} &=& 
s_{13} \left( c_{23} \tilde{\alpha}_{\mu e} - s_{23} \tilde{\alpha}_{\tau e} \right) 
+ c_{13} 
\left[ 2 c_{23} s_{23} ( \tilde{\alpha}_{\mu \mu} - \tilde{\alpha}_{\tau \tau} ) + c_{23}^2 \tilde{\alpha}_{\tau \mu}^* - s_{23}^2 \tilde{\alpha}_{\tau \mu} \right] 
= \left( K_{32} \right)^*, 
\nonumber \\
K_{33} &=& 
2 s^2_{13} \tilde{\alpha}_{ee} \left( 1 - \frac{ \Delta_{a} }{ \Delta_{b} } \right) 
+ 2 c^2_{13} 
\left[
s_{23}^2 \tilde{\alpha}_{\mu \mu} + c_{23}^2 \tilde{\alpha}_{\tau \tau} + c_{23} s_{23} \mbox{Re} \left( \tilde{\alpha}_{\tau \mu} \right) 
\right] 
\nonumber \\
&+&
2 c_{13} s_{13} \mbox{Re} 
\left( s_{23} \tilde{\alpha}_{\mu e} + c_{23} \tilde{\alpha}_{\tau e} \right). 
\label{Kij-elements}
\end{eqnarray}
\begin{eqnarray} 
&& 
\Phi_{11} =
K_{11} 
+ 2 c_{\varphi}^2 s_{\varphi}^2 \left( K_{22} - K_{11}  \right)
- c_{\varphi}^2 s_{\varphi}^2 \left( K_{22} - K_{11}  \right) 
\left\{ e^{  i ( h_{2} - h_{1} ) x } + e^{ - i ( h_{2} - h_{1} ) x } \right\} 
\nonumber \\
&-&
c_{\varphi} s_{\varphi} \cos 2 \varphi 
\left( K_{12}  e^{ - i \delta} + K_{21} e^{ i \delta}  \right) 
\nonumber \\
&+& c_{\varphi} s_{\varphi} 
\left\{ 
- \left( s_{\varphi}^2 K_{12}  e^{ - i \delta} - c_{\varphi}^2 K_{21}  e^{ i \delta} \right) e^{  i ( h_{2} - h_{1} ) x } +
\left( c_{\varphi}^2 K_{12} e^{ - i \delta} - s_{\varphi}^2 K_{21} e^{ i \delta}  \right) e^{ - i ( h_{2} - h_{1} ) x } \right\}, 
\nonumber \\
&& \Phi_{12} =
e^{ i \delta} 
\biggl[
c_{\varphi} s_{\varphi} \cos 2 \varphi \left( K_{22} - K_{11} \right) 
+ c_{\varphi} s_{\varphi} \left\{ s_{\varphi}^2 e^{  i ( h_{2} - h_{1} ) x } - c_{\varphi}^2 e^{ - i ( h_{2} - h_{1} ) x } \right\} 
\left( K_{22} - K_{11} \right) 
\nonumber \\
&+&
2 c^2_{\varphi} s^2_{\varphi} \left( K_{12} e^{- i \delta} + K_{21} e^{ i \delta} \right) 
+ s_{\varphi}^2 \left( s_{\varphi}^2 K_{12} e^{- i \delta} - c_{\varphi}^2 K_{21} e^{ i \delta} \right) 
e^{  i ( h_{2} - h_{1} ) x } 
\nonumber \\
&+&
c_{\varphi}^2 \left( c_{\varphi}^2 K_{12} e^{ - i \delta} - s_{\varphi}^2 K_{21} e^{ i \delta} \right) 
e^{ - i ( h_{2} - h_{1} ) x } 
\biggr], 
\nonumber \\
&& \Phi_{13} = 
\left( s_{\varphi}^2 K_{13} + c_{\varphi} s_{\varphi} K_{23} e^{ i \delta}  \right) e^{ - i ( h_{3} - h_{2} ) x } 
+ \left( c_{\varphi}^2 K_{13} - c_{\varphi} s_{\varphi} K_{23} e^{ i \delta} \right) e^{ - i ( h_{3} - h_{1} ) x }, 
\nonumber
\end{eqnarray}
\begin{eqnarray} 
&& \Phi_{21} =
e^{- i \delta} 
\biggl\{
c_{\varphi} s_{\varphi} \cos 2 \varphi \left( K_{22} - K_{11} \right) 
- c_{\varphi} s_{\varphi} \left\{ c_{\varphi}^2 e^{  i ( h_{2} - h_{1} ) x } - s_{\varphi}^2 e^{ - i ( h_{2} - h_{1} ) x } \right\} 
\left( K_{22} - K_{11} \right) 
\nonumber \\
&+&
2 c^2_{\varphi} s^2_{\varphi} \left( K_{12} e^{- i \delta} + K_{21} e^{ i \delta} \right)
+
c^2_{\varphi} \left( c^2_{\varphi} K_{21} e^{ i \delta} - s^2_{\varphi} K_{12} e^{- i \delta} \right) e^{  i ( h_{2} - h_{1} ) x } 
\nonumber \\
&+&
s^2_{\varphi} \left( s^2_{\varphi} K_{21} e^{ i \delta} - c^2_{\varphi} K_{12} e^{- i \delta} \right)e^{ - i ( h_{2} - h_{1} ) x } 
\biggr\},
\nonumber \\
&& \Phi_{22} =
K_{22} - 2 c^2_{\varphi} s^2_{\varphi} \left( K_{22} - K_{11} \right)
+ c_{\varphi}^2 s_{\varphi}^2 \left( K_{22} - K_{11} \right) 
\left\{ e^{  i ( h_{2} - h_{1} ) x } + e^{ - i ( h_{2} - h_{1} ) x } \right\}   
\nonumber \\
&+&
c_{\varphi} s_{\varphi} 
\biggl[
\cos 2 \varphi \left( K_{12} e^{- i \delta} + K_{21} e^{ i \delta} \right) 
+ 
\left( s_{\varphi}^2 K_{12} e^{- i \delta} - c_{\varphi}^2 K_{21} e^{ i \delta} \right) e^{  i ( h_{2} - h_{1} ) x } 
\nonumber \\
&-&
\left( c_{\varphi}^2 K_{12} e^{- i \delta} - s_{\varphi}^2 K_{21} e^{ i \delta} \right) e^{ - i ( h_{2} - h_{1} ) x } 
\biggr],
\nonumber \\
&& \Phi_{23} =
e^{- i \delta} 
\biggl[
\left( c_{\varphi} s_{\varphi} K_{13} + c_{\varphi}^2 K_{23} e^{ i \delta} \right) e^{ - i ( h_{3} - h_{2} ) x } 
- 
\left( c_{\varphi} s_{\varphi} K_{13} - s_{\varphi}^2 K_{23} e^{ i \delta} \right) e^{ - i ( h_{3} - h_{1} ) x } 
\biggr],
\nonumber
\end{eqnarray}
\begin{eqnarray} 
&& \Phi_{31} =
\left( s_{\varphi}^2 K_{31} + c_{\varphi} s_{\varphi} K_{32} e^{- i \delta} \right) e^{ i ( h_{3} - h_{2} ) x } 
+ 
\left( c_{\varphi}^2 K_{31} - c_{\varphi} s_{\varphi} K_{32} e^{- i \delta} \right) e^{ i ( h_{3} - h_{1} ) x }, 
\nonumber \\
&& \Phi_{32} =
e^{ i \delta} 
\biggl[
\left( c_{\varphi} s_{\varphi} K_{31} + c_{\varphi}^2 K_{32} e^{- i \delta} \right) e^{ i ( h_{3} - h_{2} ) x } 
- 
\left( c_{\varphi} s_{\varphi} K_{31} - s_{\varphi}^2 K_{32} e^{- i \delta} \right) e^{ i ( h_{3} - h_{1} ) x } 
\biggr],
\nonumber \\
&& \Phi_{33} = K_{33}.
\label{Phi-ij-elements}
\end{eqnarray}

\section{The first order tilde basis unitary evolution $\tilde{S}$ matrix elements}
\label{sec:tilde-S-summary}

Here, we present the result of unitary $\tilde{S}$ matrix elements which come from first order UV parameter related part of the Hamiltonian.
\begin{eqnarray}
&& \tilde{S} (x)^{ \text{EV} }_{11} 
\nonumber \\
&=& 
\Delta_{b} \left\{ 
K_{11} + 2 c_{\varphi}^2 s_{\varphi}^2 \left( K_{22} - K_{11} \right) 
- c_{\varphi} s_{\varphi} \cos 2 \varphi \left( K_{12} e^{- i \delta} + K_{21} e^{ i \delta} \right) 
\right\} 
(-ix) \left( c_{\varphi}^2 e^{ - i h_{1} x } + s_{\varphi}^2 e^{ - i h_{2} x } \right) 
\nonumber \\
&+& 
c_{\varphi} s_{\varphi} 
\left\{
c_{\varphi} s_{\varphi} \cos 2 \varphi \left( K_{22} - K_{11} \right) 
+ 2 c^2_{\varphi} s^2_{\varphi} \left( K_{12} e^{- i \delta} + K_{21} e^{ i \delta} \right) 
\right\} 
(-ix) \left( e^{ - i h_{2} x } - e^{ - i h_{1} x } \right) 
\nonumber \\
&+&
c_{\varphi} s_{\varphi}
\left[
- 2 c_{\varphi} s_{\varphi} \left( K_{22} - K_{11} \right) 
+ \cos 2\varphi \left( K_{12} e^{- i \delta} + K_{21} e^{ i \delta} \right)
\right]
\frac{1}{ h_{2} - h_{1} } 
\left( e^{ - i h_{2} x } - e^{ - i h_{1} x } \right).
\end{eqnarray}
\begin{eqnarray} 
&& 
\tilde{S} (x)^{ \text{EV} }_{12} 
\nonumber \\
&=& 
e^{ i \delta}
\Delta_{b} 
\biggl[
\left\{ 
c_{\varphi} s_{\varphi} \cos 2 \varphi \left( K_{22} - K_{11} \right) 
+ 2 c^2_{\varphi} s^2_{\varphi} \left( K_{12} e^{- i \delta} + K_{21} e^{ i \delta} \right) 
\right\}
(-ix) \left( c_{\varphi}^2 e^{ - i h_{1} x } + s_{\varphi}^2 e^{ - i h_{2} x } \right) 
\nonumber \\
&+& 
c_{\varphi} s_{\varphi} 
\left\{
K_{22} - 2 c^2_{\varphi} s^2_{\varphi} \left( K_{22} - K_{11} \right)
+ c_{\varphi} s_{\varphi} \cos 2 \varphi \left( K_{12} e^{- i \delta} + K_{21} e^{ i \delta} \right) 
\right\} 
(-ix) \left( e^{ - i h_{2} x } - e^{ - i h_{1} x } \right)
\nonumber \\
&+& 
\biggl\{
- c_{\varphi} s_{\varphi} \cos 2 \varphi \left( K_{22} - K_{11} \right) 
+ \left\{ K_{12} e^{- i \delta} 
- 2 c^2_{\varphi} s^2_{\varphi} \left( K_{12} e^{- i \delta} + K_{21} e^{ i \delta} \right) \right\} 
\biggr\}
\frac{1}{ h_{2} - h_{1} } \left( e^{ - i h_{2} x } - e^{ - i h_{1} x } \right) 
\biggr].
\nonumber \\
\end{eqnarray}
\begin{eqnarray} 
&&
\tilde{S} (x)^{ \text{EV} }_{13} =  
\Delta_{b}
\biggl[
\left( s^2_{\varphi} K_{13} + c_{\varphi} s_{\varphi} K_{23} e^{ i \delta} \right) 
\frac{1}{ h_{3} - h_{2} } 
\left( e^{ - i h_{3} x } - e^{ - i h_{2} x } \right) 
\nonumber \\
&+& 
\left( c^2_{\varphi} K_{13} - c_{\varphi} s_{\varphi} K_{23} e^{ i \delta} \right) 
\frac{1}{ h_{3} - h_{1} } 
\left( e^{ - i h_{3} x } - e^{ - i h_{1} x } \right)
\biggr].
\end{eqnarray}
\begin{eqnarray} 
&& 
\tilde{S} (x)^{ \text{EV} }_{21} 
\nonumber \\
&=& 
e^{ - i \delta}
\Delta_{b} 
\biggl[
\left\{
c_{\varphi} s_{\varphi} \cos 2 \varphi \left( K_{22} - K_{11} \right) 
+ 2 c^2_{\varphi} s^2_{\varphi} \left( K_{12} e^{- i \delta} + K_{21} e^{ i \delta} \right) 
\right\} 
(-ix) \left( s_{\varphi}^2 e^{ - i h_{1} x } + c_{\varphi}^2 e^{ - i h_{2} x } \right) 
\nonumber \\
&+&
c_{\varphi} s_{\varphi} 
\left\{ 
K_{11} + 2 c_{\varphi}^2 s_{\varphi}^2 \left( K_{22} - K_{11} \right) 
- c_{\varphi} s_{\varphi} \cos 2 \varphi \left( K_{12} e^{- i \delta} + K_{21} e^{ i \delta} \right) 
\right\}
(-ix) \left( e^{ - i h_{2} x } - e^{ - i h_{1} x } \right) 
\nonumber \\
&+&
\biggl\{
- c_{\varphi} s_{\varphi} \cos 2 \varphi \left( K_{22} - K_{11} \right) 
+ \left\{ 
K_{21} e^{ i \delta} - 2 c_{\varphi}^2 s^2_{\varphi} \left( K_{12} e^{- i \delta} + K_{21} e^{ i \delta} \right) 
\right\} 
\biggr\}
\frac{1}{ h_{2} - h_{1} }
\left( e^{ - i h_{2} x } - e^{ - i h_{1} x } \right) 
\biggr].
\nonumber \\
\end{eqnarray}
\begin{eqnarray} 
&& \tilde{S} (x)^{ \text{EV} }_{22} 
\nonumber \\
&=& 
\Delta_{b} \biggl[
c_{\varphi} s_{\varphi} 
\left\{ 
c_{\varphi} s_{\varphi} \cos 2 \varphi \left( K_{22} - K_{11} \right) 
+ 2 c^2_{\varphi} s^2_{\varphi} \left( K_{12} e^{- i \delta} + K_{21} e^{ i \delta} \right) 
\right\}
(-ix) \left( e^{ - i h_{2} x } - e^{ - i h_{1} x } \right) 
\nonumber \\
&+& 
\left\{
K_{22} - 2 c^2_{\varphi} s^2_{\varphi} \left( K_{22} - K_{11} \right)
+ c_{\varphi} s_{\varphi} \cos 2 \varphi \left( K_{12} e^{- i \delta} + K_{21} e^{ i \delta} \right) 
\right\} 
(-ix) \left( s_{\varphi}^2 e^{ - i h_{1} x } + c_{\varphi}^2 e^{ - i h_{2} x } \right) 
\nonumber \\
&+&
\biggl\{
2 c^2_{\varphi} s^2_{\varphi} 
\left( K_{22} - K_{11} \right)
- c_{\varphi} s_{\varphi} \cos 2\varphi 
\left( K_{12} e^{- i \delta} + K_{21} e^{ i \delta} \right) 
\biggr\}
\frac{1}{ h_{2} - h_{1} }
\left( e^{ - i h_{2} x } - e^{ - i h_{1} x } \right) 
\biggr].
\end{eqnarray} 
\begin{eqnarray} 
&& 
\tilde{S} (x)^{ \text{EV} }_{23} = 
e^{ - i \delta}
\Delta_{b} 
\biggl\{
\left[ c_{\varphi} s_{\varphi} K_{13} + c^2_{\varphi} K_{23} e^{ i \delta} \right]
\frac{1}{ h_{3} - h_{2} } 
\left( e^{ - i h_{3} x } - e^{ - i h_{2} x } \right)
\nonumber \\
&-& 
\left[ c_{\varphi} s_{\varphi} K_{13} - s^2_{\varphi} K_{23} e^{ i \delta} \right]
\frac{1}{ h_{3} - h_{1} } 
\left( e^{ - i h_{3} x } - e^{ - i h_{1} x } \right)
\biggr\}.
\end{eqnarray}
\begin{eqnarray} 
&&
\tilde{S} (x)^{ \text{EV} }_{31} =  
\Delta_{b} 
\biggl[
\left( s_{\varphi}^2 K_{31} + c_{\varphi} s_{\varphi} K_{32} e^{ - i \delta} \right) 
\frac{1}{ h_{3} - h_{2} } 
\left( e^{ - i h_{3} x } - e^{ - i h_{2} x } \right) 
\nonumber \\
&+&
\left( c_{\varphi}^2 K_{31} - c_{\varphi} s_{\varphi} K_{32} e^{ - i \delta} \right) 
\frac{1}{ h_{3} - h_{1} }
\left( e^{ - i h_{3} x } - e^{ - i h_{1} x } \right)
\biggr].
\end{eqnarray}
\begin{eqnarray} 
&& 
\tilde{S} (x)^{ \text{EV} }_{32} = 
e^{ i \delta} 
\Delta_{b} 
\biggl[
\left( c_{\varphi} s_{\varphi} K_{31} + c_{\varphi}^2 K_{32} e^{ - i \delta} \right) 
\frac{1}{ h_{3} - h_{2} } 
\left\{ e^{ - i h_{3} x } - e^{ - i h_{2} x } \right\} 
\nonumber \\
&-&
\left( c_{\varphi} s_{\varphi} K_{31} - s_{\varphi}^2 K_{32} e^{ - i \delta} \right) 
\frac{1}{ h_{3} - h_{1} } 
\left\{ e^{ - i h_{3} x } - e^{ - i h_{1} x } \right\} 
\biggr].
\end{eqnarray}
\begin{eqnarray} 
&& 
\tilde{S} (x)^{ \text{EV} }_{33} = 
\left( -ix \Delta_{b} \right) e^{ - i h_{3} x } K_{33}.
\end{eqnarray}

\section{The zeroth-order $\nu$SM S matrix elements}
\label{sec:SM-S-0th}

Here, we give the expressions of the flavor basis $S$ matrix elements of $\nu$SM part at zeroth order. The superscript ``$\nu$SM'' is abbreviated. 
\begin{eqnarray} 
S_{ee}^{(0)} &=& 
c^2_{13} \left( c^2_{\varphi} e^{ - i h_{1} x } + s^2_{\varphi} e^{ - i h_{2} x } \right) 
+ s^2_{13} e^{ - i h_{3} x }, 
\nonumber \\
S_{e \mu}^{(0)} &=& 
c_{23} c_{13} c_{\varphi} s_{\varphi} e^{ i \delta} 
\left( e^{ - i h_{2} x } - e^{ - i h_{1} x } \right) 
- s_{23} c_{13} s_{13} 
\left( c^2_{\varphi} e^{ - i h_{1} x } + s^2_{\varphi} e^{ - i h_{2} x } - e^{ - i h_{3} x } \right), 
\nonumber \\ 
S_{e \tau}^{(0)} &=& 
- c_{23} c_{13} s_{13} 
\left( c^2_{\varphi} e^{ - i h_{1} x } + s^2_{\varphi} e^{ - i h_{2} x } - e^{ - i h_{3} x } \right) 
- s_{23} c_{13} c_{\varphi} s_{\varphi} e^{ i \delta} 
\left( e^{ - i h_{2} x } - e^{ - i h_{1} x } \right), 
\nonumber \\
S_{\mu e}^{(0)} &=& 
c_{23} c_{13} c_{\varphi} s_{\varphi} e^{ - i \delta}
\left( e^{ - i h_{2} x } - e^{ - i h_{1} x } \right) 
- s_{23} c_{13} s_{13} 
\left( c^2_{\varphi} e^{ - i h_{1} x } + s^2_{\varphi} e^{ - i h_{2} x } - e^{ - i h_{3} x } \right) 
= S_{e \mu} (- \delta), 
\nonumber \\
S_{\mu \mu}^{(0)} &=& 
c^2_{23} 
\left( s^2_{\varphi} e^{ - i h_{1} x } + c^2_{\varphi} e^{ - i h_{2} x } \right) 
+ s^2_{23} 
\left\{ s^2_{13} 
\left( c^2_{\varphi} e^{ - i h_{1} x } + s^2_{\varphi} e^{ - i h_{2} x } \right) 
+ c^2_{13} e^{ - i h_{3} x } 
\right\} 
\nonumber \\
&-&
2 c_{23} s_{23} s_{13} c_{\varphi} s_{\varphi} \cos \delta 
\left( e^{ - i h_{2} x } - e^{ - i h_{1} x } \right), 
\nonumber \\
S_{\mu \tau}^{(0)} &=& 
s_{13} c_{\varphi} s_{\varphi} 
\left( s^2_{23} e^{ i \delta} - c^2_{23} e^{ - i \delta} \right) 
\left( e^{ - i h_{2} x } - e^{ - i h_{1} x } \right) 
\nonumber \\
&+&
c_{23} s_{23} \left[
s^2_{13} 
\left( c^2_{\varphi} e^{ - i h_{1} x } + s^2_{\varphi} e^{ - i h_{2} x } \right) 
+ c^2_{13} e^{ - i h_{3} x } 
- \left( s^2_{\varphi} e^{ - i h_{1} x } + c^2_{\varphi} e^{ - i h_{2} x } \right) 
\right], 
\nonumber \\ 
S_{\tau e}^{(0)} &=& 
- c_{23} c_{13} s_{13} 
\left( c^2_{\varphi} e^{ - i h_{1} x } + s^2_{\varphi} e^{ - i h_{2} x } - e^{ - i h_{3} x } \right)
- s_{23} c_{13} c_{\varphi} s_{\varphi} e^{ - i \delta}
\left( e^{ - i h_{2} x } - e^{ - i h_{1} x } \right) 
= S_{e \tau} (- \delta), 
\nonumber \\ 
S_{\tau \mu}^{(0)} &=& 
s_{13} c_{\varphi} s_{\varphi} 
\left( s^2_{23} e^{ - i \delta} - c^2_{23} e^{ i \delta} \right) 
\left( e^{ - i h_{2} x } - e^{ - i h_{1} x } \right) 
\nonumber \\
&+&
c_{23} s_{23} 
\left[ s^2_{13} 
\left( c^2_{\varphi} e^{ - i h_{1} x } + s^2_{\varphi} e^{ - i h_{2} x } \right) 
+ c^2_{13} e^{ - i h_{3} x } 
- \left( s^2_{\varphi} e^{ - i h_{1} x } + c^2_{\varphi} e^{ - i h_{2} x } \right)
\right] = S_{\mu \tau} (- \delta),
\nonumber \\
S_{\tau \tau}^{(0)} &=& 
s^2_{23} 
\left( s^2_{\varphi} e^{ - i h_{1} x } + c^2_{\varphi} e^{ - i h_{2} x } \right) 
+ c^2_{23} 
\left\{ s^2_{13} 
\left( c^2_{\varphi} e^{ - i h_{1} x } + s^2_{\varphi} e^{ - i h_{2} x } \right) 
+ c^2_{13} e^{ - i h_{3} x } 
\right\}
\nonumber \\
&+&
2 c_{23} s_{23} 
s_{13} c_{\varphi} s_{\varphi} \cos \delta 
\left( e^{ - i h_{2} x } - e^{ - i h_{1} x } \right).
\label{S-elements}
\end{eqnarray}

\section{The neutrino oscillation probability in the $\nu_{\mu} \rightarrow \nu_{e}$ and the other channels} 
\label{sec:Pmue-Pmutau}

In this appendix, we give the expressions of the rest of the terms of $P(\nu_{\mu} \rightarrow \nu_{e})_{ \text{ EV } }^{(1)}$ which are not presented in section~\ref{sec:general-formula-P}. We also briefly mention how to compute the neutrino oscillation probability in the $\nu_{\mu} - \nu_{\tau}$ sector.

\subsection{The neutrino oscillation probability in the $\nu_{\mu} \rightarrow \nu_{e}$ channel: Rest of the unitary evolution part} 
\label{sec:Pmue-rest}

We recapitulate the definition \eqref{P-mue-four-terms} of the four terms of $P(\nu_{\mu} \rightarrow \nu_{e})_{ \text{ EV } }^{(1)}$ again for convenience:
\begin{eqnarray} 
&& 
P(\nu_{\mu} \rightarrow \nu_{e})_{ \text{ EV} }^{(1)} =
P(\nu_{\mu} \rightarrow \nu_{e})_{ \text{ EV} }^{(1)} \vert_{ \text{D-OD} } 
\nonumber \\
&+&
P(\nu_{\mu} \rightarrow \nu_{e})_{ \text{int-UV} }^{(1)} \vert_{ \text{OD1} } 
+ P(\nu_{\mu} \rightarrow \nu_{e})_{ \text{int-UV} }^{(1)} \vert_{ \text{OD2} } 
+ P(\nu_{\mu} \rightarrow \nu_{e})_{ \text{int-UV} }^{(1)} \vert_{ \text{OD3} }, 
\label{P-mue-four-terms2}
\end{eqnarray}
where the subscripts ``D'' and ``OD'' refer to the diagonal and the off-diagonal $K_{ij}$ variables. 

The first term of eq.~\eqref{P-mue-four-terms2} is given in eq.~\eqref{P-mue-D-OD}. Now, we present the remaining three ``OD'' terms:
\begin{eqnarray} 
&& 
P(\nu_{\mu} \rightarrow \nu_{e})_{ \text{ EV} }^{(1)} \vert_{ \text{OD1} }  
\nonumber \\
&=& 
4 c_{23} c^2_{13} 
\mbox{Re} \left( K_{12} e^{- i \delta} \right) 
\nonumber \\
&\times& 
\frac{ \Delta_{b} }{ h_{2} - h_{1} } 
\biggl[
- s_{23} \cos \delta 
\left\{ \sin^2 \frac{ ( h_{3} - h_{2} ) x }{2}
- \sin^2 \frac{ ( h_{3} - h_{1} ) x }{2} 
- \cos 2\varphi \sin^2 \frac{ ( h_{2} - h_{1} ) x }{2} \right\} 
\nonumber \\
&+& 
c_{23} \sin 2 \varphi 
\sin^2 \frac{ ( h_{2} - h_{1} ) x }{2}
+ 2 s_{23} \sin \delta 
\sin \frac{( h_{3} - h_{2} ) x}{2} 
\sin \frac{( h_{2} - h_{1} ) x}{2} 
\sin \frac{( h_{1} - h_{3} ) x}{2} 
\biggr]
\nonumber \\
&+& 
4 c_{23} s_{23} c^2_{13} 
\mbox{Im} \left( K_{12} e^{- i \delta} \right)  
\nonumber \\
&\times&
\frac{ \Delta_{b} }{ h_{2} - h_{1} }
\biggl[ 
\sin \delta 
\left\{ \sin^2 \frac{ ( h_{3} - h_{2} ) x }{2}
- \sin^2 \frac{ ( h_{3} - h_{1} ) x }{2} 
- \cos 2\varphi \sin^2 \frac{ ( h_{2} - h_{1} ) x }{2} \right\} 
\nonumber \\
&+& 
2 \cos \delta 
\sin \frac{( h_{3} - h_{2} ) x}{2} 
\sin \frac{( h_{2} - h_{1} ) x}{2} 
\sin \frac{( h_{1} - h_{3} ) x}{2} 
\biggr].
\label{P-mue-OD1}
\end{eqnarray}
%
\begin{eqnarray} 
&& 
P(\nu_{\mu} \rightarrow \nu_{e})_{ \text{ EV} }^{(1)} \vert_{ \text{OD2} }  
\nonumber \\
&=& 
- 4 c^2_{23} c_{13} s_{13} c_{\varphi} s_{\varphi} 
\mbox{Re} \left( c_{\varphi} s_{\varphi} K_{31} + c_{\varphi}^2 K_{32} e^{ - i \delta} \right) 
\nonumber \\
&\times& 
\frac{ \Delta_{b} }{ h_{3} - h_{2} } 
\biggl\{
\sin^2 \frac{ ( h_{3} - h_{2} ) x }{2} 
- \sin^2 \frac{ ( h_{3} - h_{1} ) x }{2} 
+ \sin^2 \frac{ ( h_{2} - h_{1} ) x }{2} 
\biggr\} 
\nonumber \\
&+& 
4 c^2_{23} c_{13} s_{13} c_{\varphi} s_{\varphi} 
\mbox{Re} \left( c_{\varphi} s_{\varphi} K_{31} - s_{\varphi}^2 K_{32} e^{ - i \delta} \right) 
\nonumber \\
&\times& 
\frac{ \Delta_{b} }{ h_{3} - h_{1} } 
\biggl\{
\sin^2 \frac{ ( h_{3} - h_{2} ) x }{2} 
- \sin^2 \frac{ ( h_{3} - h_{1} ) x }{2} 
- \sin^2 \frac{ ( h_{2} - h_{1} ) x }{2} 
\biggr\} 
\nonumber \\
&+& 
4 c_{23} s_{23} c_{13} s^2_{13} 
\biggl\{ 
\cos \delta 
\mbox{Re} \left( c_{\varphi} s_{\varphi} K_{31} + c_{\varphi}^2 K_{32} e^{ - i \delta} \right) 
- \sin \delta 
\mbox{Im} \left( c_{\varphi} s_{\varphi} K_{31} + c_{\varphi}^2 K_{32} e^{ - i \delta} \right) 
\biggr\}
\nonumber \\
&\times&
\frac{ \Delta_{b} }{ h_{3} - h_{2} } 
\biggl[
c^2_{\varphi} 
\left\{ \sin^2 \frac{ ( h_{3} - h_{1} ) x }{2} - \sin^2 \frac{ ( h_{2} - h_{1} ) x }{2} \right\}
+ ( 1 + s^2_{\varphi} ) 
\sin^2 \frac{ ( h_{3} - h_{2} ) x }{2} 
\biggr] 
\nonumber \\
&-& 
4 c_{23} s_{23} c_{13} s^2_{13} 
\biggl\{
\cos \delta 
\mbox{Re} \left( c_{\varphi} s_{\varphi} K_{31} - s_{\varphi}^2 K_{32} e^{ - i \delta} \right) 
- \sin \delta 
\mbox{Im} \left( c_{\varphi} s_{\varphi} K_{31} - s_{\varphi}^2 K_{32} e^{ - i \delta} \right) 
\biggr\} 
\nonumber \\
&\times&
\frac{ \Delta_{b} }{ h_{3} - h_{1} } 
\biggl[
s^2_{\varphi} 
\left\{ \sin^2 \frac{ ( h_{3} - h_{2} ) x }{2} - \sin^2 \frac{ ( h_{2} - h_{1} ) x }{2} \right\} 
+ (1 + c^2_{\varphi} ) \sin^2 \frac{ ( h_{3} - h_{1} ) x }{2} 
\biggr]
\nonumber \\
&-& 
8 c_{23} c_{13} s_{13} 
\biggl[
\left(
s_{23} s_{13} c^2_{\varphi} \cos \delta 
+ c_{23} c_{\varphi} s_{\varphi} \right) 
\mbox{Im} 
\left( c_{\varphi} s_{\varphi} K_{31} + c_{\varphi}^2 K_{32} e^{ - i \delta} \right) 
\nonumber \\
&+& 
s_{23} s_{13} c^2_{\varphi} \sin \delta 
\mbox{Re} \left( c_{\varphi} s_{\varphi} K_{31} + c_{\varphi}^2 K_{32} e^{ - i \delta} \right) 
\biggr]
\frac{ \Delta_{b} }{ h_{3} - h_{2} } 
\sin \frac{( h_{3} - h_{2} ) x}{2} 
\sin \frac{( h_{2} - h_{1} ) x}{2} 
\sin \frac{( h_{1} - h_{3} ) x}{2} 
\nonumber \\
&-& 
8 c_{23} c_{13} s_{13} 
\biggl[ 
\left( 
s_{23} s_{13} s^2_{\varphi} \cos \delta 
- c_{23} c_{\varphi} s_{\varphi} 
\right)
\mbox{Im} \left( c_{\varphi} s_{\varphi} K_{31} - s_{\varphi}^2 K_{32} e^{ - i \delta} \right) 
\nonumber \\
&+& 
s_{23} s_{13} s^2_{\varphi} \sin \delta 
\mbox{Re} \left( c_{\varphi} s_{\varphi} K_{31} - s_{\varphi}^2 K_{32} e^{ - i \delta} \right) 
\biggr]
\frac{ \Delta_{b} }{ h_{3} - h_{1} } 
\sin \frac{( h_{3} - h_{2} ) x}{2} 
\sin \frac{( h_{2} - h_{1} ) x}{2} 
\sin \frac{( h_{1} - h_{3} ) x}{2}. 
\nonumber \\
\label{P-mue-OD2}
\end{eqnarray}
%
\begin{eqnarray} 
&& 
P(\nu_{\mu} \rightarrow \nu_{e})_{ \text{ EV} }^{(1)} \vert_{ \text{OD3} }  
\nonumber \\
&=& 
- 4 c_{23} s_{23} c_{13} c_{\varphi} s_{\varphi} 
\biggl\{ \cos \delta 
\mbox{Re} \left[ s^2_{\varphi} 
\left( c^2_{13} K_{13} - s^2_{13} K_{31} \right) 
+ c_{\varphi} s_{\varphi} 
\left( c^2_{13} K_{23} e^{ i \delta} - s^2_{13} K_{32} e^{ - i \delta} \right) \right] 
\nonumber \\
&+&
\sin \delta 
\mbox{Im} \left[ s^2_{\varphi} 
\left( c^2_{13} K_{13} - s^2_{13} K_{31} \right) 
+ c_{\varphi} s_{\varphi} 
\left( c^2_{13} K_{23} e^{ i \delta} - s^2_{13} K_{32} e^{ - i \delta} \right) \right] 
\biggr\}
\nonumber \\
&\times&
\frac{ \Delta_{b} }{ h_{3} - h_{2} } 
\biggl\{
\sin^2 \frac{ ( h_{3} - h_{2} ) x }{2} 
- \sin^2 \frac{ ( h_{3} - h_{1} ) x }{2} 
+ \sin^2 \frac{ ( h_{2} - h_{1} ) x }{2} 
\biggr\} 
\nonumber \\
&-& 
4 c_{23} s_{23} c_{13} c_{\varphi} s_{\varphi} 
\biggl\{ \cos \delta 
\mbox{Re} \left[
c^2_{\varphi} 
\left( c^2_{13} K_{13} - s^2_{13} K_{31} 
\right) 
- c_{\varphi} s_{\varphi} 
\left( c^2_{13} K_{23} e^{ i \delta} - s^2_{13} K_{32} e^{ - i \delta} 
\right) \right] 
\nonumber \\
&+& 
\sin \delta 
\mbox{Im} \left[
c^2_{\varphi} 
\left( c^2_{13} K_{13} - s^2_{13} K_{31} 
\right) 
- c_{\varphi} s_{\varphi} 
\left( c^2_{13} K_{23} e^{ i \delta} - s^2_{13} K_{32} e^{ - i \delta} 
\right) \right] 
\biggr\} 
\nonumber \\
&\times& 
\frac{ \Delta_{b} }{ h_{3} - h_{1} }
\biggl\{
\sin^2 \frac{ ( h_{3} - h_{2} ) x }{2} 
- \sin^2 \frac{ ( h_{3} - h_{1} ) x }{2} 
- \sin^2 \frac{ ( h_{2} - h_{1} ) x }{2} 
\biggr\} 
\nonumber \\
&-& 
4 s^2_{23} c_{13} s_{13} 
\mbox{Re} \left[ s^2_{\varphi} 
\left( c^2_{13} K_{13} - s^2_{13} K_{31} \right) 
+ c_{\varphi} s_{\varphi} 
\left( c^2_{13} K_{23} e^{ i \delta} - s^2_{13} K_{32} e^{ - i \delta} \right) 
\right] 
\nonumber \\
&\times&
\frac{ \Delta_{b} }{ h_{3} - h_{2} } 
\biggl[
c^2_{\varphi} 
\left\{ - \sin^2 \frac{ ( h_{3} - h_{1} ) x }{2} 
+ \sin^2 \frac{ ( h_{2} - h_{1} ) x }{2}
\right\}
- ( 1 + s^2_{\varphi} ) 
\sin^2 \frac{ ( h_{3} - h_{2} ) x }{2} 
\biggr]
\nonumber \\
&-& 
4 s^2_{23} c_{13} s_{13} 
\mbox{Re} 
\left[
c^2_{\varphi} 
\left( c^2_{13} K_{13} - s^2_{13} K_{31} 
\right) 
- c_{\varphi} s_{\varphi} 
\left( c^2_{13} K_{23} e^{ i \delta} 
- s^2_{13} K_{32} e^{ - i \delta} \right) 
\right] 
\nonumber \\
&\times&
\frac{ \Delta_{b} }{ h_{3} - h_{1} } 
\biggl[
s^2_{\varphi} 
\left\{ - \sin^2 \frac{ ( h_{3} - h_{2} ) x }{2} + \sin^2 \frac{ ( h_{2} - h_{1} ) x }{2} \right\} 
- (1 + c^2_{\varphi} ) \sin^2 \frac{ ( h_{3} - h_{1} ) x }{2} 
\biggr]
\nonumber \\
&+& 
8 s_{23} c_{13} 
\biggl\{
\left(
c_{23} c_{\varphi} s_{\varphi} \cos \delta - s_{23} s_{13} c^2_{\varphi} 
\right)
\mbox{Im} \left[ s^2_{\varphi} \left( c^2_{13} K_{13} - s^2_{13} K_{31} \right) + c_{\varphi} s_{\varphi} \left( c^2_{13} K_{23} e^{ i \delta} - s^2_{13} K_{32} e^{ - i \delta} \right) \right]
\nonumber \\
&+& 
c_{23} c_{\varphi} s_{\varphi} 
\sin \delta \mbox{Re} \left[ s^2_{\varphi} 
\left( c^2_{13} K_{13} - s^2_{13} K_{31} \right) 
+ c_{\varphi} s_{\varphi} 
\left( c^2_{13} K_{23} e^{ i \delta} - s^2_{13} K_{32} e^{ - i \delta} \right) \right] 
\biggr\}
\nonumber \\
&\times&
\frac{ \Delta_{b} }{ h_{3} - h_{2} } 
\sin \frac{( h_{3} - h_{2} ) x}{2} 
\sin \frac{( h_{2} - h_{1} ) x}{2} 
\sin \frac{( h_{1} - h_{3} ) x}{2} 
\nonumber \\
&+& 
8 s_{23} c_{13} 
\biggl\{
\left( 
c_{23} c_{\varphi} s_{\varphi} \cos \delta 
+ s_{23} s_{13} s^2_{\varphi} 
\right) 
\mbox{Im} \left[
c^2_{\varphi} \left( c^2_{13} K_{13} - s^2_{13} K_{31} \right) 
- c_{\varphi} s_{\varphi} 
\left( c^2_{13} K_{23} e^{ i \delta} - s^2_{13} K_{32} e^{ - i \delta} 
\right) \right] 
\nonumber \\
&+& 
c_{23} c_{\varphi} s_{\varphi} \sin \delta 
\mbox{Re} \left[
c^2_{\varphi} 
\left( c^2_{13} K_{13} - s^2_{13} K_{31} \right) 
- c_{\varphi} s_{\varphi} 
\left( c^2_{13} K_{23} e^{ i \delta} - s^2_{13} K_{32} e^{ - i \delta} 
\right) \right] 
\biggr\}
\nonumber \\
&\times&
\frac{ \Delta_{b} }{ h_{3} - h_{1} } 
\sin \frac{( h_{3} - h_{2} ) x}{2} 
\sin \frac{( h_{2} - h_{1} ) x}{2} 
\sin \frac{( h_{1} - h_{3} ) x}{2}.
\label{P-mue-OD3}
\end{eqnarray}

\subsection{The neutrino oscillation probability in the $\nu_{\mu} - \nu_{\tau}$ sector} 
\label{sec:nu-mu-nu-tau}

We refrain from explicit computation of the oscillation probabilities in $\nu_{\mu} - \nu_{\tau}$ sector. The reason is that the expression is too lengthy and not particularly structure revealing beyond that we have discussed in this paper with the explicit expressions of $P(\nu_{\mu} \rightarrow \nu_{e})^{(1)}$. If one still needs these expressions of the probabilities, one can readily calculate them following the instruction given in section~\ref{sec:general-formula-P}. For general readers, we recommend to use e.g., mathematica software to perform computation of the oscillation probability using \eqref{P-three-types} due to its complexity even at first order. Also, we note again that the exact formula \cite{Fong:2017gke} exists to fulfill needs for accurate numerical computation.

\end{document}